\documentclass[fleqn,usenatbib]{mnras}

\usepackage{newtxtext,newtxmath} 
\usepackage{hyperref}
\usepackage{orcidlink}
\usepackage{bm}
\usepackage{stfloats}

\usepackage[T1]{fontenc}

\DeclareRobustCommand{\VAN}[3]{#2}
\let\VANthebibliography\thebibliography
\def\thebibliography{\DeclareRobustCommand{\VAN}[3]{##3}\VANthebibliography}

\usepackage{graphicx}	
\usepackage{amsmath}	
\usepackage{float} 
\usepackage{enumitem}
\usepackage{subcaption}

\title[Hot Jupiters with tilted magnetic dipoles]{Exploring the impact of tilted magnetic dipoles on the atmospheric dynamics of hot Jupiters: towards an improved magnetohydrodynamic framework}

\author[J.S. Fecanin et al.]{
James S. Fecanin\orcidlink{0009-0000-5641-6644},$^{1, 2}$\thanks{E-mail: j.s.fecanin@sms.ed.ac.uk}
Hayley Q. Beltz\orcidlink{0000-0002-6980-052X},$^{3,4}$
John R. T. Allen\orcidlink{0009-0006-1653-3524}$^{2}$
and Thaddeus D. Komacek\orcidlink{0000-0002-9258-5311}$^{2,4}$
\\
$^{1}$ Institute for Astronomy, University of Edinburgh, Royal Observatory, Edinburgh EH9 3HJ, UK \\
$^{2}$Department of Physics (Atmospheric, Oceanic and Planetary Physics), University of Oxford, Oxford OX1 3PU, UK\\
$^{3}$Department of Physics and Astronomy, University of Kansas, Lawrence, KS 66045, USA \\
$^4$Department of Astronomy, University of Maryland, College Park, MD 20742, USA
}

\date{Accepted 2026 July 21. Received 2026 July 11; in original form 2026 April 28}

\pubyear{\the\year{}}

\begin{document}
\label{firstpage}
\pagerange{\pageref{firstpage}--\pageref{lastpage}}
\maketitle

\begin{abstract}
The atmospheres of hot Jupiters lie in a dynamical regime without a solar system analogue. The strongly irradiated daysides reach temperatures sufficiently hot for substantial thermal ionization of atmospheric species, resulting in flows that can interact with the planetary magnetic field. These magnetic effects can significantly impact wind speeds, atmospheric temperatures, and large-scale circulation patterns. Previous work combining 3D atmospheric models and magnetic prescriptions has shown the impact of magnetic effects on temperature and velocity profiles are dependent on local atmospheric properties as well as the set of assumptions employed by the magnetic prescription. In this work, we examine a commonly employed magnetic model---a perfectly aligned dipole---in 3D General Circulation Models (GCMs) and extend this framework to allow for tilting of the deep-seated internal magnetic dipole field relative to the axis of rotation. We find that the inclusion of a tilted dipole introduces pronounced north–south asymmetries into the temperature profile leading to latitudinally shifted hotspots and deflection of winds that would otherwise be axially symmetric. We additionally simulate \textit{James Webb Space Telescope}/NIRSpec phase curves. We find that the strength of the magnetic field has the most significant effect on the simulated phase curves, with stronger magnetic fields increasing the amplitude of the phase curve and reducing the hotspot offset. Our model can provide qualitative insight into how the magnetic dipole strength or orientation may influence the large scale atmospheric dynamics and represents one of the most sophisticated incorporations of magnetic effects in GCMs for hot Jupiter atmospheres to date. 

\end{abstract}

\begin{keywords}
{exoplanets – planets and satellites: atmospheres – planets and satellites: gaseous planets – planets and satellites: magnetic fields. }
\end{keywords}



\section{Introduction}
\label{sec:intro}

Hot Jupiters are one of the most well-documented classes of exoplanets due to their observationally advantageous characteristics: as massive gas giants orbiting extremely close to their host stars \citep{Parmentier_2018_Phase_Curves}, they yield high signal-to-noise ratios and favourable planet-to-star flux ratios. Due to their proximity to their host star, hot Jupiters receive a stellar flux of tens of thousands of times greater than that received by our Jupiter \citep{Showman_2020_Review, HJ_review_origins_properties_Tad}, have orbital periods on the order of days, and are expected to be tidally locked---resulting in extreme day-to-night temperature gradients. As a result, their atmospheric dynamics are heavily characterized by the advection of the hot dayside air to the far cooler nightside.

Since the discovery of the first hot Jupiter \citep{first_hotJupiter}, these planets’ atmospheres have been extensively studied through both observations and theoretical modelling. These works have produced a broadly consistent picture of the main dynamical features of their atmospheres. Multidimensional general circulation models (GCMs) find that at pressures $\gtrsim$10 mbar, a dominant dynamical feature emerges: a strong superrotating (west-to-east) equatorial jet \citep{Showman_2011_EastWest_Equatorial_Jet, Heng_2012, Rauscher_Menou_2013, Tsai_2014, Hammond_2018}. This zonal jet is associated with the corresponding advection of the planetary hotspot eastward of the sub-stellar point \citep{Showman_2020_Review, HJ_review_origins_properties_Tad}. This hotspot offset is a well-established observational feature seen in phase curve observations of many of these planets \citep{Bell_2021, May_2022}. The strength of the jet depends on many factors, including planetary rotation rate, strength of stellar irradiation, and, of particular relevance to this work, the presence of a planetary magnetic field and Lorentz forces resulting in atmospheric drag. Numerical models that include a stronger drag, regardless of the underlying mechanism, generally predict a reduction in the strength of the equatorial jet, and consequently a smaller hotspot offset \citep{Rauscher_Menou_2013, Komacek_2016_constant_drag, Beltz_2022}.

Even with these promising features, significant discrepancies remain between observations and theoretical predictions of the characteristics of hot Jupiter atmospheres. Notably, recent eclipse-mapping observations have revealed possible northward or southward hotspot offsets \citep{challener2024latitudinalasymmetrydaysideatmosphere, hammond2024twodimensionaleclipsemappinghot}, which have not yet been satisfactorily explained by existing simulations. Additionally, several hot Jupiters have almost negligible eastward hotspot offsets, or even westward hotspot offsets, rather than the expected eastern offset \citep{Keating_2020,Bell_2021,May_2022}, potentially indicating stronger damping of the atmospheric flows \citep{Hindle_2021}. Many hot Jupiters are also observed to have radii larger than predicted by interior and atmospheric models \citep{guillot1996}. This discrepancy, commonly referred to as the `radius anomaly’ \citep{HJ_review_origins_properties_Tad}, further suggests that key physical processes may be missing or inadequately modelled in current hot Jupiter simulations.

One such mechanism is magnetism, which is the focus of this work. The close proximity of hot Jupiters to their host stars ensures that their dayside temperatures are sufficiently high for atmospheric species with low ionization potentials (such as K, Na) to thermally ionize in bulk \citep{ionisation}. As a result, the atmospheric dynamics in these regimes fall within the domain of magnetohydrodynamics (MHD), wherein partially ionized flows interact non-negligibly with the magnetic field of the planet, altering the resulting atmospheric circulation. \cite{Perna2010a, Perna2010b} were among the first to consider the impact of magnetic forcing, particularly Lorentz forces and the associated Ohmic dissipation, on the atmospheric dynamics and thermal structure of hot Jupiters by parametrizing magnetic effects as a dissipation term analogous to a Rayleigh drag. Subsequently, \cite{Rogers:2014aa} and \cite{Rogers_2014} conducted anelastic MHD simulations of hot Jupiter atmospheres and demonstrated that above $\sim$1500 K, magnetic Reynolds stresses (magnetic resistivity) become comparable to or exceed hydrodynamic Reynolds stresses, making magnetic effects crucial to include in 3D circulation models in these regimes.

Accounting for MHD effects has been predicted to influence the atmospheric dynamics of hot Jupiters in a variety of ways. One such effect is that the magnetic diffusion may significantly slow winds while depositing heat into the atmosphere \citep{Perna2010b, Rauscher_Menou_2013}. Reduced wind speeds decrease the efficiency of heat advection, leading to hotter daysides and cooler nightsides \citep{Perna2010a, Komacek_2016_constant_drag, Beltz_2022}. Magnetism is also predicted to weaken the superrotating zonal jet in the mid-atmosphere (pressures of \(\approx\)10 mbar--1 bar), and may even deflect it if the planetary magnetic field is not coaxial with the rotation axis \citep{Batygin_2014}. Furthermore, heating due to Ohmic dissipation deposits energy deep within the atmosphere, yet this effect may not be fully captured by common model approximations, such as ideal or semi-ideal MHD alongside hydrostatic balance \citep{Rogers_2014, HJ_radanom_mechanisms, Vigan_2025_1d_deep_atmos_mhd}.

The treatment of magnetism in 3D GCMs of hot Jupiter atmospheres has evolved greatly over time. Early studies represented magnetic effects using idealized drag prescriptions to focus on their global impact, either through constant Rayleigh drag formulations (e.g. \citealp{Liu_Showman_2013, Komacek_2016_constant_drag}), or as a pressure-dependent Rayleigh drag \citep{Perna2010a, Perna2010b}. Later work introduced spatially and temporally varying drag \citep{Rauscher_Menou_2013, Beltz_2022}, calculated kinematically from local atmospheric conditions. More recently, \cite{christie2025geometricconsiderationshotjupiter} applied a {quasistatic, neutral-dominant} approximation, based on \cite{zhu2004}, particularly focusing on incorporating radial and meridional currents that were previously neglected. \cite{blocker2026inhomogeneousmagneticcouplingexoplanets} extended this framework to include the Hall effect and a more complete treatment of plasma parameters. In parallel, several groups explored the role of induced magnetic fields generated by atmospheric currents under simplified dynamics, demonstrating that the effects of a fully self-consistent MHD treatment may be non-negligible in certain regimes \citep{rogers2017,Boening2025}.

Changing the form of the magnetic drag prescription in 3D models can significantly alter the resulting atmospheric dynamics of the simulated hot Jupiters. Constant drag timescales \citep{Liu_Showman_2013, Komacek_2016_constant_drag} are simple to implement numerically and reproduce reduced hotspot offsets and increased day–night temperature contrasts, but they cannot capture the complex 3D structure of magnetic interactions present in kinematic MHD models with spatially varying drag \citep{Rauscher_Menou_2013, Beltz_2022}. This is particularly relevant for ultrahot Jupiters (UHJs), where the magnetic resistivity can change by many orders of magnitude between the day and night side \citep{Rauscher_Menou_2013}. Furthermore, applying uniform drag to UHJ models results in magnetic fields orders of magnitude stronger on the nightside than the dayside, which is physically unrealistic \citep{Beltz_2022}.
The specific treatment of resistivity also impacts the final result. For example, accounting for the full Pedersen resistivity, which also implicitly includes the ambipolar term, rather than just the Ohmic component leads to weaker wind damping and day–night temperature contrasts at the same magnetic field strength \citep{christie2025geometricconsiderationshotjupiter}. Including the Hall term, which deflects currents perpendicular to the magnetic field, has been shown by \cite{blocker2026inhomogeneousmagneticcouplingexoplanets} to further modify the flow, causing it to be more day-to-night (a greater divergent component relative to rotational, \citealp{2021PNAS..11822705H}). This contrasts with models that neglect these terms and favour eastward equatorial zonal flow on the nightside and meridional flows along magnetic field lines elsewhere (e.g., \citealp{Beltz_2022}). Furthermore, more complete MHD treatments (e.g., \citealp{rogers2017}) introduce non-trivial magnetic induction feedbacks that can give rise to previously unexplored features, such as formation of highly space- and time-dependent transient and turbulent jets on daysides where magnetic effects are the strongest \citep{Boening2025}.

Much remains unknown about the strength, geometry, and orientation of hot Jupiter magnetic fields; therefore, hot Jupiter magnetic fields are most often modelled as an aligned dipole for simplicity, and strengths within the range of roughly 0.01–100G are employed in accordance with scaling arguments based on our Solar system \citep{Christensen2010,Yadav_2017}. Furthermore, although faster rotation periods tend to favour the formation of stable Taylor columns aligned with the axis of rotation \citep{Davidson_2016, Dynamo_review}, an aligned dipolar magnetic field itself is an unstable configuration according to Cowling’s antidynamo theorem \citep{rudiger2006magnetic, Dynamo_review}. This means that even if the magnetic dipole is aligned with the rotation axis for a short time, such a position is unstable and it will change due to the turbulent nature of the dynamics that keep the dynamo operating \citep{wicht2010}. 

Inferences on hot Jupiter magnetic fields are made extrapolating from the planets in our Solar system \citep{Elias_L_pez_2025}, simulations of hot Jupiter dynamos \citep{magneticsim}, scaling arguments \citep{Yadav_2017}, and inferences from signals from hot Jupiter star-planet interactions \citep{Cauley_2019}. Jupiter's magnetic field has a strength of \(\approx\)10G and the dipole is tilted by roughly 10\(^\circ\) relative to the planet's rotation axis. For the most part, the magnetic field is dipolar but has non-insignificant higher order multipole contributions \citep{Connerney2017, connerney2018}. By contrast, Saturn's magnetic field is substantially weaker [\(O( 0.1\)G)], with a very small tilt \(\approx 0.01^\circ\) and is North-South asymmetric \citep{saturn}. By analogy with the gas giants in our own Solar system, and based on scaling arguments \citep{Yadav_2017}, it is therefore plausible that a hot Jupiter could possess a relatively strong field with modest obliquities of up to \( O(10^\circ)\). \cite{Batygin_2014} developed an analytic theory describing how an oblique dipolar magnetic field may perturb an otherwise axially symmetric simple west-to-east circulation. However, the influence of an oblique magnetic dipole has not yet been investigated using fully 3D General Circulation Models (GCMs) for hot Jupiter atmospheres.

Here, we revisit the theoretical framework used to model magnetic effects in hot Jupiter atmospheres, extending the formulation first introduced by \cite{Perna2010a} to include oblique magnetic dipoles (Section \ref{sec:theory}). We then describe our atmospheric model and its implementation within the \texttt{MITgcm} in the ADvanced Atmospheric \texttt{MITgcm} (\texttt{ADAM}) modelling framework in Section \ref{sec:methods}. In Section \ref{sec:results} we present GCM simulations, for which we used WASP-121b as a case study for a typical ultra-hot Jupiter. After post-processing our model output, we display simulated phase curves for different field strengths and tilt cases to predict whether magnetic properties could be inferred through phase curve observations. Finally, Section \ref{sec:discussion} discusses the advantages and limitations of our model, scrutinizing the validity of the approximations made, and outlining avenues for further work. We summarize key results in Section \ref{sec:conclusion}.

\section{Theory}
\label{sec:theory}
In this work, we use the \texttt{SPARC/MITgcm} \cite{adcroft2004} within the \texttt{ADAM} framework \citep{Kataria_2016,Parmentier_2018, Steinrueck2021,Murphy_2025} to simulate the atmospheres of ultra-hot Jupiters. The \texttt{MITgcm} is an atmospheric GCM that solves the primitive meteorological equations. Additional details of the numerical algorithm are provided in Section \ref{sec:methods_algorithm}. For each model time-step, there exists a 3D grid of atmospheric parcels which contains values for the local temperature, pressure, isobaric velocity, pressure velocity, and metallicity. Here, we describe how we represent magnetic effects as external forcing and energy transfer terms in a form that can be implemented smoothly into the existing GCM framework.

\subsection{Magnetic field model}
\label{sec:theory_Bfieldmodel}


In accordance with previous models of hot Jupiter magnetism (e.g. \citealp{Batygin_2014,Rauscher_Menou_2013, Beltz_2022}), we choose to model our magnetic field as a dipole for simplicity. Importantly, we take both the field strength and orientation to be free parameters. In spherical polar coordinates, the magnetic field due to a dipole \textit{aligned with the axis of rotation} (polar angle \(\theta=0\)) is given by

\begin{equation}
\label{equ:B_aligned}
    \bm{B}=B\frac{R^3}{r^3}\left(\cos\theta,\frac{1}{2}\sin\theta,0\right)\equiv(B_r,B_\theta,B_\phi).
\end{equation}

Here,  \(R(\theta,\phi)\) is the radius at the top of the model atmosphere, and $B$ denotes the characteristic magnetic field strength (in Gauss) at the planetary poles, evaluated at \(R\). This can also be expressed in terms of a magnetic moment \(\bm{m}\), the unit vector representing the dipole orientation and \(\bm{r}\) is the position vector

\begin{equation}
    \label{equ:B_moment}
    \bm{B}=B\frac{R^3}{2\,r^3}\left(3\frac{\bm{r\cdot m}}{r^2}\,\bm{r}-\bm{m}\right).
\end{equation}

In a general case, the magnetic dipole is oriented  at colatitude \(\delta\) and longitude \(\varphi\). We define $\Delta\phi\equiv \phi-\varphi$. We obtain the magnetic field corresponding to such an oblique dipole by substituting the unit vector of the tilted dipole: \(\bm{m}=(\sin\delta\cos\varphi,\,\,\sin\delta\sin\varphi,\,\,\cos\delta)\) into equation (\ref{equ:B_moment}), and converting back into spherical polars:

\begin{align}
\label{equ:Btilt}
    \bm{B}_{\mathrm{tilt}}=B\frac{R^3}{2\,r^3}\left[\cos\delta\begin{pmatrix}
2\cos\theta  \\
\sin\theta \\
0
\end{pmatrix}+\sin\delta\begin{pmatrix}
2\sin\theta\cos\Delta\phi  \\
-\cos\theta\cos\Delta\phi \\
-\sin\Delta\phi
\end{pmatrix}\right]\equiv\begin{pmatrix}
B_r  \\
B_\theta \\
B_\phi 
\end{pmatrix}.
\end{align}

This formula gives the magnetic field strength and direction of a tilted dipole at any location on the planet.

\subsection{Full Induction equation}
\label{sec:theory_inductionequ}
We adopt the full non-ideal magnetic induction equation, following the approach of \cite{Perna2010a}:

\begin{equation}
    \label{equ:induction_full2}
    \frac{\partial \bm{B}}{\partial t} = \nabla \times \left[\bm{u}\times\bm{B}-\frac{4\pi\eta\bm{J}}{c}-\frac{\bm{J}\times\bm{B}}{en_e}+\frac{(\bm{J}\times\bm{B})\times\bm{B}}{c\gamma\rho_i\rho}\right] .
\end{equation}

Here, \(n_e\) is the electron number density, \(\rho_i\) is the ion mass density, {$c$ is the speed of light, $e$ is the elementary charge constant,} and \(\rho\) is the total mass density. \(\gamma\) is the ambipolar drag coefficient between ionised and neutral particles. \(\eta\) is the Ohmic magnetic diffusivity, which is proportional to the inverse of conductivity, but we refer to it henceforth as the resistivity as has been done in other works on this topic \citep{Rauscher_Menou_2013, Beltz_2022}. {For the sake of clarity, \(\eta\) formally relates to the full magnetic resistivity \(\rho_r\) and to the conductivity \(\sigma\) as}:

\begin{equation}
    \eta= \frac{c^2\rho_r}{4\pi}\qquad\qquad\rho_r=\frac{1}{\sigma} .
\end{equation}

The first term on the right-hand side of equation (\ref{equ:induction_full2}) represents magnetic advection by the bulk flow. The remaining terms describe diffusive processes arising from particle–particle interactions: Ohmic dissipation due to electron–neutral collisions, the Hall effect arising from differences in mass between positive and negative charge carriers, and ambipolar diffusion associated with ion–neutral coupling in partially ionized plasmas. See \cite{plasma_book, Davidson_2016, Ballester2018Plasmas} for a discussion on the physical origins of these terms.

\subsection{Magnetic resistivity}
\label{sec:theory_eta}
Density \(\rho\), the specific gas constant \(R_a\), and resistivity \(\eta\) are all local quantities that depend on the velocities, temperatures, and the H tracer within each grid cell, tracking hydrogen dissociation \citep{Tan_Komacek_2019,Roth_2021}. Following \cite{LIU_2008,Menou2012,Rauscher_Menou_2013}, we calculate the ionisation fraction \(x\) and resistivity \(\eta\) based on classical ionization theory, such as \cite{Draine_resistivity}, evaluating the ionization fraction and hence the magnetic resistivity using the Saha equation for the first 28 elements. We take 

\begin{equation}
\label{equ:eta}
    \eta=230\frac{\sqrt{T}}{x}\quad \mathrm{cm^{2}s^{-1}},\qquad \mathrm{ where:}
\end{equation}
\begin{equation}
    x=\sum_{i=1}^{28}\frac{a_i}{a_H}x_{i},\qquad\quad\frac{x_i^2}{1-x_i^2}=\left(\frac{2\pi m_ekT}{h^2}\right)^{3/2}\frac{a_H}{a_i}\frac{\mu}{\rho}e^{-\varepsilon_i/kT} .
\end{equation}

Here, \(T\) is the temperature and \(\rho\) is the mass density. \(m_e\) is the electron mass, \(\mu\) is the mean molecular mass, and \(x_i\) is the ionization fraction for element \(i\). \(k\) is the Boltzmann constant, \(h\) is the Planck constant. \(\varepsilon_i\) and \(a_i\) are the first ionisation potentials and relative abundances for element \(i\). The relative abundances \(a_i\) are assumed to be solar. The prefactor in Equation (\ref{equ:eta}) comes from various fundamental constants (see \citealp{Menou2012}).
This method was previously used to model ionization fraction \(x\) in \cite{Rauscher_Menou_2013, Beltz_2022}, and other works.

\citet{Rauscher_Menou_2013} showed that for typical temperature and density ranges on a hot Jupiter, the resistivity spans many orders of magnitude from \(10^9\) cm$^2$s$^{-1}$ in the hottest and upper parts of the atmosphere, to \(10^{28}\) cm$^2$s$^{-1}$ in the coldest regions (see their Fig. 1). Thus, \(\eta\) is large and strongly location dependent, therefore cannot be removed from any integrals.

\subsection{Neglecting Hall and ambipolar effects}
\label{sec:hall_ambipolar}
Now, we use order of magnitude estimates and scaling arguments to approximate the ratio of the Hall and ambipolar terms to the Ohmic term

\begin{equation}
\label{equ:HO_AO}
    \mathrm{\frac{Hall}{Ohmic}}\approx\frac{cB}{4\pi\eta en_e}\qquad\qquad\mathrm{\frac{Ambipolar}{Ohmic}}
    \approx\frac{B^2}{4\pi\eta\gamma\rho_i\rho} .
\end{equation}

Using the ideal gas law, equation (\ref{equ:eta}) for $\eta$, the expression $n_e=n x$, and $\rho=n\mu$, {where $n$ represents the number density, }these ratios can be written as

\begin{equation}
\label{equ:HO_AO_BTp}
    \mathrm{\frac{Hall}{Ohmic}}\approx\frac{ck_B}{230\,(4\pi e)}\frac{B\sqrt{T}}{p}
    \qquad\mathrm{\frac{Ambipolar}{Ohmic}}\approx\frac{k_B^2}{230\,(\pi\gamma\mu_i\mu)}\frac{B^2T^{3/2}}{p^2}
\end{equation}

To estimate their magnitudes, we adopt the representative values $B\approx 10$G, a mean molecular mass $\mu\approx 2 \times10^{-24}$g (corresponding to H$_2$), mean ion mass of $\mu\approx 30 \times 10^{-24}$g \citep{Rauscher_Menou_2013}, and $T\approx$ 2400K for a WASP-121b-like planet \citep{Sing_2024}. The drag coefficient $\gamma$ is calculated following \cite{Draine_resistivity} and \cite{Perna2010a}. With these values, the ratios scale as

\begin{equation}
\label{equ:HO_AO_approx}
    \mathrm{\frac{Hall}{Ohmic}}\approx\frac{10^3}{p}
    \qquad\qquad\mathrm{\frac{Ambipolar}{Ohmic}}\approx\frac{2 \times 10^5}{p^2}
\end{equation}

Where $p$ is in dyn cm$^{-2}$. These ratios remain less than unity for $p \gtrsim1$ mbar and $p \gtrsim0.1$ mbar, respectively, and are therefore likely negligible throughout most of the atmosphere. We henceforth neglect the Hall and ambipolar diffusion terms relative to the Ohmic term, as in \cite{Perna2010a}. Under this approximation, our induction equation (\ref{equ:induction_full2}) becomes

\begin{equation}
    \label{equ:induction_equation}
    \frac{\partial \bm{B}}{\partial t} = \nabla \times \left[\bm{u}\times\bm{B}-\frac{4\pi\eta\bm{J}}{c}\right] .
\end{equation}

Henceforth, we work in the regime where the Ohmic effect is dominant. We will re-examine this approximation in more detail in Section \ref{sec:discussion_HOAO}. 

\subsection{Energy transfer}
\label{sec:theory_energy}

The addition of MHD introduces a new component of heat dissipation that transfers kinetic energy to thermal energy. This is primarily due to the Ohmic dissipation term in equation (\ref{equ:induction_equation}). To determine the magnitude of the Ohmic dissipation term, we take the magnetic energy density $u_\mathrm{B}\equiv B^2/8\pi$ (e.g. \citealp{energy_density_txtbk}) and differentiate it with respect to time. By applying the chain rule and substituting in equation (\ref{equ:induction_equation}) for $\partial\bm{B}/\partial t$, we can write


\begin{equation}
\label{equ:magn_energy}
    \frac{\partial}{\partial t}u_\mathrm{B}=\frac{\partial}{\partial t}\left(\frac{B^2}{8\pi}\right)=\frac{\bm{B}}{4\pi}\cdot\nabla\times\left[\bm{u}\times \bm{B}-\frac{4\pi\eta \bm{J}}{c}\right] 
\end{equation}

The first term is the energy transfer between the magnetic and velocity fields. We specifically wish to focus on the latter term, which is a heat transfer component, so we define \(u_\mathrm{B}=u_\mathrm{v}+u_\mathrm{ohm}\), where\footnote{For the complete treatment of energy transfer in an MHD system, as well as the physical reasoning for why advection transfers energy into the bulk kinetic energy, and Ohmic dissipation into heat, see \cite{energy_density_txtbk}.}

\begin{equation}
\label{equ:bulk_KE_ohm_energy}
    \frac{\partial}{\partial t}u_\mathrm{v}=\frac{\bm{B}}{4\pi}\cdot\nabla\times\left[\bm{u}\times \bm{B}\right]\qquad\quad  \frac{\partial}{\partial t}u_\mathrm{ohm}=-\frac{\bm{B}}{4\pi}\cdot\nabla\times\left[\frac{4\pi\eta \bm{J}}{c}\right]
\end{equation}

After performing the vector manipulations with the heat‐dissipation (Ohmic) term (\(u_\mathrm{ohm}\)), the change in specific heat due to Ohmic dissipation is obtained by taking the negative of the Ohmic-heating contribution to the change in magnetic energy density and dividing it by the mass density \citep{zhu2004}

\begin{equation}
\label{equ:q_ohm}
    q_{\mathrm{ohm}}= -\frac{1}{\rho}\frac{\partial \,u_{\mathrm{ohm}}}{\partial t}=\frac{1}{\rho}\frac{\bm{B}}{4\pi}\cdot\left[\nabla\times\frac{4\pi\eta \bm{J}}{c}\right]=\frac{1}{\rho}\frac{4\pi\eta}{c^2}|\bm{J}|^2.
\end{equation}

This represents the rate of heat dissipation due to the Ohmic resistivity, which will be passed into the heat transfer equation.

\subsection{Neglecting induced fields}
\label{sec:theory_non_self_consistency}
Here we justify approximating the $\bm{B}$ in the first term of equation (\ref{equ:induction_equation}) as the background dipole field ($\bm{B}_0$), effectively neglecting the induced field ($\delta\bm{B}$). Henceforth, only the second $\bm{J}$ term would have a dependence on the unknown $\delta\bm{B}$, greatly simplifying the maths. We roughly follow the approach of \cite{Batygin_2014} in this subsection, with a few model-specific modifications. We decompose the general steady state magnetic field as \(\bm{B}=\bm{B}_0+\delta \bm{B}\), where \(\bm{B}_0\) denotes the time-constant background dipole field generated by the planet's interior, and \(\delta\bm{B}\) captures any time-variation and induced magnetic field components. The corresponding current density is \(\bm{J}=\bm{J}_0+\delta \bm{J}=\delta \bm{J}\). \(\bm{J}_0\) vanishes because the curl of the dipole field is 0. \(\bm{J}\) is related to \(\bm{B}\) through Ampere's law

\begin{equation}
    \label{equ:ampere}\bm{J}=\frac{c}{4\pi}\nabla\times\bm{B}\quad=\quad\frac{c}{4\pi}\nabla\times\delta\bm{B}=\delta\bm{J} .
\end{equation}

This allows us to write the induction equation (\ref{equ:induction_equation}) as

\begin{equation}
    \label{equ:induction_expanded}
    \frac{\partial \mathbf{\delta B}}{\partial t} = \nabla \times \left[\mathbf{u}\times\mathbf{B}_0+\mathbf{u}\times\mathbf{\delta B}-\eta\,\nabla\times\mathbf{\delta B}\right] .
\end{equation}

We seek the steady-state solution, in which case the left-hand side vanishes. In principle, the steady-state of equation (\ref{equ:induction_expanded}) could be inverted to solve exactly for \(\delta\bm{B}\) (and hence \(\delta \bm{J}\)), given that \(\bm{B}_0\) is fixed, and \(\bm{u},\,\eta\) are calculated dynamically based on the local atmospheric conditions at a given gridpoint and time. In practice, this solution is highly non-trivial and a more efficient approach would be to treat the \(\delta\bm{B}\) magnetic field as a time-varying parameter, coupling an MHD approach with a GCM. Several groups have done preliminary work on this in 2D or even 3D models (e.g. \citealp{rogers2017,Hindle_2021, Boening2025}). Instead, we employ a commonly used approximation and assume that the induced atmospheric magnetic field is weak compared to the background field generated by the internal planetary dynamo. This is equivalent to assuming that the magnetic Reynolds number (Re\(\mathrm{_m}\)) is small. The magnetic Reynolds number is formally defined as the ratio of the advective to the diffusive terms in the magnetic induction equation \citep{KnaepenReynoldsnumber}

\begin{equation}
    \mathrm{Re_{m}}\equiv\frac{\left|c\,\nabla\times(\bm{u}\times \bm{B})\right|}{\left|4\pi\,\nabla\times(\eta\bm{J})\right|} .
\end{equation}

As discussed in \cite{sorianoguerrero2025nonidealmhdsimulationshot}, this exact magnetic Reynolds number approaches unity as equation (\ref{equ:induction_expanded}) reaches steady state, which is not very useful practically. Instead, following similar works on magnetism in hot Jupiter atmospheres, we define the magnetic Reynolds number in terms of the induced field alone, as the ratio of the advection to the diffusion of \(\delta \bm{B}\)

\begin{equation}
\label{equ:Rem_definition}
    \mathrm{Re_m}\equiv\frac{\left|c\,\nabla\times(\mathbf{u}\times \delta \bm{B})\right|}{\left|4\pi\,\nabla\times(\eta\,\delta \bm{J})\right|}.
    \end{equation}

If \(\mathrm{Re_m}<<1\), the steady state of equation (\ref{equ:induction_expanded}) implies \(c\,\nabla\times(\bm{u}\times \bm{B}_0)\approx 4\pi\,\nabla\times(\eta\,{\delta \bm{J}})\), so the condition \(\mathrm{Re_m}<<1\) is equivalent to

\begin{equation}
\label{equ:before_Rem}
    \mathrm{Re_m}\approx\frac{\left|\nabla\times(\bm{u}\times \delta \bm{B})\right|}{\left|\nabla\times(\bm{u}\times \bm{B}_0)\right|}\approx{\frac{\left|\delta \bm{B}\right|}{\left|\bm{B}_0\right|}}<<1 .
\end{equation}

Therefore, a small magnetic Reynolds number as defined in equation (\ref{equ:Rem_definition}) indicates that the advection of the induced field is negligible compared to the diffusion of the induced field. In this limit, equation (\ref{equ:induction_expanded}) reduces to a system of three equations for the three components of current density

\begin{align}
\label{equ:MHD_ideal}\nabla\times\left[\bm{u}\times\bm{B}_0-\frac{4\pi\eta}{c}\bm{J}\right] = 0\quad\mathrm{(a),}\qquad\qquad\nabla\cdot\bm{J}=0\quad\mathrm{(b)} .
\end{align}

Boundary conditions are required to determine the integration constants; we impose \(\bm{J}\) = 0 at the top of the model atmosphere. To justify this approximation, we require that \(\mathrm{Re_m}\lesssim O\)(1) ensuring that advective effects of the induced magnetic fields remain negligible compared to the other terms \citep{Menou2012}. This assumption reflects a limitation of using a GCM model, rather than a fully coupled MHD model, to computationally solve this system. 

We apply a simple scaling argument to the induction equation (equation \ref{equ:MHD_ideal}a) and Ampere’s law (equation \ref{equ:ampere}): \(4\pi \,\delta J/c\approx \mathcal{U}B_0/\eta\) and \(4\pi \,\delta J/c\approx\delta B/\mathcal{H}\),
respectively. Combining these scalings gives:

\begin{equation}
\label{equ:Rem_raw}
    \mathrm{Re_m}\approx\frac{\delta B}{B_0}\approx\frac{\mathcal{UH}}{\eta} .
\end{equation}

We estimate a priori the approximate size of magnetic Reynolds number to justify making this approximation. We approximate the scale for velocity as \(\mathcal{U}\approx 1\)km s$^{-1}$ \(\approx 10^5\) cm s$^{-1}$ \citep{Roth_2021}, and take the atmospheric scale height \(\mathcal{H} \approx R_a T/g\), where \(R_a\) is the specific gas constant of the atmosphere and $g$ is the acceleration due to gravity. On daysides of hot Jupiters these values are of the order \(T \approx\) 2500 K, \(g \approx\) 1000 cm s\(^{-2}\) and \(R_a\approx\) 5\(\times10^7\) cm\(^2\)\ s\(^{-2}\) K$^{-1}$ \citep{Bell_2018,Tan_Komacek_2019,Roth_2021}. As a result, \(H\approx 10^8\) cm \citep{Rauscher_Menou_2013,Beltz_2022}. We calculate \(\eta\) for a range of temperatures, pressures, and densities on the dayside, finding \(\eta\) can be as low as \(10^{10}\)cm$^2$ s$^{-1}$, but can reach as high as \(10^{28}\)cm$^2$ s$^{-1}$. Re\(_\mathrm{m}\) is therefore highly location dependent, so we write:

\begin{equation}
\label{equ:Rem}
    \mathrm{Re_m}=O\left(\frac{10^{13}\mathrm{cm^2/s}}{\eta}\right)\lesssim 1 .
\end{equation}

\cite{Beltz_2022} calculated this quantity for various temperatures and pressures in the ultra-hot Jupiter regime in their Fig. 1, showing that for the much of the atmosphere \(\mathrm{Re_m}\) is \(\lesssim\)1, with values of \(O\)(1) occurring in the hotter, upper parts of the atmosphere. Generally speaking, for certain regimes of ultra-hot Jupiters neglecting induced magnetic field effects may not be an accurate assumption \citep{rogers2017}, but we will re-examine this in more detail in Section \ref{sec:discussion_Rem}.

\subsection{Derivation of a magnetic drag timescale}
\label{sec:theory_tau}
Equation (\ref{equ:MHD_ideal}a) consists of 2 independent equations, so we choose to focus on the \({\theta}\) and \({\phi}\) components. The third component is fixed by the continuity equation (\ref{equ:MHD_ideal}b). We now invoke {what we refer to as the `thin atmosphere'} assumption, in which the planetary radius $R$ is much larger than the radial length scale \(\Delta r\) over which relevant quantities vary, i.e. \(R>>\Delta r\). To leading order, this allows us to approximate \(r\simeq R\) and therefore treat \(r\) as a constant. We can henceforth rename these lone-standing \(r\)'s to \(R\) for simplicity. Since horizontal scales are \(O(R)\) and vertical scales are \(O(\Delta r)\), a simple scaling analysis of the continuity equation (\ref{equ:MHD_ideal}b) gives:

\begin{equation}
    \label{equ:cont_scaling}
    0 \approx\frac{J_r}{\Delta r}+\frac{J_\theta+J_\phi}{R} .
\end{equation}

\(\Delta r<<R\) then implies \(J_r<<J_{\theta},\,\,J_{\phi}\).

Therefore, the thin atmosphere approximation allows us to neglect \(J_r\). Solving for \(J_{\theta}\) and \(J_{\phi}\) by uncurling equation (\ref{equ:MHD_ideal}), neglecting \(J_r\), and integrating over $r$ from top of the atmosphere \(R_{\mathrm{max}}\) to $r$, using our boundary conditions that \(\bm{J}=\bm{0}\) and \(\bm{u}=\bm{0}\) at the top of the atmosphere

\begin{equation}
    \label{equ:J_phi_full}
    J_{\phi} = \frac{c}{4\pi\eta}\left[u_{r}B_{\theta}-u_{\theta}B_{r}-\frac{1}{r\sin\theta}\frac{\partial}{\partial\phi}\int_{R}^{r}(u_{\theta}B_{\phi}-u_{\phi}B_{\theta})dr\right] ,
\end{equation}

\begin{equation}
    \label{equ:J_theta_full}
    J_{\theta} = \frac{c}{4\pi\eta}\left[u_{\phi}B_{r}-u_{r}B_{\phi}-\frac{1}{r}\frac{\partial}{\partial\theta}\int_{R}^{r}(u_{\theta}B_{\phi}-u_{\phi}B_{\theta})dr\right].
\end{equation}

For \(\Delta r<<R\), we can approximate the magnetic field as constant with respect to $r$ within the integral, and therefore each of these terms in equation (\ref{equ:J_phi_full}) scale roughly as:

\begin{equation}
    J=O\left( \frac{c}{4\pi\eta}\left[\mathcal{W}B-\mathcal{U}B+\frac{\Delta r}{R}\mathcal{U}B\right]\right) .
\end{equation}

Where \(\mathcal{W}\) is the characteristic scale for the radial (vertical) velocity, and \(\mathcal{U}\) is the characteristic scale for the horizontal or 2D components of the velocities on an isobar. \(\Delta r<<R\) also allows us to assume isobaric surfaces are approximately flat, in which case the continuity equation reduces to the incompressibility equation (\(\nabla\cdot\bm{u}\approx 0\)), and thus, using the same argument we used to neglect radial current, \(\mathcal{W}<<\mathcal{U}\). Therefore, using this scaling argument, we approximate equations (\ref{equ:J_phi_full}) and (\ref{equ:J_theta_full}) as

\begin{equation}
\label{equ:J}
    J_{\phi}\approx-\frac{cu_{\theta}B_r}{4\pi\eta},\qquad\qquad\qquad J_{\theta}\approx\frac{cu_{\phi}B_r}{4\pi\eta}.
\end{equation}

A more rigorous method is used in Section \ref{sec:discussion_pert} to examine the regions where these assumptions breakdown.

The force applied to the fluid is given by Lorentz force in the MHD regime:

\begin{equation}
\label{equ:Lorentz}
    \bm{f}=\frac{\bm{J}\times \bm{B}}{c\rho}.
\end{equation}

Now that we have our current in terms of quantities that our GCM can easily calculate, we can substitute equations (\ref{equ:J}),
using our assumption \(J_r<<J_{\theta},\,\,J_{\phi}\), into the Lorentz force (equation \ref{equ:Lorentz})

\begin{align}
\label{equ:f_tau_theta}
    f_{\theta}=\frac{1}{c\rho}[J_{\phi}B_r-J_rB_{\phi}]\approx \frac{J_{\phi}B_r}{c\rho}\approx-\frac{u_\theta B_r^2}{4\pi\eta\rho},\\f_{\phi}=\frac{1}{c\rho}[J_rB_{\theta}-J_{\theta}B_r]\approx -\frac{J_{\theta}B_r}{c\rho}\approx -\frac{u_\phi B_r^2}{4\pi\eta\rho}.
\label{equ:f_tau_phi}
\end{align}

\cite{Perna2010a} introduced a magnetic drag timescale found from a discretized approximation of specific force \(f_i\approx u_i/\tau_i\). From equations (\ref{equ:f_tau_theta}) and (\ref{equ:f_tau_phi}), we can read off our magnetic drag timescale as

\begin{equation}
\label{equ:tau}    \boxed{\tau_\theta\approx\tau_\phi\approx\tau_\mathrm{mag}\equiv\frac{4\pi\eta\rho}{B_r^2}=\frac{4\pi\eta\rho}{B^2(\cos\delta\cos\theta+\sin\delta\sin\theta\cos\Delta\phi)^2}.}
\end{equation}

And so

\begin{equation}
\label{equ:GCM_force}    \boxed{\bm{f}_{p,\,\mathrm{Lorentz}}=-\frac{\bm{u}_p}{\tau_\mathrm{mag}},}
\end{equation}

where $\bm{u}_p$ is the velocity vector within an isobar. The negative sign indicates that the force is dissipative for all space, that is, it acts against the direction of motion. We neglect the radial component of the force using the hydrostatic approximation, which we re-examine in Section \ref{sec:discussion_Fr}. Lastly, we find an expression for \(q_\mathrm{ohm}\) in equation (\ref{equ:q_ohm}) by substituting in \(|\bm{J}|^2\approx J_\theta^2+ J_\phi^2\) from equation (\ref{equ:J})

\begin{equation}
\label{equ:GCM_energy}
    \boxed{q_{\mathrm{ohm}}=\frac{B_r^2\,(u_\theta^2+u_\phi^2)}{4\pi\eta\rho}=\frac{|\bm{u}_p|^2}{\tau_\mathrm{mag}}.}
\end{equation}

Here $\bm{u}_p$ is calculated by the GCM, \(B_r\) is given by equation (\ref{equ:Btilt}), \(\eta\) is given by equation (\ref{equ:eta}), and \(\rho\) is given by the ideal gas law. Now, we have a self consistent magnetic forcing term and energy transfer that is free of all integrals and dependent only on atmospheric conditions at a given point and can be implemented into the GCM.

\section{Numerical methods}

\label{sec:methods}
Here, we introduce some of the numerics underlying the \texttt{SPARC/MITgcm} \citep{adcroft2004} within the \texttt{ADAM} framework and lay out the model parameters and assumptions used in our simulations. We describe the alterations to the typical \texttt{MITgcm} we implemented in our model and discuss how the post-processing to generate spectroscopic phase curves was accomplished.

We make the assumption that the part of the atmosphere we are simulating behaves as an ideal gas and is in hydrostatic balance. From these values, we can calculate the local densities, mean molecular mass, ionization ratios, and other dynamical constants which allow us to determine the magnetic forcing and energy transfer terms.

\subsection{Algorithm for the altered momentum equation}

\label{sec:methods_algorithm}
The 3D \texttt{MITgcm} \citep{adcroft2004} is an atmospheric model that evolves the primitive meteorological equations forward in time. Within each time-step, a 3D grid of atmospheric parcels stores values for the temperature, pressure, isobaric velocity, pressure velocity, and local composition assuming chemical equilibrium. Local hydrostatic equilibrium
is assumed so that altitude maps to pressure bijectively, and we work in pressure coordinates henceforth. At each time-step, the dynamical quantities are incremented according to the primitive meteorological equations, and according to the model parameters and external heating and forcing terms that are implemented into the model, including those due to magnetism derived in the previous section shown in equations (\ref{equ:GCM_force}) and (\ref{equ:GCM_energy}).

Instead of typical spherical polar topology, the \texttt{MITgcm} has the option to use a different topology. We choose the cubesphere grid option for each 2D isobaric layer in order to avoid singularities at the poles. The pressure dimension is then formed from these 2D cubesphere meshes stacked on top of each other. Since this is not a spherical polar grid, Euclidean transformations are implemented to transform any equation into the proper form. We discuss this topology and the coordinate transformation (relevant for our magnetic drag calculation) in more depth in Appendix \ref{app:topology}.

A few countermeasures have to be implemented in order to avoid numerical instabilities. This can arise if the timescale on which a phenomenon works is of the same order or less than the timescale on which the model is incremented, then this could cause issues.

In order to avoid these issues, we implement a `ramp-up' for the drag timescale in early times. For example, if the simulation time is less than the ramp-up time, then the strength of the drag force is scaled down by the ratio of the current simulation time to the ramp-up time. We also add in a maximum magnetic drag, or equivalently, a minimum drag timescale \(\tau_{\mathrm{min}}\) such that \(\tau_{\mathrm{mag}} = \mathrm{max}(\tau_{\mathrm{mag}},\,\,\tau_{\mathrm{min}})\), that holds for all time in order to avoid numerical instabilities at the few regions where there is large horizontal divergence and therefore the drag takes a much larger local value. We implement a similar ramp-up in the heating subroutine.

We choose the minimum drag timescale to be 1000s for a magnetic field strength of 3G, 100s for 10G, and 50s for a strength of 30G. This is consistent with the order of magnitude estimates of minimum drag timescales that \cite{Rauscher_Menou_2013} calculated. In theory, the minimum drag timescale must be at least the same order of magnitude as the time-step. The dynamical time-step could also be reduced, but at the expense of longer computation times. The larger the minimum drag timescale, the more we are risking missing important physics especially in the uppermost atmosphere. In Appendix \ref{app:min_timescale}, we examine how varying the minimum drag timescale changes the resulting circulation.

Lastly, around the magnetic equator in the drag timescale in Equation (\ref{equ:tau}), \(B_r\) can cause numerical issues if it is too small. This results in the drag timescale being calculated inaccurately despite the expectation that it should be large in these regions. Thus, we also implement a minimum angular function \(f_{r,\mathrm{min}}\), so \(B_r = B\,\,\mathrm{max}(f_r,\,\,f_{r,\mathrm{min}})\), which we set to $10^{-5}$ for all runs. \(f_r\) is the angular function corresponding to the radial component of magnetic field, as defined in Equation (\ref{equ:Btilt}). Although this is a very minor numerical adjustment, we highlight this here because increasing \(f_{r,\mathrm{min}}\) could possibly be one way to parameterise the effects of higher order currents we neglected in equations (\ref{equ:J_phi_full}) and (\ref{equ:J_theta_full}). We discuss this further in Section \ref{sec:discussion_pert}.

\begin{table}
\centering
\begin{tabular}{l|l}
 \hline
\multicolumn{2}{c}{$\mathbf{Planetary\,\,\, parameters}$} \\
\hline
 Parameter & Value\\
 \hline
 Bottom boundary planetary radius & 130,384 km\\
 Surface gravity & 8.43 m/s\(^2\)\\
 Semimajor axis & 0.02544 AU\\
 Orbital/rotation period & 1.27 days\\
 Bottom boundary temperature & 4100 K\\
 Bottom boundary pressure & 34 bars\\
 Top boundary pressure & 5.8 $\mu$bars\\
 Stellar irradiation & $6.79 \times 10^{6}$ W m${^{-2}}$ \\
 Atmospheric specific heat capacity $c_p$ & 13000 J kg$^{-1}$ K$^{-1}$ \\
 \hline
 \multicolumn{2}{c}{$\mathbf{Numerical\,\, parameters}$} \\
 \hline
 Model timestep & 10 s \\
 Number of pressure levels & 53 \\
 Number of tiles within a cubesphere face & 64\\
  \hline
\end{tabular}
\caption{Planetary parameters for WASP-121b used in the simulation. (e.g. \citep{W121b_params_2016, W121b_params_2020})}
\label{tab:HJ}
\end{table}

\subsection{Simulation parameters and runs}
\label{sec:methods_simulation}
WASP-121b is an ultra-hot Jupiter orbiting its host star WASP-121, an F-star around 260 pc away. We consider this a representative ultra-hot Jupiter and note that it has been the subject of observations \citep{W121b_params_2016, W121b_params_2020, WASP121b_Splinter_2025} and theoretical works \citep{Lee2022,wardenier2024phaseresolvingabsorptionsignatureswater, Seidel2025}.  We assume the planet has solar metallicity, and list our assumed planetary parameters in Table \ref{tab:HJ}. 

We use the standard \texttt{MITgcm} \citep{adcroft2004} with a correlated-k radiative transfer treatment (see \citealp{lee2021} for a comparison of different radiative transfer schemes). To isolate the influence of our tilted dipole treatment, we do not include clouds in our model, consistent with  recent studies \citep{Beltz_2022,christie2025geometricconsiderationshotjupiter}.  This is a reasonable approximation for ultra-hot Jupiters as they are predicted to have largely cloud-free daysides and only patchy nightside clouds \citep{Komacek_2022_Nightside_Clouds}.

One key difference between hot and ultra-hot Jupiters is role of hydrogen dissociation and subsequent recombination on ultra-hot Jupiters. The daysides of ultra-hot Jupiters are hot enough for H\(_2\) to thermally dissociate. Once advected to the cooler nightside, recombination of atomic to molecular hydrogen occurs. This effect has a cooling effect on the dayside and a heating effect on the nightside, thus reducing the day–night temperature contrast \citep{Bell_2018, Tan_Komacek_2019}. Notably, this will be in tension with MHD effects, which work to increase the day–night temperature contrast. We therefore implement the hydrogen dissociation scheme as laid out in \cite{Tan_Komacek_2019} and \cite{Komacek_2022_Nightside_Clouds}. This treats atomic hydrogen as a thermodynamically active tracer in the GCM, allowing for localised heating or cooling where the mass mixing ratio of atomic hydrogen decreases or increases due to recombination or thermal dissociation. 



\begin{figure}
    \includegraphics[width=0.8\linewidth]{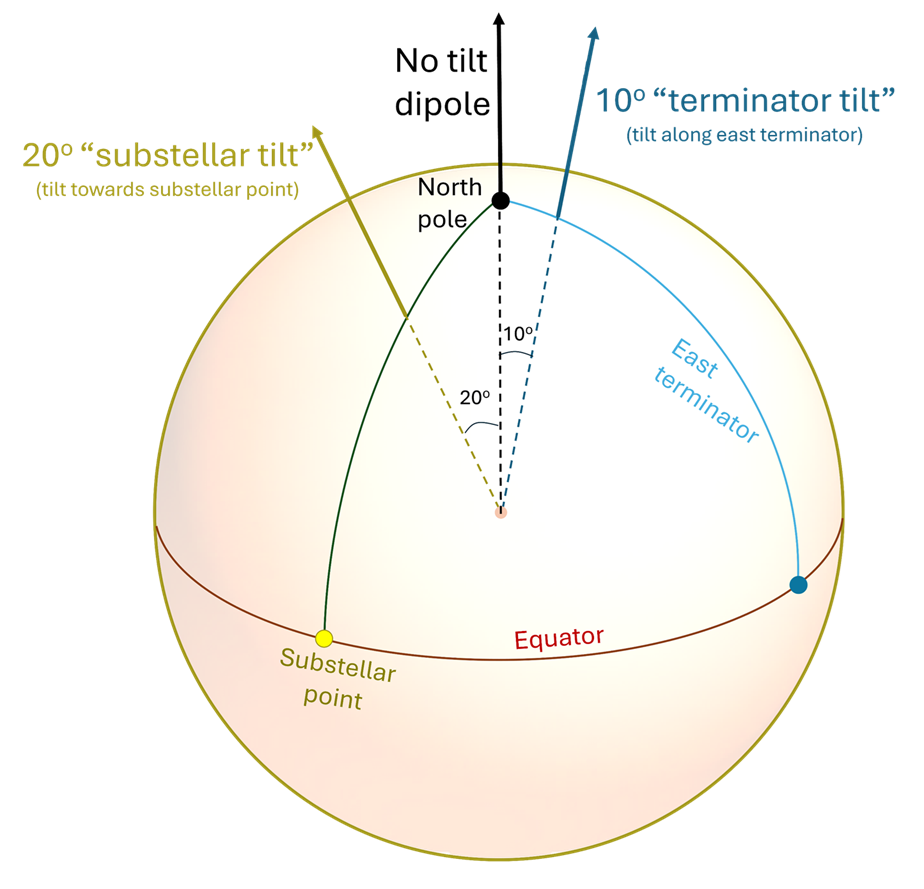}
  \caption{\label{fig:schematic} Schematic illustrating the geometry of the magnetic field dipole orientations used in this work.}
\end{figure}

All parameters are kept fixed between the runs aside from the three magnetic field parameters: the magnetic field strength and the two degrees of freedom that define the orientation. In Table \ref{tab:B-field_params} we describe the magnetic field and numerical parameters for the non-tilted runs, and in Table \ref{tab:B-field_params_tilted},we list the magnetic field parameters for the runs with an oblique magnetic field. The minimum drag timescale listed in Table \ref{tab:B-field_params} correspond to the smallest allowed timescale in each run. A minimum timescale is included in the model for numerical stability. We run one of each non-tilted case, for a total of 4 non-tilted simulations with varying magnetic field strength, and then a combination of each of the listed parameters in Table \ref{tab:B-field_params_tilted}, giving a total of 18 tilted simulations. The magnetic field strengths correspond to the strengths of the magnetic field at the poles. Our surface magnetic field is not uniform so this results in weaker field strengths at the equator.

\begin{table}
\centering
\begin{tabular}{c|c}
\hline
  Magnetic field strength & Minimum drag timescale\\
 \hline
  0G & --- \\
  3G & 1000s \\
  10G & 100s\\
  30G & 50s \\
 \hline
\end{tabular}
\caption{
\label{tab:B-field_params} Minimum drag timescale corresponding to the different magnetic field strength runs.}
\end{table}

\begin{table}
\centering
\begin{tabular}{|l|l|}
 \hline
 Parameter & Values\\
 \hline
 Magnetic field strength & 3G, 10G, 30G\\
 Latitudinal tilt & 10\(^\circ\), 20\(^\circ\), 50\(^\circ\)\\
 Longitudinal tilt direction& Towards substellar point (0\(^\circ\)),\\
 & Along east terminator (90\(^\circ\))\\
 \hline
\end{tabular}
\caption{
\label{tab:B-field_params_tilted} This table summarises our simulations with a tilted magnetic field: Simulations for all combinations of magnetic field parameter values listed were conducted, giving a total of {18} runs. The minimum drag timescale used corresponds to the magnetic field strength, as in Table \ref{tab:B-field_params}.}
\end{table}

We initially ran each model for 1 million time-steps, corresponding to just under 100 planetary orbits. To check the convergence of the runs, we examined the kinetic energy and the thermal energy as a function of pressure and time. We consider the model to be sufficiently converged when these quantities become roughly constant with time. The 3G and 10G runs were not completely converged especially in the deeper pressure layers (pressures above about 1 bar) and so they were run for another million time-steps.

\subsection{Post-processing}
\label{sec:methods_post_processing}
The temperatures, velocities and particle tracers as a function of latitude, longitude and pressure, time-averaged over the last 20,000 time-steps{--approximately 2 planetary orbits--}are used in our analyses. 


We simulate white-light phase curves  using \texttt{gCMCRT}---an advanced, open-source, 3D Monte-Carlo radiative-transfer code specifically designed for post-processing GCM outputs to simulate observations of exoplanets \citep{2022ELeegcmcrt}. White light spectroscopic phase-curves are calculated for the 3G and 30G scenarios for 10$^{\circ}$ and 50$^{\circ}$ tilted dipole scenarios, toward the substellar point and terminator (see Fig. \ref{fig:schematic}), as well as no-tilt cases. We calculate the thermal emission from the atmosphere at 24 different viewing angles. The \texttt{gCMCRT} code calculates this by simulating the path of 10,240,000 individual photon packets through the full 3D atmosphere \citep{2022ELeegcmcrt}. We calculate the emission spectrum of the planet for the wavelength range $0.3 \rightarrow 5.5 \,\mu m$, at a resolution $R \sim 500$, {and weight these emission data according to the throughput of NRS1 and NRS2 \citep{2024rodrigofilterthroughput}. This allows for comparisons to \citep{w121_PhaseCurve}}. We consider the following opacity sources: Na, K, H$_2$O, CH$_4$, CO, CO$_2$, NH$_3$, TiO, VO, FeH, Fe, and HCN. Opacities are computed with the correlated-k method, with opacity table sourced from the \textit{hi}gh-resolution \textit{tran}smission molecular absorption database (HITRAN). Continuum opacity due to collision induced absorption (CIA) is calculated from the HITRAN tabulated values for H$_2$-H$_2$, He-H$_2$, H$_2$-H, He-H, H$^{-}$, H$_2 ^{-}$ (free-free), and He$^{-}$ (free-free).

\section{Results}
\label{sec:results}
Here, we cover the simulation results by discussing the three main impacts on features: the impact on the temperature profile, the mid-atmospheric strong zonal equatorial jet, and the hotspot offset. We additionally show white-light phase curves calculated using \texttt{gCMCRT}.

\begin{figure*}
\vspace{0cm}
  \centering
  \makebox[\textwidth][c]{%
    \includegraphics[
        width=.95\textwidth,
        trim=0cm 0cm 0cm 0cm,   
        clip
    ]{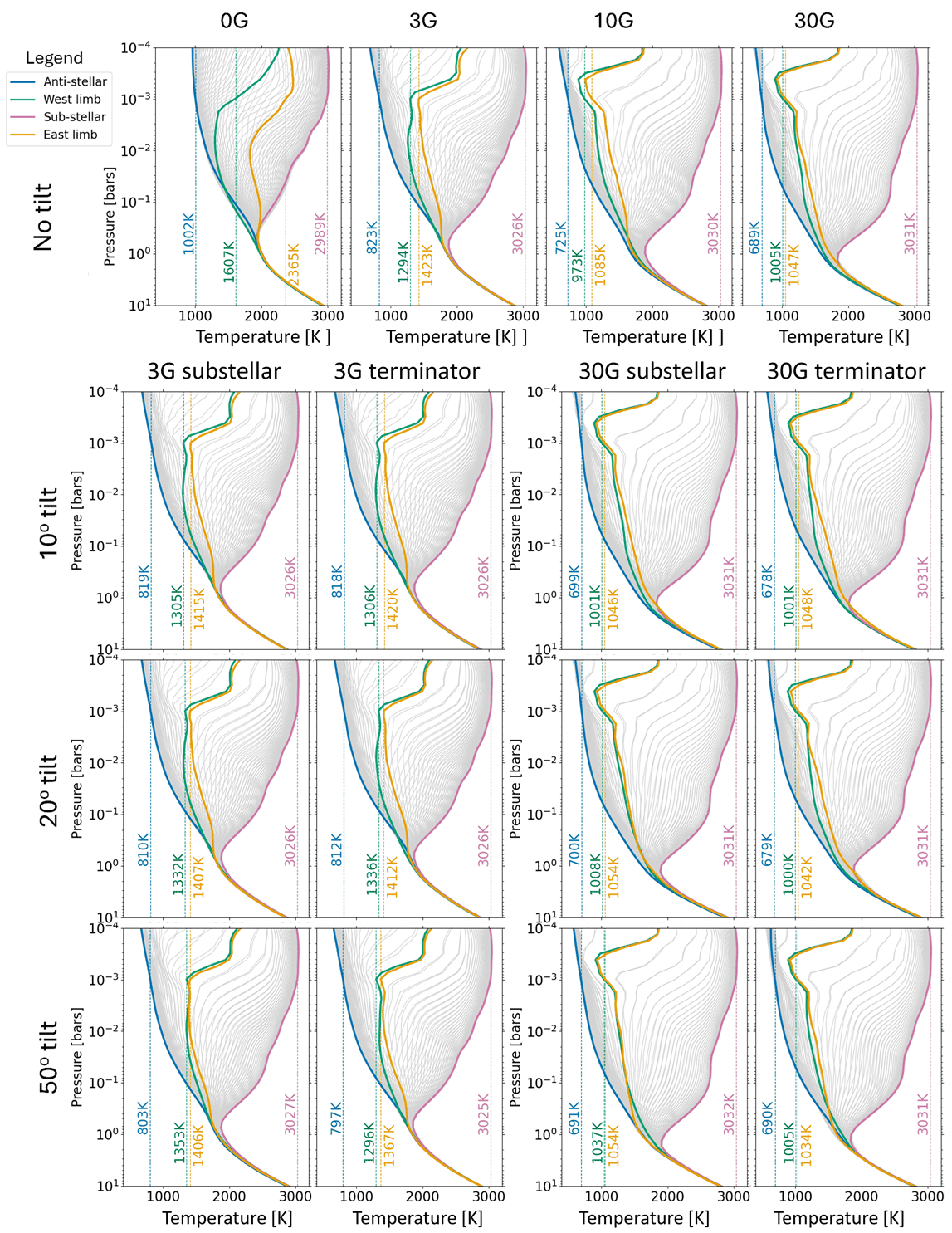}}
  \caption{\label{fig:general_TP} Pressure versus temperature profiles for our grid of models. The bolded lines from left to right represent the latitudinally averaged temperature–pressure profiles at the substellar point (longitude 180\(^\circ\)), the west limb (-90\(^\circ\)), the east limb (90\(^\circ\)), and the substellar point (0\(^\circ\)). The 1 mbar temperatures for each of these four profiles are listed for reference. For models without a tilt, we see that the day–night temperature contrast increases with increasing field strength, in agreement with previous works. For our tilted dipole models, we note that the tilt strength/direction has a minimal effect on the global day–night temperature contrast.   }
\end{figure*}

\begin{figure*}
\vspace{0cm}
  \centering
  \makebox[\textwidth][c]{%
    \includegraphics[
        width=\textwidth,
        trim=4cm 0.5cm 8cm 33cm,   
        clip
    ]{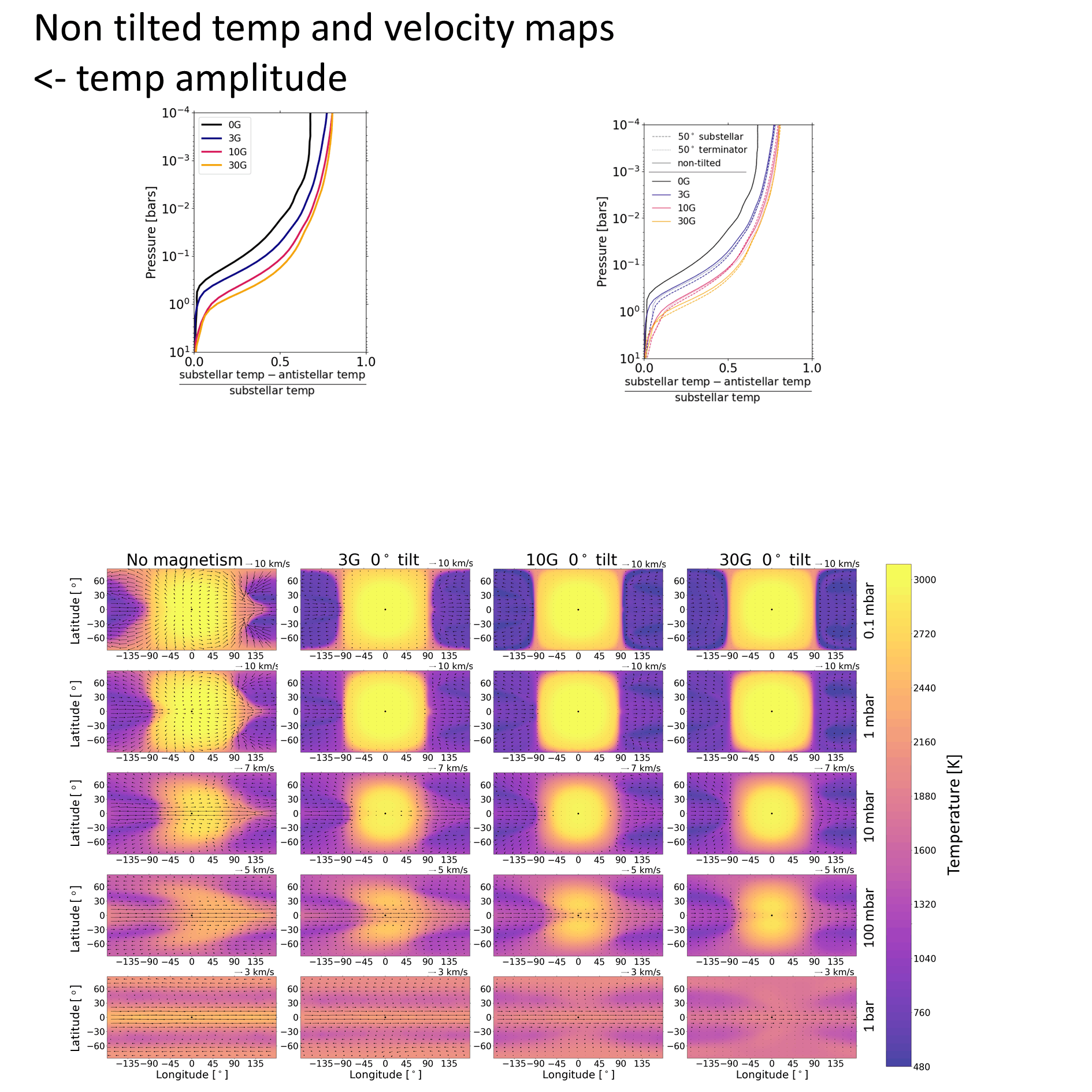}}
  \caption{\label{fig:temp_wind_notilt} Temperature maps for simulations with an aligned dipole (non-tilted) magnetic field plotted at various pressure levels. The arrows represent the horizontal (zonal and meridional) velocities. Note that the velocity vectors are not all the same magnitude, decreasing in scale with increasing pressure. As magnetic field strength increases, the equatorial jet is weakened and the hotspot offset moves closer to the substellar point. At the deepest pressures, where our drag timescale is the longest, the models are extremely similar. }
\end{figure*}

\begin{figure}
  \centering
  \includegraphics[width=0.6\linewidth]{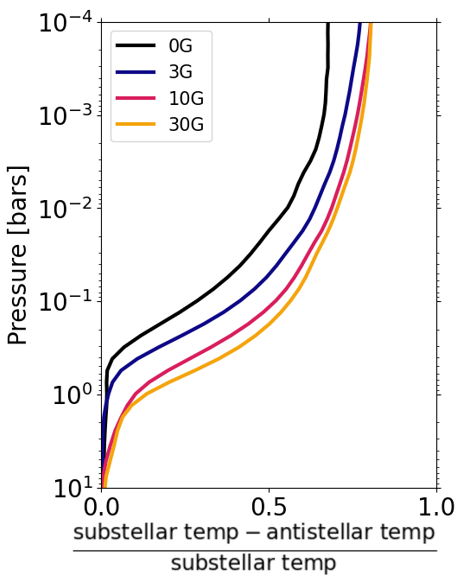}
  \caption{\label{fig:TP_diff_daynight} Substellar-antistellar temperature contrast normalized by the substellar temperature as a function of pressure. Each line represents a non-tilted case. At all pressures, increasing field strength increases the day-night contrast, but otherwise the normalised substellar-antistellar temperature contrast has a similar dependence on pressure. }
\end{figure}

\subsection{Temperature Structure}
\label{sec:results_tp}
\subsubsection{Aligned dipole magnetic field}

We first show the latitudinally averaged temperature versus pressure (T–P) profiles from our cases with an aligned dipole and varying magnetic field strength in the top row of Fig. \ref{fig:general_TP} with selected longitudes highlighted for various magnetic field parameter runs. Each of our aligned dipole cases exhibit strong dayside temperature inversions which decrease in magnitude moving further away from the substellar point, eventually becoming completely non-inverted on the nightside. Consistent with previous works exploring MHD effects in hot Jupiter atmospheres \citep{Rauscher_Menou_2013, Komacek_2016_constant_drag, Beltz_2022}, Figure \ref{fig:general_TP} shows that the day–night temperature contrast increases with increased magnetic field strength. Stronger magnetic fields correspond to stronger Lorentz forces, resulting in less effective heat redistribution between hot dayside and cooler nightsides. This also leads to slower wind speeds, as we discuss in more detail in Section \ref{sec:results_velocity}. The day–night asymmetry in the drag causes the axial asymmetry between longitudes to persist deeper in the atmosphere; for example, the 0G temperature–pressure profiles for different longitudes converge to one temperature at around 500 mbar, whereas in the 30G case this occurs at a few bars. Increasing the magnetic field strength causes the east and west longitudes become more symmetric in their temperature structure, which is also a consequence of the damping of the otherwise predominantly west-to-east winds. This can be seen in the temperature–pressure profiles in Fig. \ref{fig:general_TP}, where the east and west longitudes of the same degree become closer together in the upper atmosphere. 


The temperature and wind vector maps shown in Fig. \ref{fig:temp_wind_notilt} illustrate how the 3D temperature profile at different pressure levels depends on magnetic field strength. As in Fig. \ref{fig:general_TP}, these maps clearly display the increase in the day–night temperature contrast with increasing field strength. Fig. \ref{fig:temp_wind_notilt} lso reveals a significant difference in the wind speeds between the daysides and nightsides. In the upper atmospheres at pressures of \(\lesssim\)10mbar, dayside flow is characterized by isobaric divergence away from the sub-stellar point, carrying hot air directly away from the hotspot. In the no-magnetism case, this flow is asymmetric, with eastward advection being preferred due to planetary rotation. This flow becomes more and more symmetric with greater magnetic field strength, with horizontal winds slowing and being directed symmetrically away from the substellar point on the 2D isobaric plane. Nightside flow is characterized by a nightside-only west-to-east equatorial jet in upper atmospheres, accompanied by weaker counter jets in the mid latitudes. At pressures of 10 mbar~\(\lesssim p\lesssim\)~500 mbar, a fully formed superrotating equatorial jet starts to emerge. With increased magnetic field strength, this jet becomes both weaker and thinner, spanning a reduced range of pressures. This pattern is most pronounced on the dayside and goes in concert with the increased temperature contrasts, as both effects are a direct consequence of increasing the drag strength within our model.

In the 10–100 mbar pressure range, Fig. \ref{fig:temp_wind_notilt} shows that a narrow equatorial strip of slightly faster winds persists. This feature arises because Ohmic drag is weakest near the magnetic equator, and our simplified model exaggerates this effect as horizontal currents nearly completely vanish where the radial magnetic field is small. As a consequence of this, deeper in the atmosphere we observe a slight splitting of the hotspot. Generally, as the magnetic field strength is increased, we find a greater east–west symmetry, and the hot Jupiter planetary-scale `chevron’ pattern as seen in \cite{showman2009} becomes less apparent.

To more thoroughly quantify the day–night temperature contrast, we consider the day–night temperature difference amplitude, given by \(\Delta T(p)_{ss-as}=\frac{T_s(p)-T_a(p)}{T_s(p)}\), where \(T_s(p)\) and \(T_a(p)\) represent the substellar and antistellar latitudinally averaged pressure versus temperature profiles, respectively. This quantity is plotted as a function of pressure in Fig. \ref{fig:TP_diff_daynight} for non-tilted cases, and in Fig. \ref{fig:TP_diff_daynight_tilted} for tilted cases as well. As expected, the day–night temperature contrast increases with increasing magnetic field strength and follows a very similar functional form for all field strengths: the contrast is very small at deeper pressure levels indicating axial symmetry, and increases rapidly around 1 bar. For stronger magnetic field runs, this increase begins at deeper pressures and persists into the upper atmosphere.


\subsubsection{Oblique dipole magnetic field}
\label{sec:results_tp_oblique}

Next, we analyse how tilting the magnetic field impacts the temperature versus pressure profiles, shown previously in Fig. \ref{fig:general_TP} for the non-tilted scenario.
Generally, tilting the dipole has very little impact on the latitudinally averaged temperature versus pressure profiles, with the most substantial change being that the nightside temperatures may decrease by \(\approx20\)K between the aligned dipole and largest tilt case because the magnetic equator, where the drag is weakest, is displaced from the substellar point. The latitudinally averaged T–P profiles show that the day–night temperature contrast is roughly constant regardless of field orientation.

\begin{figure*}
\vspace{0cm}
  \centering
  \makebox[\textwidth][c]{%
    \includegraphics[
        width=\textwidth,
        trim=4cm 2.5cm 4.5cm 8.1cm,   
        clip
    ]{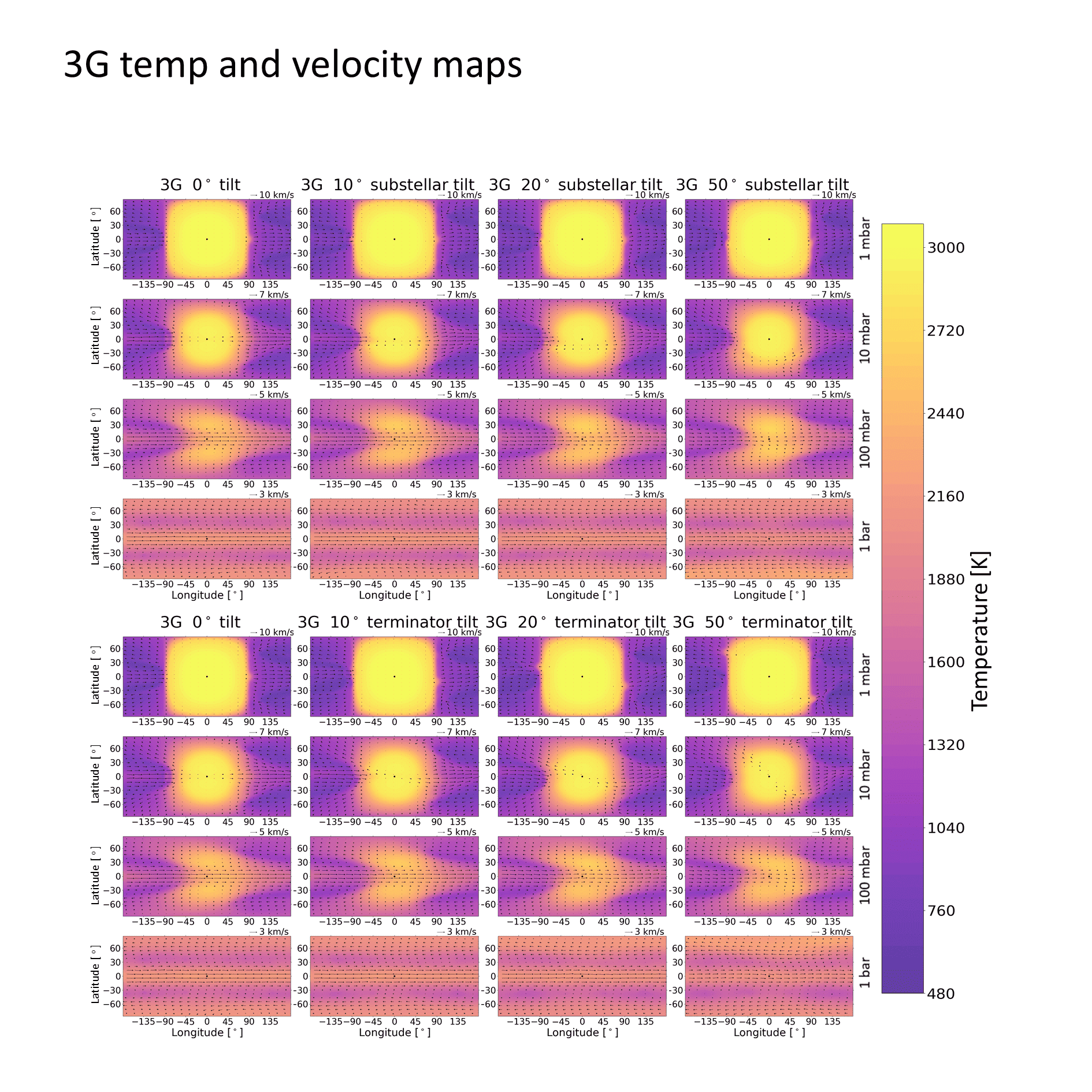}}
  \caption{\label{fig:temp_wind_3G} Temperature maps for magnetic field runs of various dipole orientations with strength 3G plotted at various pressure levels. The arrows represent the horizontal (zonal and meridional) velocities. Note that the velocity vectors are not all the same magnitude. Around 10-100 mbar, the differences in the velocity and temperature profiles are most significant. At the deepest pressures, where our drag timescale is the longest, the models are extremely similar.}
\end{figure*}

Selected temperature and isobaric velocity maps for various orientations of magnetic field strengths of 3G and 30G are shown in Figs \ref{fig:temp_wind_3G} and \ref{fig:temp_wind_30G}, respectively. The corresponding maps for 10G runs are presented in Appendix \ref{app:figs}. Consistent with the T-P profiles, both Figures \ref{fig:temp_wind_3G} and \ref{fig:temp_wind_30G} indicate that the average day to night temperature contrasts do not differ substantially for different magnetic field orientations, but in contrast the 3D mapping illuminates differences that are not evident in the temperature–pressure profiles shown in Fig. \ref{fig:general_TP}. 

For the 3G cases in Fig. \ref{fig:temp_wind_3G}, tilting the magnetic dipole most noticeably impacts the 10–100 mbar pressure levels. The drag is weakest along the magnetic equator, so when the dipole is tilted and the magnetic equator is displaced from the geographic equator, the location of the strongest winds shifts accordingly. This in turn changes the features seen in the temperature field as well, becoming less North–South symmetric as the tilt is increased. At pressures \(\lesssim 10\)~mbar on the dayside, the terminator tilt cases remain nearly point-symmetric about the sub-stellar point, while the substellar tilt cases lose this symmetry. 

Lastly, although the stronger magnetic field (30G) cases generally produce greater symmetry in the uppermost atmosphere (\(\lesssim 5\)~mbar), regardless of tilt, in the mid-atmosphere (10~mbar - 1~bar) a large tilt induces significant north–south asymmetries in the temperature and velocity fields with this effect being stronger in the stronger magnetic field cases. We discuss these results further in the context of how varying magnetic field strength and tilt impact the strong mid-atmosphere equatorial jet in the following Sections \ref{sec:results_velocity} and \ref{sec:results_hotspot}.

\begin{figure*}
\vspace{0cm}
  \centering
  \makebox[\textwidth][c]{%
    \includegraphics[
        width=\textwidth,
        trim=3cm 1.5cm 2.5cm 5cm,   
        clip
    ]{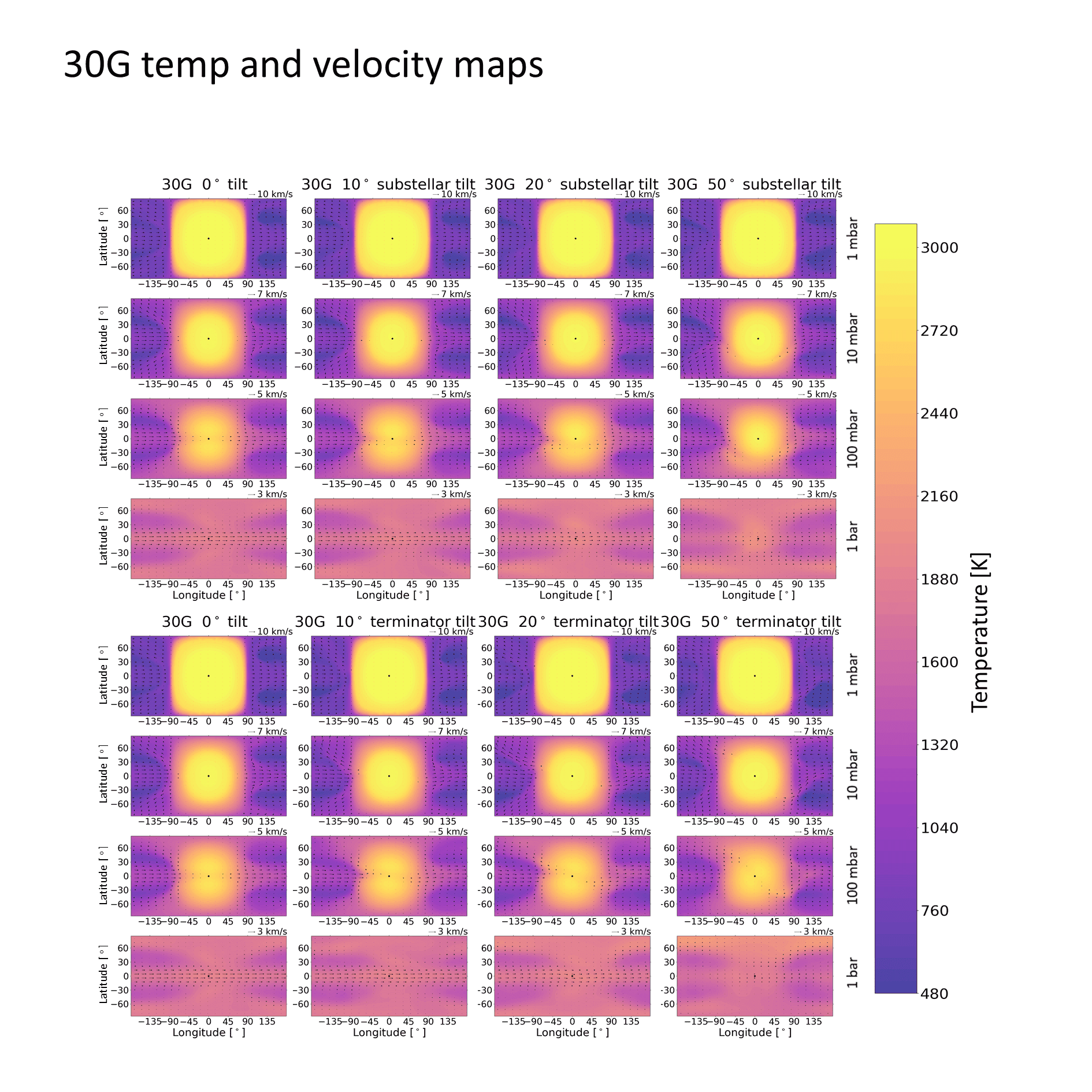}}
  \caption{\label{fig:temp_wind_30G} Similar to Figure \ref{fig:temp_wind_3G}, but showing temperature and velocity maps for runs with magnetic field strengths of 30G.}
\end{figure*}

\subsection{Equatorial winds}
\label{sec:results_velocity}
\begin{figure*}
\vspace*{0cm}
  \centering
  \makebox[\textwidth][c]{%
    \includegraphics[width=1.1\textwidth]{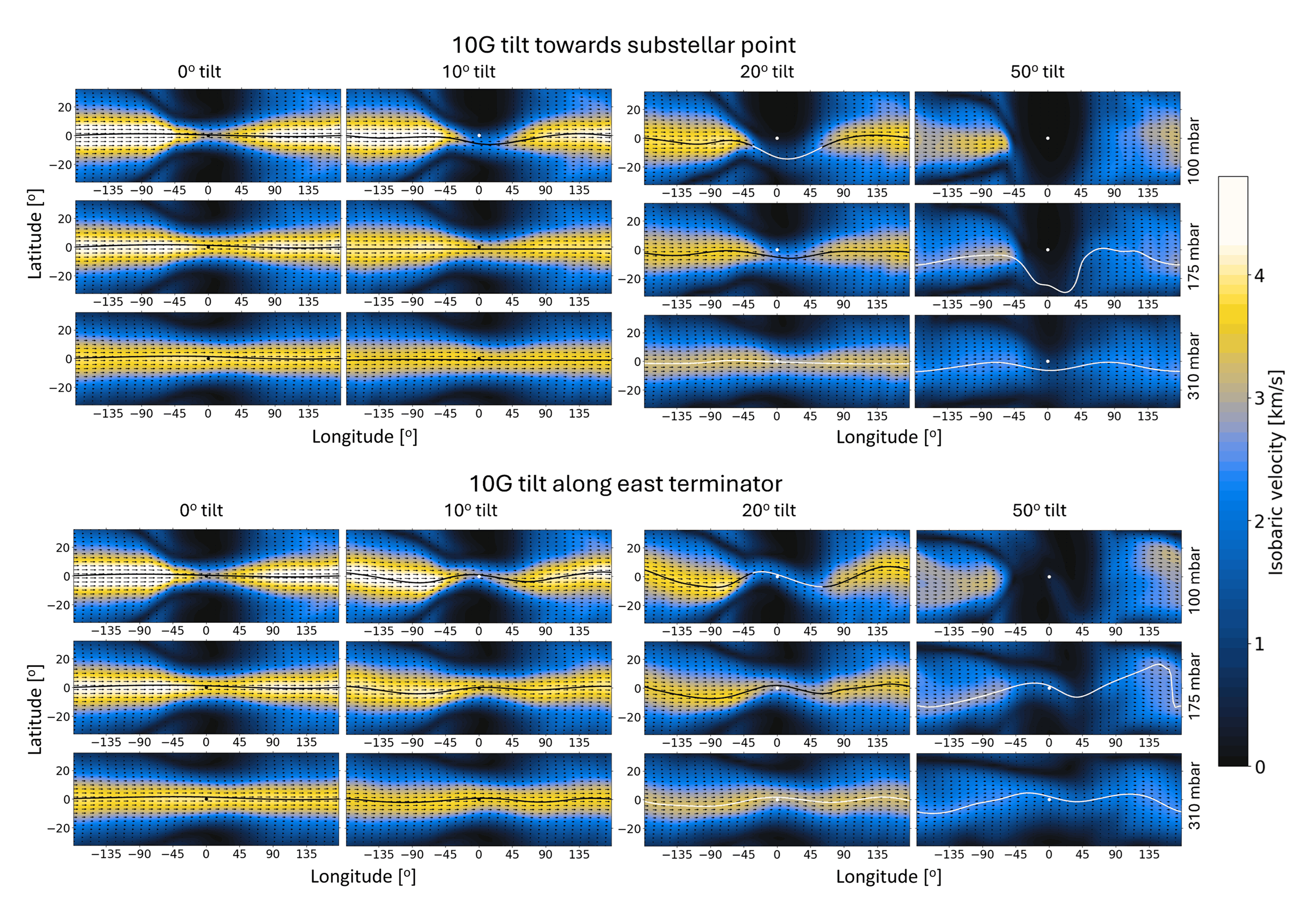}}
  \caption{\label{fig:vel_defl_10G} Atmospheric velocity on isobaric surfaces for 10G runs at various pressure levels. Arrows indicate flow direction; colours denote the magnitude of isobaric velocity. The plotted line marks the latitude where this isobaric velocity is greatest at each pressure and longitude. The line is shown only if a superrotating jet is sustained around the planet. The colour of this line changes from white to black for visibility purposes. The dot indicates the substellar point. We find that tilting the magnetic dipole can cause a deflection in the equatorial jet and this deflection is stronger for larger tilts.}
\end{figure*}

\begin{figure*}
\vspace*{0cm}
  \centering
  \makebox[\textwidth][c]{%
    \includegraphics[width=0.8\textwidth]{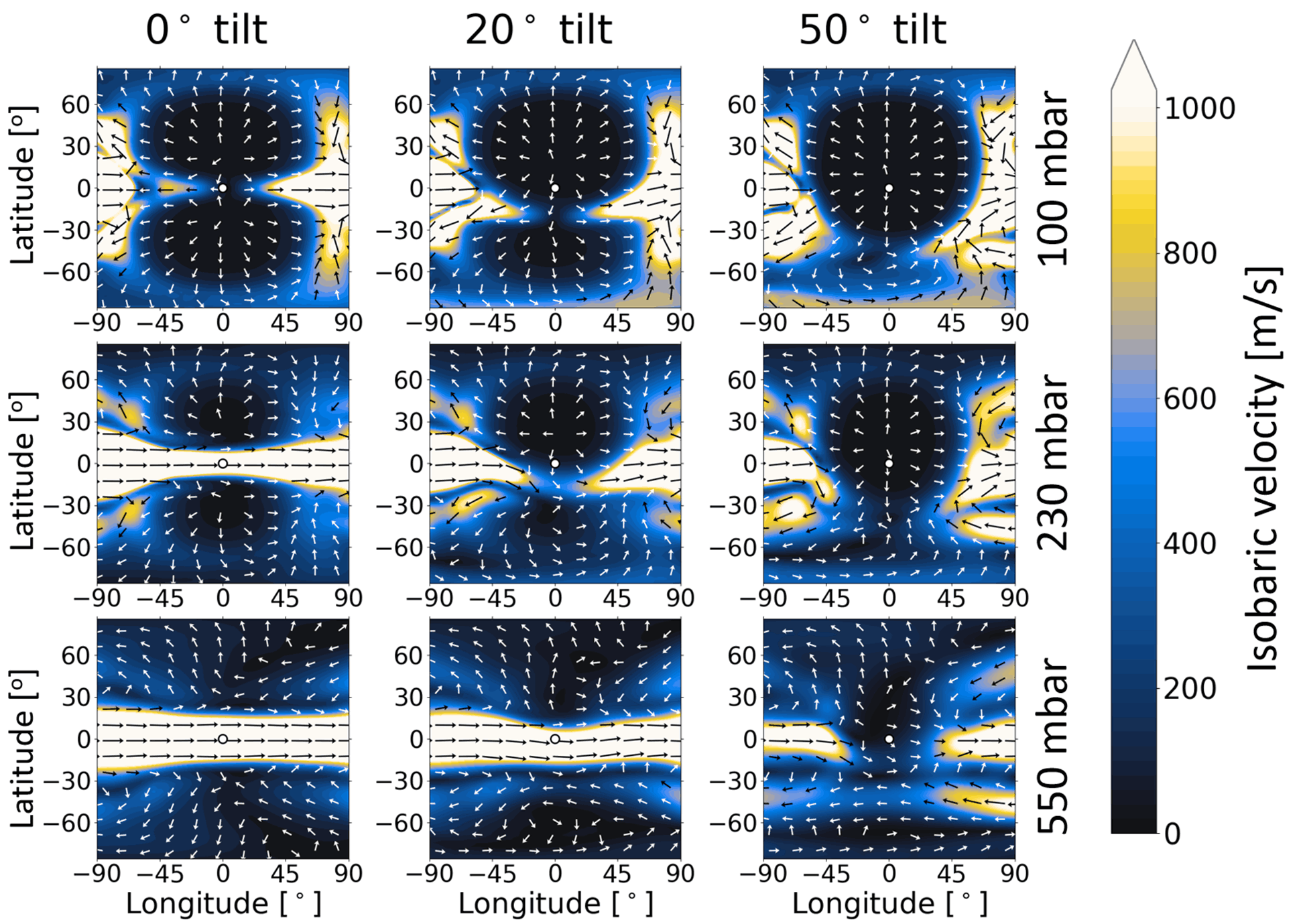}}
  \caption{\label{fig:zoomvel} {Atmospheric velocity on isobaric surfaces for 30G runs with dipole tilts towards the substellar point. Arrows indicate flow direction; colours denote the magnitude of isobaric velocity. The arrow size is not indicative of the magnitude, and the minimum arrow size was capped at 500 m/s to highlight flow direction for low speeds--these arrows are plotted in white. The dot indicates the substellar point. }}
\end{figure*}

Figs \ref{fig:vel_defl_10G} and \ref{fig:vel_defl_20deg} present the magnitude and direction of the 2D horizontal velocity on isobaric surfaces, which we hereafter refer to as `isobaric velocity’ for simplicity. Fig. \ref{fig:vel_defl_10G} shows cases with a magnetic field strength of 10G, demonstrating how the characteristics of the equatorial jet change with different dipole orientations. Fig. \ref{fig:vel_defl_20deg} displays runs with a fixed polar tilt angle of 20\(^\circ\) and a variety of magnetic field strengths. In both figures, the colour scale represents isobaric velocity magnitude, while the arrows indicate the direction. The overplotted line marks the latitude where the equatorial jet’s isobaric velocity is greatest, though we note that this feature becomes increasingly tenuous for extreme tilts and is therefore less informative.

Figure \ref{fig:vel_defl_10G} most clearly illustrates that tilting the magnetic dipole both weakens and deflects the mid-atmospheric equatorial jet. The magnetic drag, regardless of whether or not radial currents are included in the model, would be weakest along the magnetic equator. Therefore, displacing the magnetic equator from the geographic equator shifts the preferred latitude for jet formation. The jet does not align perfectly with the magnetic equator because of the complex interplay of atmospheric dynamical processes, and instead forms a roughly sinusodal pattern. As the tilt is increased, the deflection of the jet becomes more pronounced, and the jet weakens further due to being located farther away from its otherwise preferred equatorial position.

Figure \ref{fig:vel_defl_20deg}  further shows that stronger magnetic fields cause greater deflection of the equatorial jet, while the overall wind speeds decrease. Furthermore, at higher magnetic field strengths, tilting the magnetic dipole also introduces new dynamical features. We present similar isobaric velocity plots in Figure \ref{fig:zoomvel}, focusing on the 30G substellar tilt cases with enhanced low-speed arrows for visibility. In the 30G \(50^\circ\) substellar tilt case in Figure \ref{fig:zoomvel}, the 100 mbar-1 bar zonal jet is completely disrupted and replaced, from west to east, by a southwestward advection of air, leading into an anticlockwise circulation pattern in the Southern hemisphere. By contrast, the non-tilted 30G case at the same pressures exhibits nearly stagnant flow, with the equatorial jet a dominant feature at pressures of $\sim 0.5$ bar.

This result complements the analytic predictions of \cite{Batygin_2014}, though there are a few notable differences. In our simulations, such as in the 20\(^\circ\) tilt case of Figure \ref{fig:vel_defl_20deg}, the roughly sinusoidal perturbed jet exhibits two full periods rather than just one, likely because our model accounts for full hydrodynamic effects and interactions with counterjets. By contrast, the analytic model in \cite{Batygin_2014} treated the entire planet as a simple zonal flow and did not capture these complex secondary structures. We find the jet to be perturbed more strongly on the daysides, where magnetic effects are strongest. The model presented in \cite{Batygin_2014} only considered the velocity structure, not the temperature profile, so their model did not capture the dayside thinning, or even disappearance, of the jet that is evident in our simulations.

\subsection{Hotspot Offsets}
\label{sec:results_hotspot}
\begin{figure*}
\vspace*{-1cm}
\makebox[\textwidth][c]{%
    \includegraphics[
        width=.95\textwidth,
        trim=0cm 0cm 0cm 0cm,   
        clip
    ]{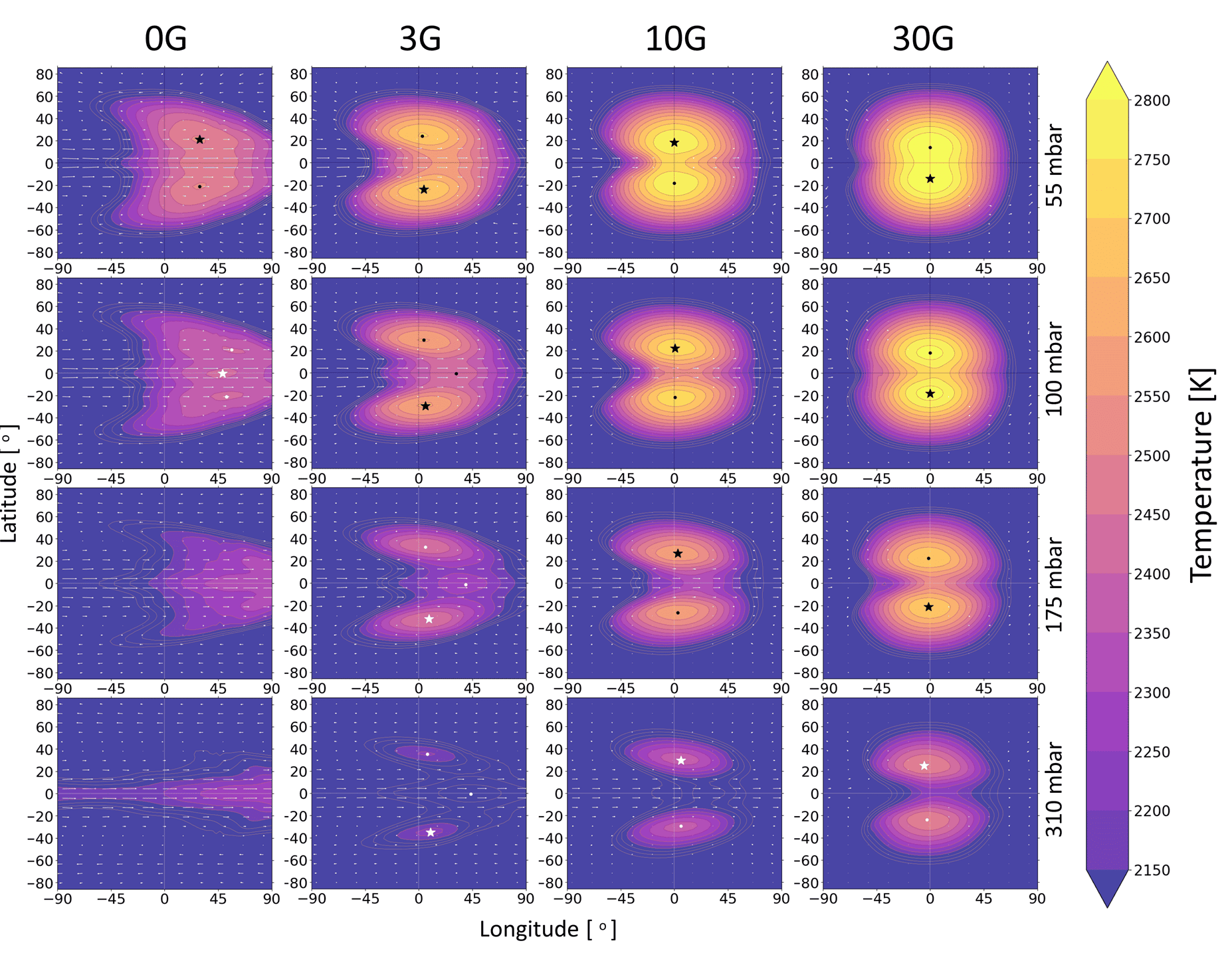}}
\vspace*{-0.8cm}
  \caption{\label{fig:Hot_notilt} High-contrast temperature contour maps for simulations with aligned magnetic dipole. Multiple pressure levels are shown between 50 and 300 mbar, with the isobaric global maxima indicated by the star, and local maxima by smaller dots. The left-hand column represents the 0G runs, and we see the classical chevron shaped hotspot feature, whereas on the right-hand 30G column, it has become a much more symmetric and regular shape due to the increased damping of the flow.}
\end{figure*}

\begin{figure*}
\vspace*{-1cm}
\makebox[\textwidth][c]{%
    \includegraphics[
        width=.95\textwidth,
        trim=4cm 10cm 19cm 4cm,   
        clip
    ]{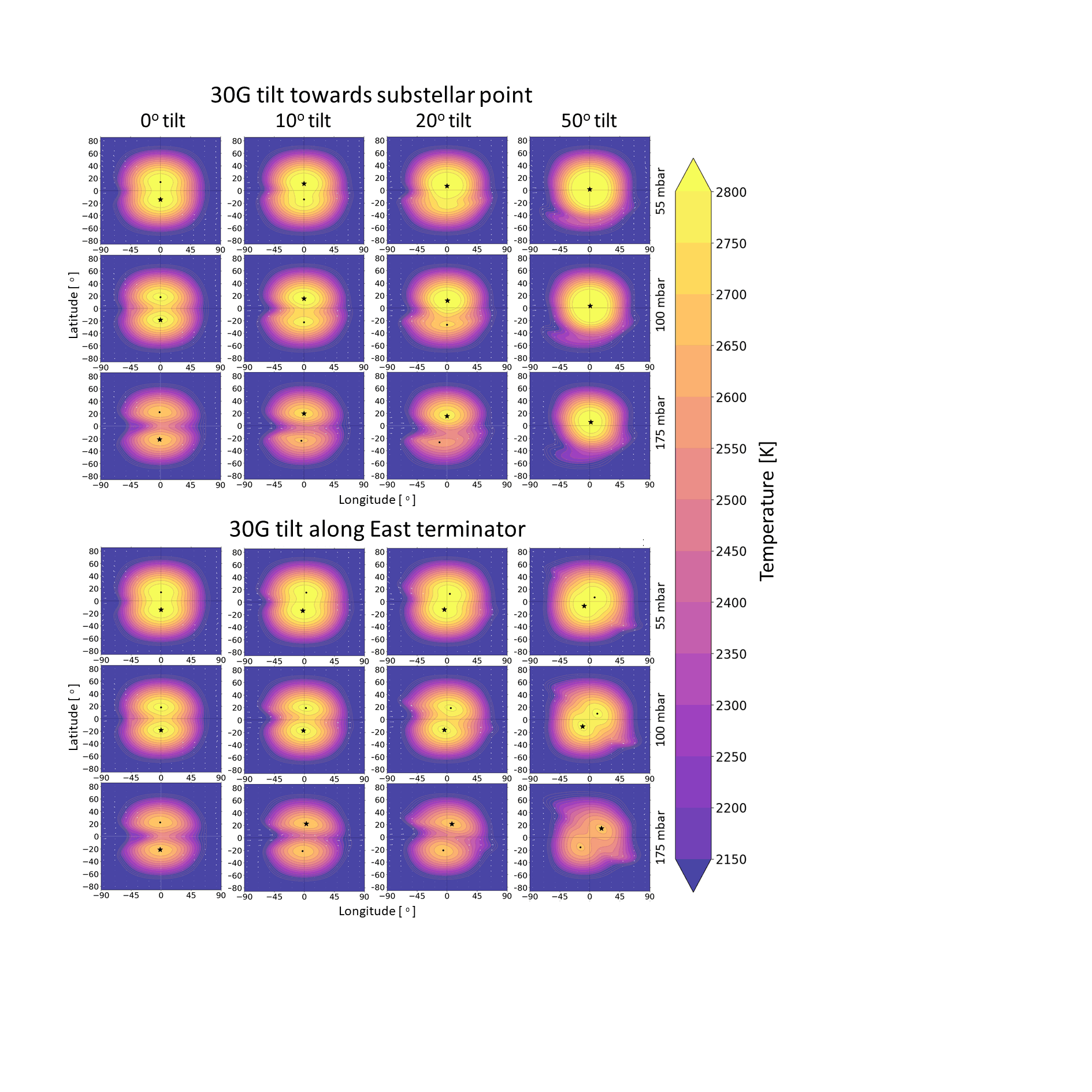}}
\vspace*{-0.5cm}
  \caption{\label{fig:hot_30G_big} High-contrast temperature contour maps for simulations with 30G magnetic field strength at various dipole orientations. Multiple pressure levels are shown between 50 and 300 mbar, with the isobaric global maxima indicated by the star, and local maxima by smaller dots. The top half displays runs with dipole tilts directed towards the substellar point, and the bottom half shows runs with dipole tilts along the east terminator. With the inclusion of a dipole tilt, the hotspot location can move latitudinally for some combinations of field strength and tilts.}
\end{figure*}

To examine the 3D temperature profile in more detail and identify hotspot location and shape, we present temperature contour maps plotted on isobars for various runs. Local maxima are marked with a dot, with the absolute maxima indicated by the star for clarity. The plotted location and value of the maxima were determined by averaging over adjacent cells with temperatures within 0.9996, about 1 K, of the maximum. This method smooths over the discretization and maxima to occur at 0$^\circ$ latitude and longitude. Fig. \ref{fig:Hot_notilt} displays these maps for the aligned-dipole cases, and Fig. \ref{fig:hot_30G_big} shows the 30G field strength simulations at a variety of orientations. The corresponding 3G and 10G cases are shown in \ref{fig:hot_3G_big} and \ref{fig:hot_10G_big} in Appendix \ref{app:figs}.

\begin{figure*}
  \centering
\vspace{-0.3cm}
  \includegraphics[width=1.0\linewidth]{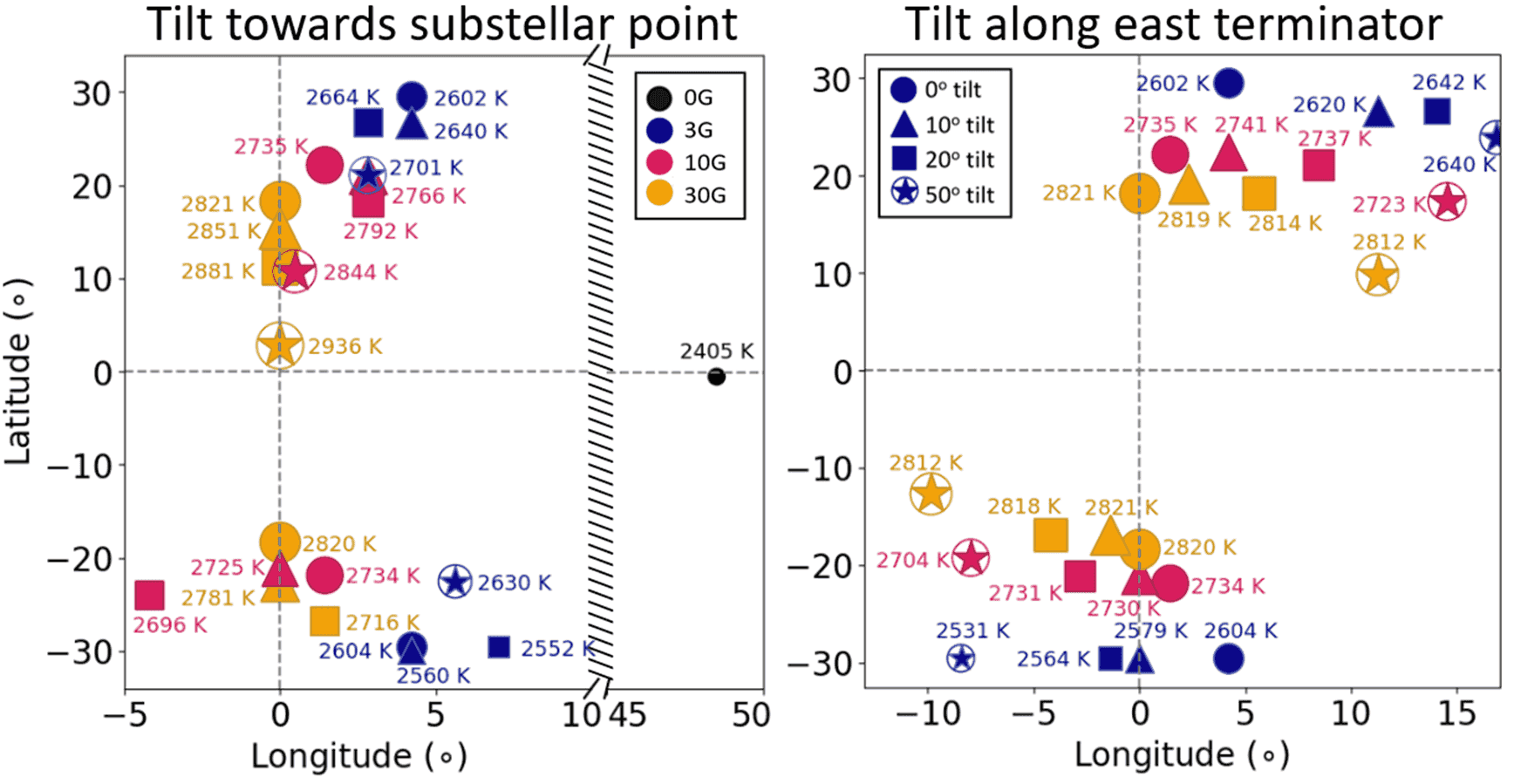}
  \vspace{-0.5cm}
  \caption{\label{fig:hotspot_locations}  Latitudinal and longitudinal locations and temperatures of the hotspots at the 100 mbar pressure level for various magnetic field strengths and tilts. The left plot shows cases of tilt away from the substellar point, and the right plot shows the runs with tilt away from the terminator. There are three independent variables being illustrated in this plot: the magnetic field strength (colour), polar tilt angle (shape), with increasing tilt/angle corresponding to increasing marker edges (0 = circle, then triangle, square, star), and azimuthal angle (substellar left, and terminator right). The size of the marker roughly corresponds to the temperature, the scaling is linear but not proportional to make the contrast between temperatures more evident. Note the break in the longitude axis on the left-hand plot. The inclusion of a dipole tilt can result in significant latitudinal hotspot offsets in both the north and south direction. }
  \vspace{-0.5cm}

\end{figure*}


These figures demonstrate the hotspot offset location and shape varies based on the magnetic dipole field strength and orientation. Generally, as magnetic field strength is increased, the eastward offset of the hotspot is decreased due to the stronger drag. This result is consistent with previous work \citep{Komacek_2016_constant_drag,Beltz_2022}. Fig. \ref{fig:Hot_notilt} shows that for the non-tilted cases, as the magnetic field strength is increased, the hotspot also becomes more symmetric and centred closer to the substellar point. Since the magnetic drag is weakest at the magnetic equator, as explained in Section \ref{sec:results_velocity}, the tilt of the field will also impact the general shape as well as the offset as shown in Fig. \ref{fig:hot_30G_big}. As a result, if the magnetic dipole is not aligned, the north–south asymmetry in the magnetic effects induces an asymmetry in the temperature profile. This is notable because most hydrodynamical effects cannot impart a north-south asymmetry. Shear instabilities \citep{Fromang2016} or storms \citep{SkinnerCho2022} could cause transient asymmetries. The MHD effects modelled here however, show persistent latitudinal asymmetries. 

To compliment these contour plots, Fig. \ref{fig:hotspot_locations} indicates the locations of the temperature local maxima at the 100 mbar level for various magnetic field strengths and orientations. 
Firstly, as seen in Figs \ref{fig:hot_30G_big} and  \ref{fig:hotspot_locations}, we note that the Northern hemisphere hotspots tend to be warmer than the Southern hemisphere hotspots for the substellar tilt runs. This is solely due to how we chose to orient our magnetic field (see the schematic in Fig. \ref{fig:schematic}). For the substellar tilt direction, the magnetic equator is located in the global South and so the Ohmic drag is stronger in the North. Were we to orient the dipole in the opposite direction, we would get a result that is a north–south mirror image of this. For most of the cases, Fig. \ref{fig:hotspot_locations} shows that as the obliquity of the magnetic field is increased, this also generally increases the temperature contrast between the Northern and Southern hemispheres, especially so for the substellar tilt cases.

It is clear from Figs \ref{fig:Hot_notilt}-\ref{fig:hotspot_locations} that the greatest effect on the 3D temperature and wind profile comes from including magnetic drag in the first place: this shifts the offset of the mid-latitude temperature maxima from almost 40\(^\circ\) to around 5\(^\circ\) longitude. As expected, for every substellar-directed tilt, the longitudinal hotspot offset is smaller for the stronger magnetic field cases.

Figs \ref{fig:Hot_notilt}-\ref{fig:hotspot_locations} demonstrate how the mid-atmospheric equatorial jet may cause a slight splitting of the hotspot using our magnetic drag scheme. We note however that in many of the simulated cases, distinct hotspots may not be detectable observationally
, such as the 55 mbar pressure level hot spots of Fig. \ref{fig:hot_30G_big}, where the difference between the maxima and saddle point temps are $\sim$3-20 K. In other cases, the temperature of the second maxima is much lower than the first, so in effect the second hotspot is absorbed into the first, such as the case of \(20^\circ\) substellar tilt in Fig \ref{fig:hot_30G_big}. In addition, the hotspot locations tend to be symmetric, with only relatively small temperature differences between the north and south hotspot. This causes the hemispheric centre of brightness to be similar to cases without hotspot splitting, which we demonstrate further in Section \ref {sec:results_phase_curves}. Lastly, we briefly note that due to our approximate magnetic drag scheme, this splitting effect in our simulation may be exaggerated. We discuss this in more depth in Section \ref{sec:discussion}.

For both the substellar and terminator tilt, there is a regime change between the 3G and the stronger 10G and 30G magnetic field cases. The 3G magnetic field case shows attributes of both the strong magnetic field models and the no-drag model; this is most clearly seen in the hotspot temperature map shown in Fig. \ref{fig:Hot_notilt}. Now, we consider the substellar and the terminator tilt cases individually. In the case of tilting the magnetic field towards the east terminator, we see in Fig. \ref{fig:hotspot_locations} that the 3G displays a mix of behaviours between the advective and strongly dragged regimes. Then, the 10G and 30G nicely show a pattern in Fig. \ref{fig:hotspot_locations} that with increasing magnetic field strength, not only does the latitudinal offset of the hotspot decrease, but the longitudinal offset of the two hotspots relative to each other increases more and more with increased dipole tilt.

When the dipole is tilted towards the substellar point, we see that we increase the magnitude of tilt, the northern hotspot in Figures \ref{fig:hot_3G_big} and \ref{fig:hot_30G_big} remains relatively fixed. However, the southern hotspot is advected slightly eastward for the 3G case and slightly westward for the 30G case. For the 3G case, this is because for greater magnitude of tilt away from the substellar point, drag in the northern hemisphere becomes weaker, and therefore the hotspot is able to be advected more effectively by the eastward equatorial jet that is the dominant feature of the atmosphere at these pressure levels. The mid-atmospheric equatorial jet does not form in our simulations with a field strength of 30G, so for the most extreme tilts or magnetic field strengths, the hotspot merges back into one, as there is no slice reaching the prescribed value of minimum drag through the dayside of the planet (e.g., see the 50\(^\circ\) tilt case of Fig. \ref{fig:hot_30G_big}). 

\begin{figure*}
  \centering
  \includegraphics[width=1.0\linewidth]{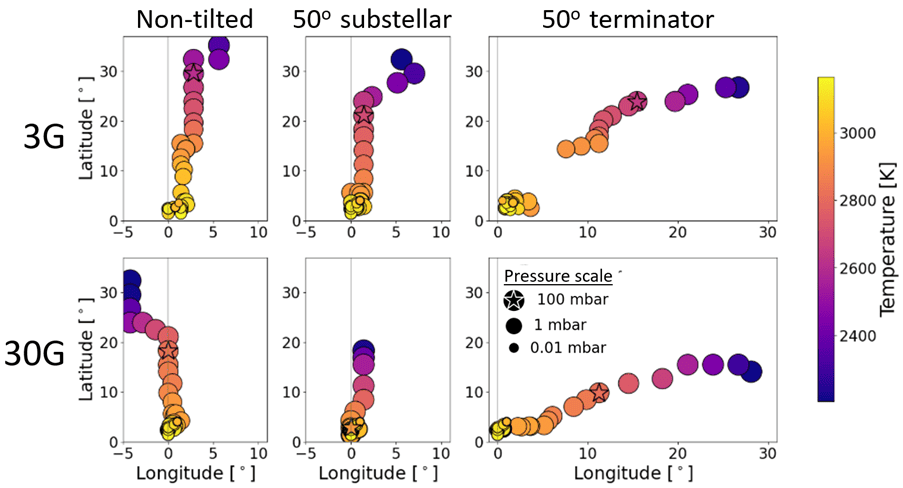}
  \vspace{-0.5cm}
  \caption{\label{fig:hotspot_pressure} Latitudinal and longitudinal locations and temperatures (marker colour) of the northern hotspots of various runs, as a function of pressure (marker size). Smaller marker sizes correspond to lower pressures, and the 100 mbar pressure level is indicated by a star for reference. The rows display different field strengths: 3G (top), and 30G (bottom). The columns correspond to different orientations: non-tilted (left), 50$^\circ$ towards the substellar point (middle), and 50$^\circ$ along east terminator (right).}
  \vspace{-0.5cm}
\end{figure*}

Lastly, in Fig. \ref{fig:hotspot_pressure} we show how the hotspot location and temperature vary with pressure. Grid point locations within $\sim 0.75$K of the temperature maximum are averaged due to the sparse gridding, yielding a more physically meaningful location and reducing overlap of points. At lower pressures, stronger instellation causes stronger magnetic drag, leading to minimal hotspot offsets. With increasing pressure, both the latitudinal and longitudinal location of the hotspot offset increases, until the temperature profile becomes nearly axially symmetric. At these depths, the hotspot could occur at any location near the poles or the equator, corresponding to the warmer latitudes. For all magnetic field strengths, tilting the dipole towards the terminator greatly increases the longitudinal hotspot offset. Increasing tilt magnitude generally decreases the latitudinal offset for both the substellar and terminator cases.  Stronger magnetic fields tend to decrease both the latitudinal and longitudinal hotspot offset at most pressure levels. In the non-tilted case, stronger magnetic fields preferentially cause westward longitudinal hotspot offsets at some levels. This is consistent with \cite{Hindle_2021}, predicting westward offsets in some regimes for stronger magnetic field strengths and warm enough equilibrium temperatures.

\subsection{Phase Curves}

\label{sec:results_phase_curves}

\begin{figure*}
\centering

\begin{subfigure}{.46\textwidth}
  \centering
  \includegraphics[width=\linewidth]{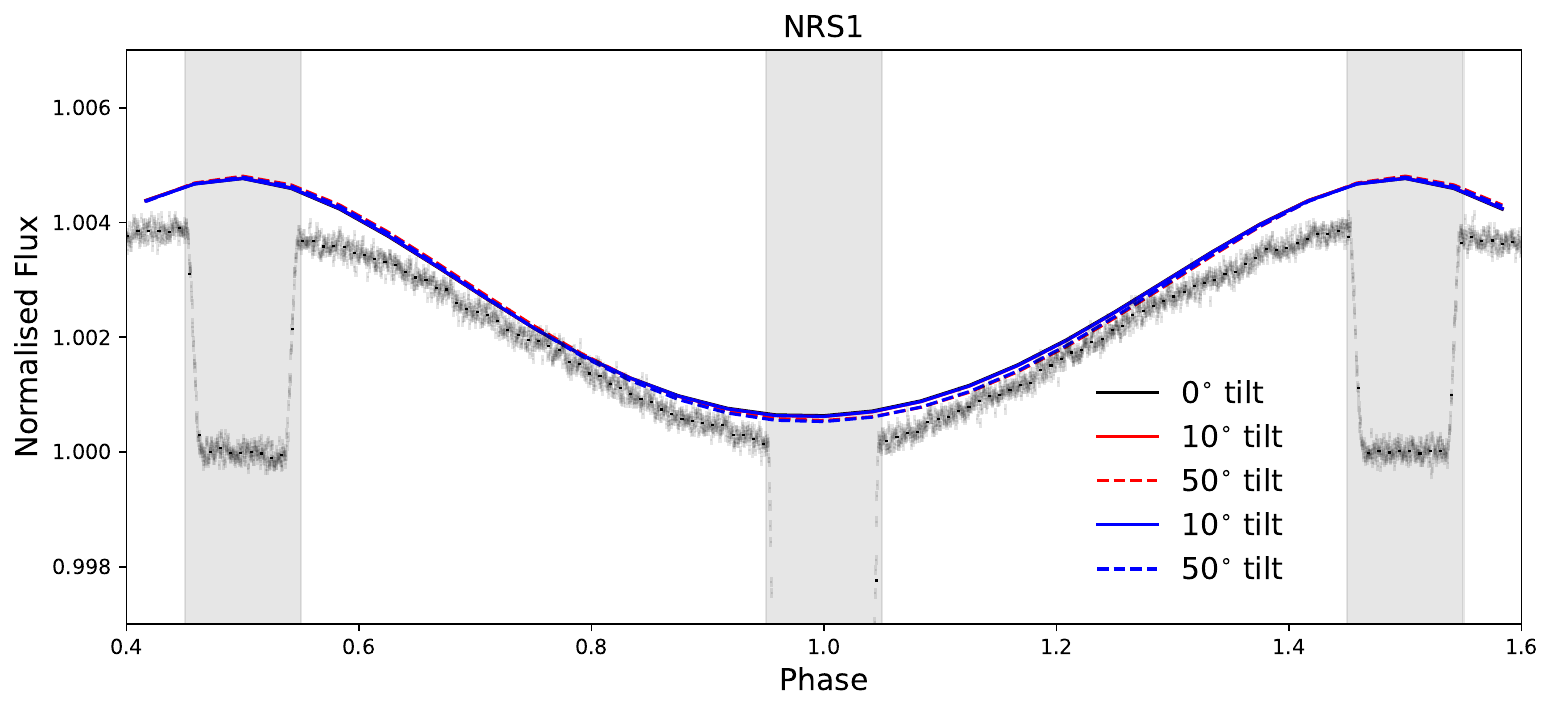}

  \label{fig:3G_nrs1}
\end{subfigure}%
\begin{subfigure}{.46\textwidth}
  \centering
  \includegraphics[width=\linewidth]{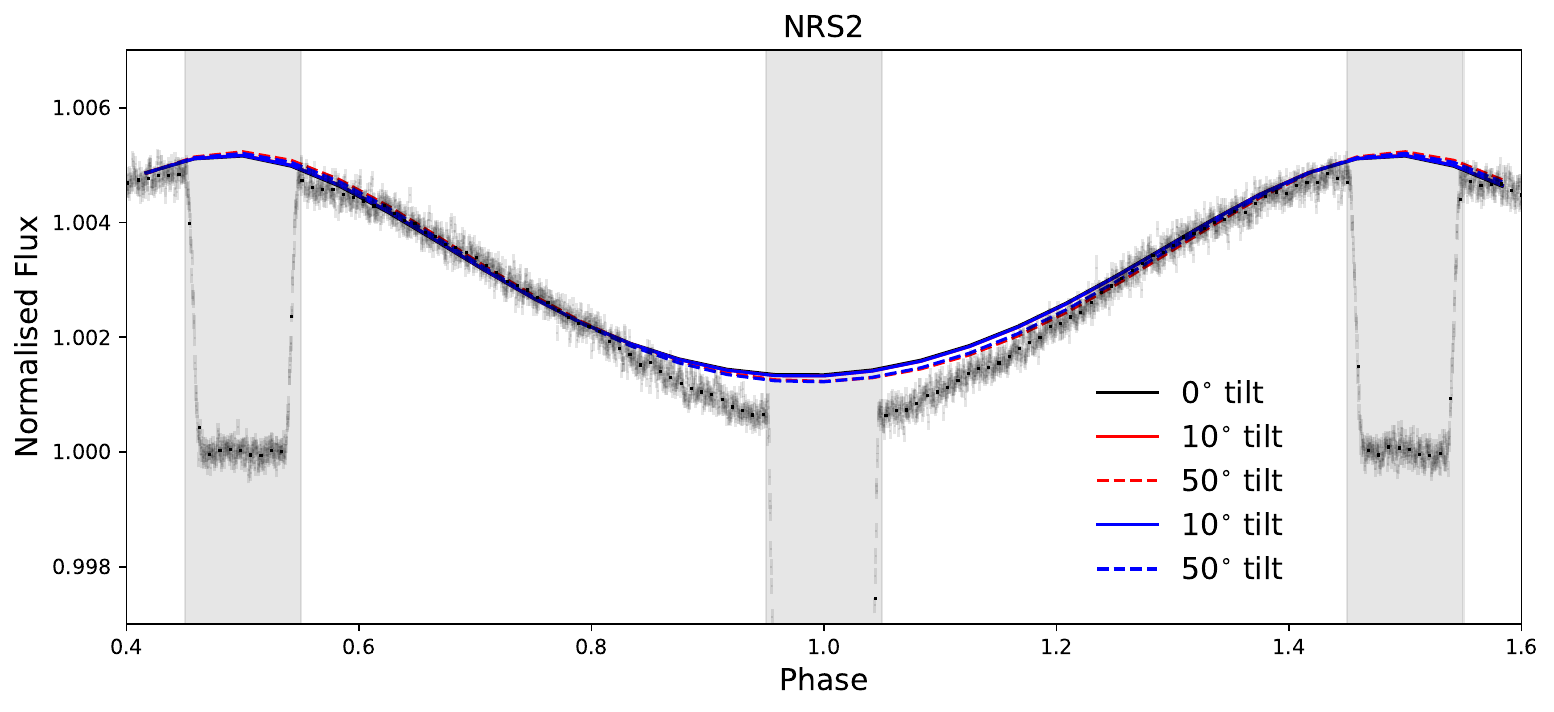}

  \label{fig:3G_nrs2}
\end{subfigure}

\begin{subfigure}{.46\textwidth}
  \centering
  \includegraphics[width=\linewidth]{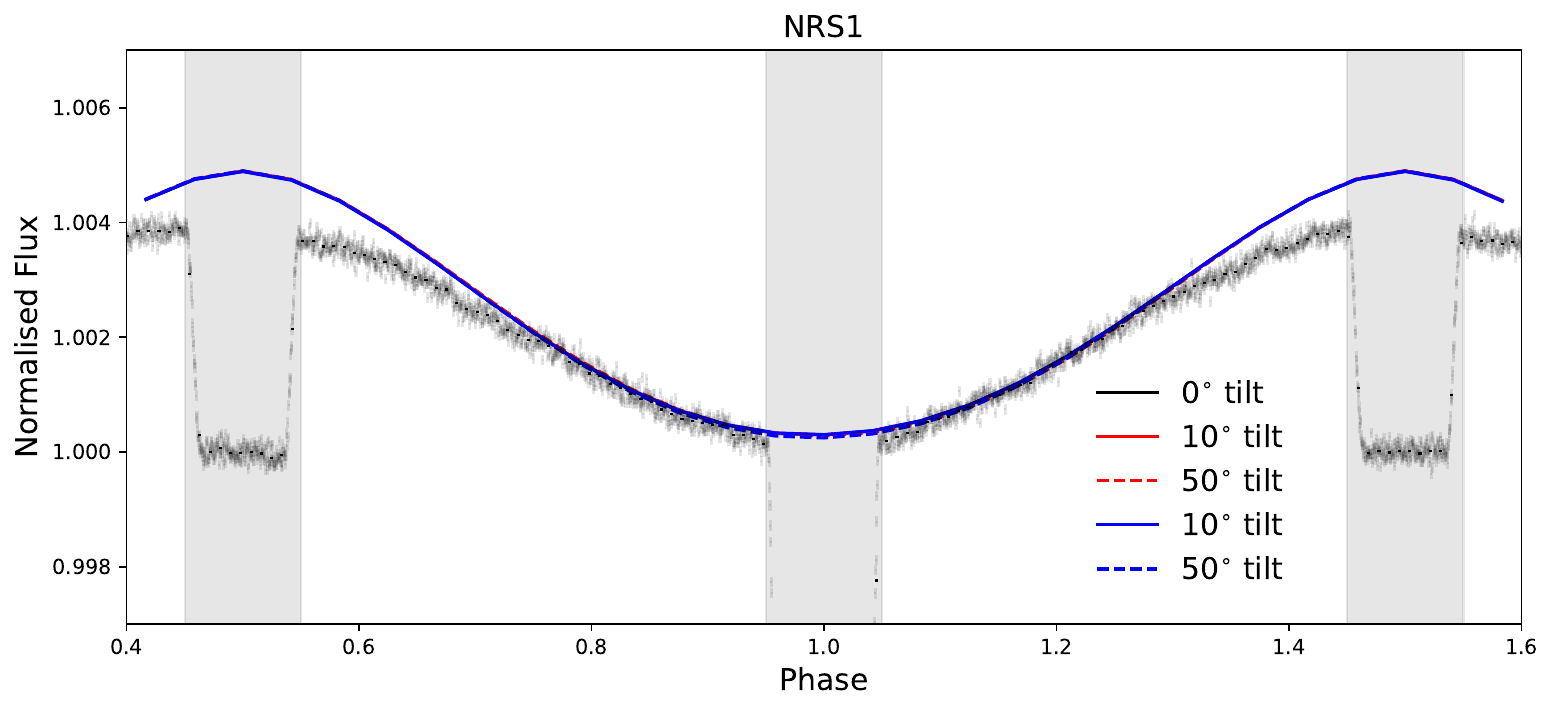}

  \label{fig:30G_nrs1}
\end{subfigure}%
\begin{subfigure}{.46\textwidth}
  \centering
  \includegraphics[width=\linewidth]{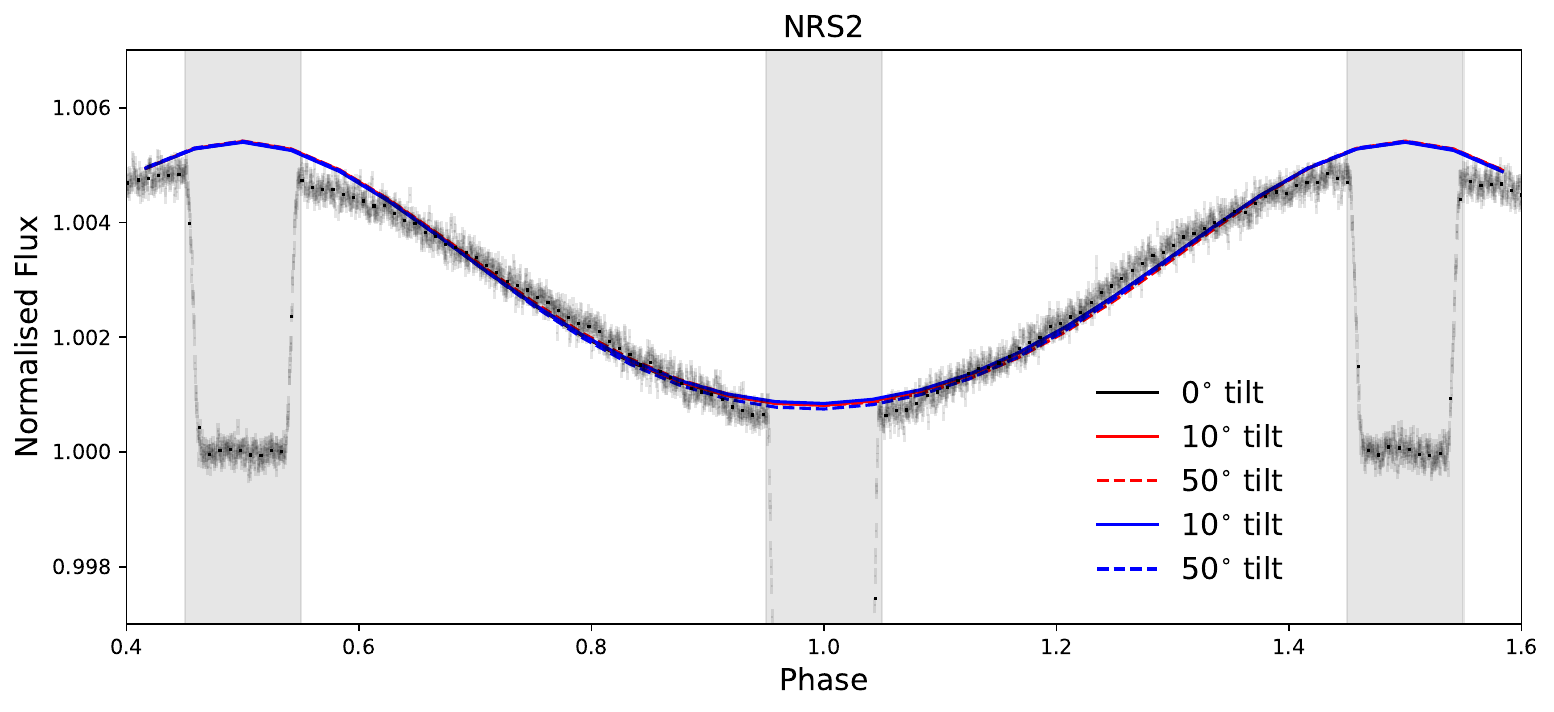}

  \label{fig:30G_nrs2}
\end{subfigure}
\caption{Top: Phase curves for planets with a magnetic field strength of 3G, at varying tilts, for both the \textit{JWST} NIRSpec detectors: NRS1 (left, $2.87 \rightarrow 3.69 \,\mu \mathrm{m}$) and NRS2 (right, $3.79 \rightarrow 5.14 \,\mu \mathrm{m}$). Bottom: The same, but for a planetary magnetic field strength of 30G. White-light phase curves generated using the \texttt{gCMCRT} 3D Monte-Carlo radiative transfer code. The observational differences between tilt cases are minimal, and are more pronounced for the 3G field strength cases. It is seen as unlikely therefore that these tilt effects would be observable with only white-light phase curves. The difference between planetary field strengths is far more pronounced. The no-tilt phase curve amplitudes (in normalised flux units $F_{\mathrm{p}}/F_{\star}$) for the 30G case are $4.6 \times 10^{-3}$ for NRS1 and $4.7 \times 10^{-3}$ for NRS2, and for the 3G case the amplitudes are $4.1 \times 10^{-3}$ for NRS1 and $3.9 \times 10^{-3}$ for NRS2. The simulated phase-curves are compared to \textit{JWST} NIRSpec/G395H observational data in grey, data sourced from  \protect\cite{Sing_2024}.}

\label{fig:phasecurves}
\end{figure*}

Fig. \ref{fig:phasecurves} shows the potential observational differences in white-light spectroscopic phase curves for extremes in magnetic field strength and tilt scenarios. The wavelength ranges chosen are for the \textit{James Webb Space Telescope (JWST)} NIRSpec NRS1 and NRS2 detectors, as discussed in Section \ref{sec:methods_post_processing}, given that this wavelength range has been studied in existing \textit{JWST} observations \citep{w121_PhaseCurve}. 

Fig. \ref{fig:phasecurves} shows clear differences in day–night heat transport between 3G and 30G magnetic field strength GCM runs; the mean difference in normalized flux across both detectors in amplitude for no tilt cases is $6.5 \times 10^{-4} \, F_{\mathrm{p}}/F_{\star}$. This is the expected result, as stronger magnetic drag would suppress dynamics and therefore the advection of heat around the planet from the hot dayside to the cool nightside \citep{Beltz_2022, sorianoguerrero2025nonidealmhdsimulationshot}. The impact of the tilt is less clear in white-light phase curves. In the lower field strength GCM runs, some differences are seen on the nightside part of the phase curve, however these effects are minimal (and in a real observation would be partly hidden by the transit of the planet itself). In the stronger magnetic field cases, all tilt scenarios are extremely similar. We therefore expect it would be challenging to observationally constrain the magnetic dipole tilt purely with white-light phase curves. However, the strength of the field could potentially be inferred from the phase curve, due to the differing amounts of day–night heat transport (as in e.g., \citealp{Arcangeli:2019aa}). 

\begin{figure*}
\centering

\begin{subfigure}{.46\textwidth}
  \centering
  \includegraphics[width=\linewidth]{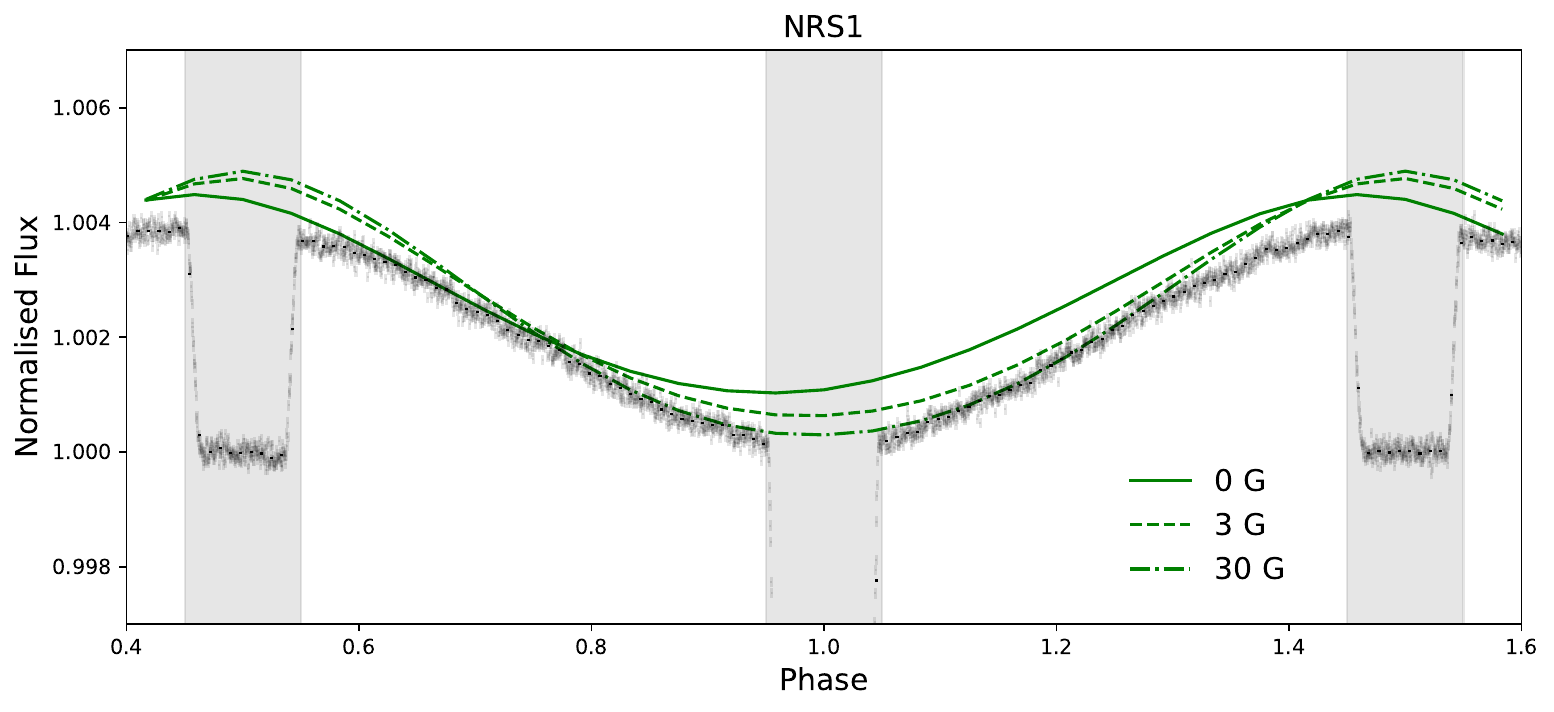}

  \label{fig:comp_nrs1}
\end{subfigure}%
\begin{subfigure}{.46\textwidth}
  \centering
  \includegraphics[width=\linewidth]{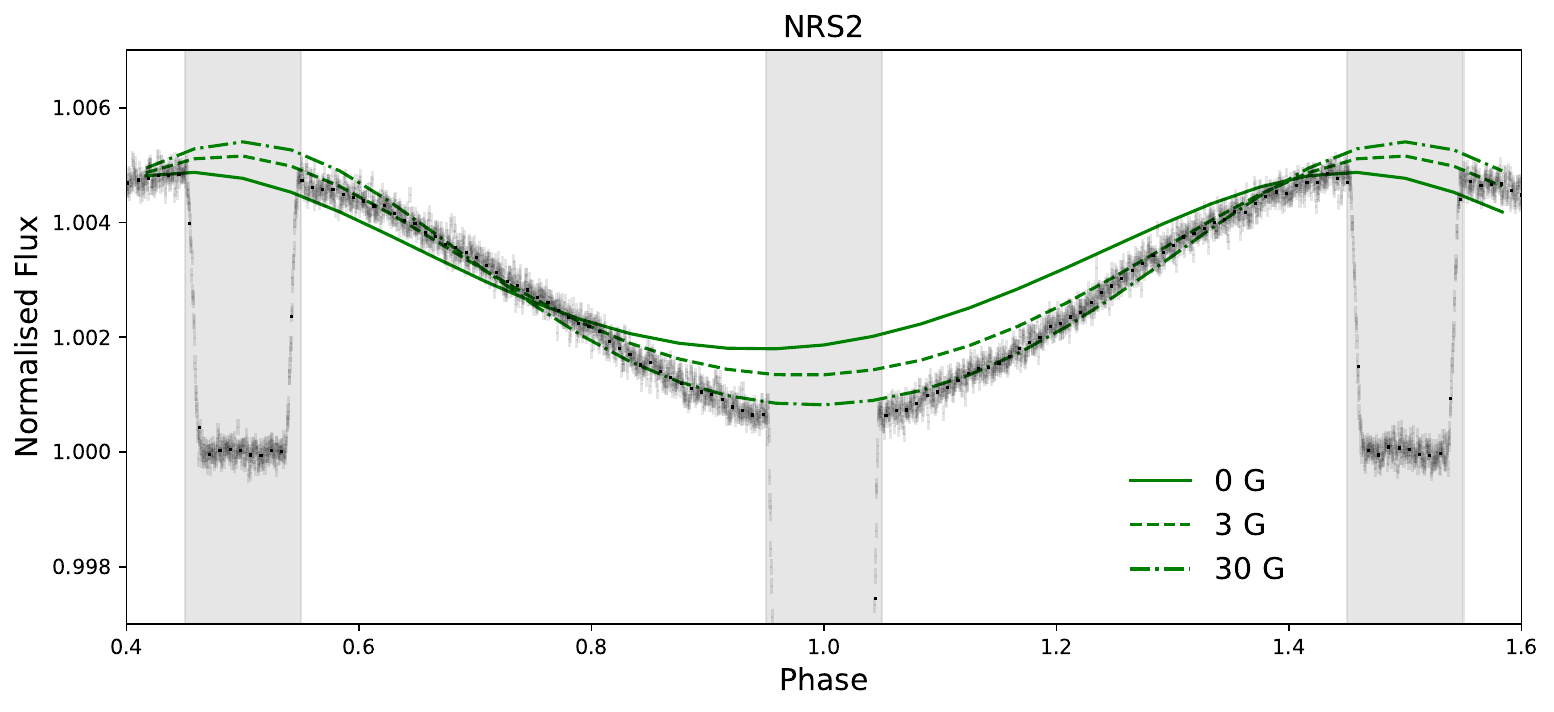}

  \label{fig:comp_nrs2}
\end{subfigure}

\caption{$0^{\circ}$ tilt cases phase curves for field strengths of 0G, 3G, and 30G. These are shown for both the NRS1 (left) and NRS2 (right) detector ranges. In the NRS1 wavelength range the amplitudes of the phase curves (in normalised flux units $F_{\mathrm{p}}/F_{\star}$) for the 0G, 3G, and 30G magnetic field scenarios are $3.2 \times 10^{-3}$, $3.9 \times 10^{-3}$, and $4.7 \times 10^{-3}$. In the NRS2 wavelength range the corresponding amplitudes are $3.5 \times 10^{-3}$, $4.2 \times 10^{-3}$, and $4.6 \times 10^{-3}$. The simulated phase-curves are compared to \textit{JWST} NIRSpec/G395H observational data in grey, data sourced from \protect\cite{Sing_2024}.}

\label{fig:mag_comparison}
\end{figure*}

{Comparing the simulated phase-curves to  observational data (sourced from \citealp{Sing_2024}), as shown in Figures \ref{fig:phasecurves} and \ref{fig:mag_comparison}, we can see that the presence of some drag mechanism is required to fit the data. It is difficult to evaluate whether strong or weak magnetic fields are required, however, as both cases overpredict the dayside flux in the NRS1 wavelength range. The observational data appear to favour the 30G field scenarios, and there is no way to distinguish between dipole tilts with white-light phase curve data alone. }

Fig. \ref{fig:mag_comparison} shows the zero tilt phase curves for fields of 0G, 3G, and 30G. The phase curve amplitudes vary significantly, with a maximum difference in amplitude (in units of normalised flux) being $1.5 \times 10^{-3} \,F_{\mathrm{p}}/F_{\star}$, or 150 ppm. The error in the JWST NIRSpec observations for normalised flux ($F_{\mathrm{p}} / F_{\star}$) are approximately $\sim$ 10 ppm \citep{w121_PhaseCurve}, therefore this difference in phase curve amplitude due to magnetic field strength would be detectable.

\section{Discussion}

\label{sec:discussion}

In this section, we delve into an analysis of the assumptions made in these models: neglecting the Hall and ambipolar term, assuming the magnetic Reynolds number is small, the thin atmosphere approximation, and the hydrostatic approximation. We also discuss some other sources of error or uncertainty, such as boundary currents, or the influence of the solar magnetosphere in shaping the magnetic field of the hot Jupiter. We conclude this section with a discussion on the validity of this modelling approach as well as future avenues of research.

\label{sec:discussion}

\subsection{Validity of Approximations}
\subsubsection{Neglecting the Hall and ambipolar terms}
\label{sec:discussion_HOAO}
The Hall and ambipolar terms were neglected in Section \ref{sec:hall_ambipolar} on the basis that they are expected to be small compared to the Ohmic term, allowing a simplified calculation for the magnetic drag. The Hall term acts perpendicularly to the field and is non-dissipative, and therefore does not contribute to heating. The ambipolar effect also acts perpendicular to the field but is dissipative, and may therefore have a more direct impact on the temperature and velocity profiles. Neglecting the ambipolar term is equivalent to simplifying the full Pedersen effect to Ohmic drag. Here, we briefly return to the full induction equation (\ref{equ:induction_full2}), and find the relative magnitudes of the Ohmic, Hall, and ambipolar terms using the scaling introduced in equation (\ref{equ:HO_AO}):

\begin{equation}
\label{equ:HO_AO2}
    \mathrm{\frac{Hall}{Ohmic}}\approx\frac{cB}{4\pi\eta en_e}\qquad\qquad\mathrm{\frac{Ambipolar}{Ohmic}}\approx\frac{B^2}{4\pi\eta\gamma\rho_i\rho}
\end{equation}

\begin{figure}
  \centering
  \includegraphics[width=1\linewidth]{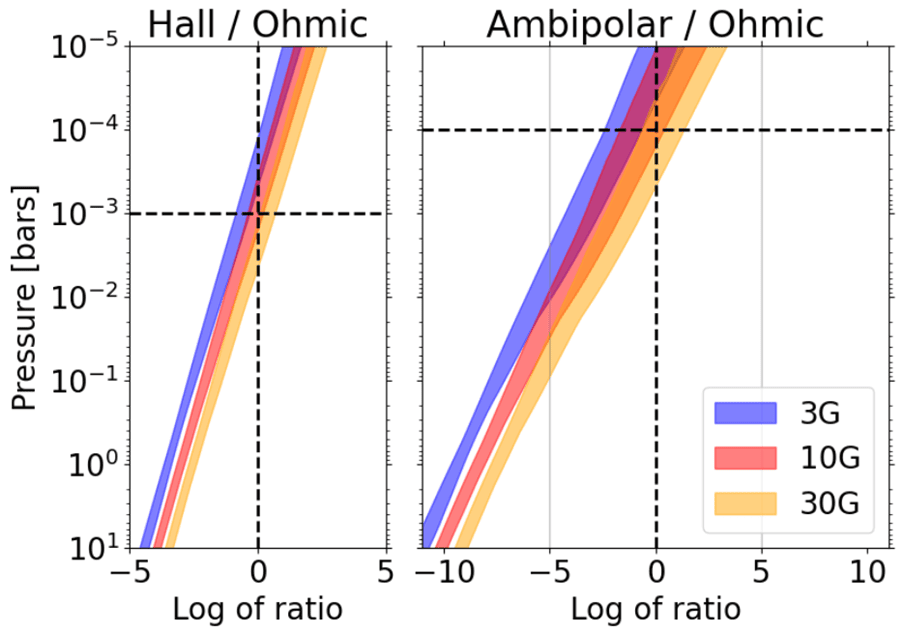}
\vspace{-0.3cm}
  \caption{\label{fig:HO_AO} Hall/Ohmic (left) and ambipolar/Ohmic (right) ratios from non-tilted runs for various magnetic field strengths (colours). Within a given run, these ratios vary most strongly with pressure. The width of each line captures the range of values these ratios take within a fixed pressure level, encompassing a range of temperatures. The horizontal dotted line indicates the order of magnitude estimate made in Section \ref{sec:hall_ambipolar}.}
\end{figure}

We calculate these ratios in every grid point for all the simulations. As they show little dependence on the dipole orientation, we focus on the 3G, 10G, 30G non-tilted cases. Fig. \ref{fig:HO_AO} displays the log of the ratios in equation (\ref{equ:HO_AO2}) as a function of pressure for the non-tilted runs. Consistent with equation (\ref{equ:HO_AO_BTp}), the gradient of these ratios scales most strongly with \(\mathrm{\frac{H}{O}}\propto B\sqrt{T}/p\) and \(\mathrm{\frac{A}{O}}\propto B^2\,T^{3/2}/p^2\). These results confirm that the order of magnitude estimates derived in Section \ref{sec:hall_ambipolar} are remarkably accurate for a hydrostatic atmosphere: if the equilibrium temperature of the planet is known, such scaling can reliably estimate the pressures at which the Hall and ambipolar terms may start to matter. 

\begin{figure*}
  \centering
  \includegraphics[width=1\linewidth]{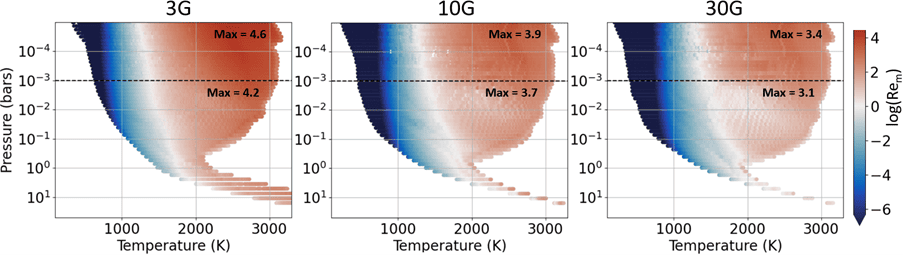}
\vspace{-0.3cm}
  \caption{\label{fig:Rem} Magnetic Reynolds number as a function of pressure and temperature for non-tilted runs of each magnetic field strength. Red values represent Re\(_m>1\), where the induced magnetic field and resulting effects may start to matter. We additionally report the value (in log space) of largest $Re_{m}$ above and below 1 millibar. In our analysis, we averaged over the \(\mathrm{Re_m}\) that are within 10 K of each other, which slightly amplified large magnetic Reynolds numbers, as seen deeper in the atmosphere. Although the magnetic Reynolds number is high in the deeper atmosphere in some places, this is likely less important since the ionization fraction is smaller here and magnetic effects are weaker.} 
\end{figure*}

Overall, the Ohmic dissipation term is dominant for much of the atmosphere. However, in the upper pressure layers p~\(\lesssim\)~1mbar and p~\(\lesssim\)~0.1mbar, the Hall/Ohmic and ambipolar/Ohmic ratios, respectively, become larger than unity on the daysides. As noted previously, the GCM is less accurate at the lowest pressures simulated due to artificial top boundary conditions. Therefore, the approximation of neglecting the Hall and ambipolar terms starts to break down in the portion of that atmosphere where numerical errors also become significant. A more physically consistent treatment of the upper atmosphere would require both the implementation of the Hall and ambipolar effects, and also increasing the number of pressure layers or decreasing the top boundary pressure in the simulation.

Our calculated ratios are similar to those estimated by \cite{Perna2010a} using scaling arguments, although they estimated these ratios to be smaller by \(\approx\)1-3 orders of magnitude. This discrepancy is likely a result of differences in planetary temperatures, as here we focus on ultra-hot Jupiters rather than the cooler hot Jupiters. The behaviour of the plotted ratios in Fig. \ref{fig:HO_AO} are similar to those in the work of \cite{christie2025geometricconsiderationshotjupiter}, who used a frozen-in magnetic field model and assumed neutral-species-dominant dynamics. \cite{christie2025geometricconsiderationshotjupiter} worked with a more complete calculation of conductivities, accounting for the full Pedersen conductivities, rather than simplifying to the Ohmic effect. In general, the full Pedersen effect would damp the winds slightly less compared to the Ohmic effect alone \citep{christie2025geometricconsiderationshotjupiter}. 

More recently, \cite{blocker2026inhomogeneousmagneticcouplingexoplanets} incorporated the Hall term within {a similar framework to \cite{christie2025geometricconsiderationshotjupiter}}. In their simulations, they showed that the Hall effect causes the isobaric flow to slightly more symmetrically diverge from the sub-stellar point compared to models that neglect it. For our particular simulated case of WASP-121b, the Hall effect only becomes significant around 1 mbar, where the isobaric flow already exhibits this diverging character (see Figs \ref{fig:temp_wind_3G}, \ref{fig:temp_wind_30G}), suggesting that additional dynamical impact is likely minimal in this regime.

\subsubsection{Small $\mathrm{Re_m}$, neglecting induced magnetic field}
\label{sec:discussion_Rem}

To compare to similar analyses of previous works, in Fig. \ref{fig:Rem} we plot the magnetic Reynolds number, approximated by equation (\ref{equ:Rem_raw}):

\begin{equation}
    \mathrm{Re_m}\approx\frac{\mathcal{UH}}{\eta}\approx\frac{\sqrt{u^2+v^2}\,R_aT}{g\eta}
\end{equation}

\(\mathcal{H}\) is the local scale height and \(\mathcal{U}\) is the horizontal velocity on the isobaric surface, with \(\eta\) calculated as in \cite{Rauscher_Menou_2013}. We calculate \(\mathrm{Re_m}\) using simulation outputs for the 3G, 10G and 30G non-tilted cases. As the magnetic field strength is increased, \(\mathrm{Re_m}\) decreases due to the increased damping of the winds, meaning that stronger internal magnetic fields induce proportionally weaker atmospheric magnetic fields. The orientation of the internal dipole does not significantly impact \(\mathrm{Re_m}\). While our focus here is on how different background magnetic fields influence the bulk \(\mathrm{Re_m}\), we note that many other factors can also affect it. For example, \cite{christie2025geometricconsiderationshotjupiter} showed that planets with greater instellation would have greater \(\mathrm{Re_m}\) because this corresponds to smaller resistivity on the daysides.

In general, the pattern observed is similar to Fig. 1 in \cite{Beltz_2022}: \(\mathrm{Re_m}\) is much smaller on the nightsides, and largest at high temperatures and shallow pressures, corresponding to the upper atmosphere daysides. But, our magnetic Reynolds number is a few orders of magnitude larger than that of \cite{Beltz_2022}. This is likely due to a few key differences between our models. First, our magnetic drag models were different: our model accounted for zonal and meridional drag due to zeroth order horizontal currents, whereas theirs only focused on zonal drag, due to their assumption of a completely aligned dipole and small meridional velocities. We also simulated a hotter planet: WASP-121b, T\(_{\mathrm{eq}}\sim 2400\)K \citep{Sing_2024}, while \cite{Beltz_2022} focused on WASP-76b, T\(_{\mathrm{eq}}\sim 2200\)K \citep{Ehrenreich_2020}. Additionally, we used a correlated-k radiative transfer scheme, whereas they used a double-grey. Thus, our upper atmosphere has a stronger thermal inversion due to absorption by metal oxide species and therefore has a higher ionisation fraction (and much lower resistivity). 

\cite{rogers2017} and \cite{Dietrich2022} argued that in the upper atmospheres of some hot Jupiters, Re\(\mathrm{_m}\) is expected to be $  > O(1).$ \cite{rogers2017} used an analytic model to quantify that the critical Re\(\mathrm{_m}=UL/\eta\) required for the onset of dynamo growth and instability of the induced field is \(\gtrsim O(10)\), values of which we do see in the uppermost atmospheres of our models (see \citealp{Dynamo_review}). \cite{Boening2025} further showed that in the upper atmosphere on the dayside, the strength of the induced magnetic field due to atmospheric currents may be up to many times stronger than the intrinsic magnetic field. In order to properly quantify the atmospheric dynamics of of ultra-hot Jupiters, it is very likely that a fully coupled MHD + GCM model will be required as the feedback between magnetic and hydrodynamic effects is highly non-trivial and non-analytic. Furthermore, a self-sustaining atmospheric dynamo could couple with the internal dynamo of the planet, further altering the shape and strength of the magnetic field. 

\subsubsection{Neglecting \(F_r\): the hydrostatic approximation}
\label{sec:discussion_Fr}

As discussed in Section \ref{sec:theory_tau}, in order to justify neglecting the radial Lorentz force, the hydrostatic timescale should be smaller than the radial magnetic timescale (${\tau_\mathrm{pres}}/{\tau_{\mathrm{mag,}r}}<<1$). This ratio is roughly given by:

\begin{equation}
    \frac{\tau_\mathrm{pres}}{\tau_{\mathrm{mag,}r}}=\left|\frac{F_{\mathrm{mag,}r}}{F_\mathrm{pres}}\right|=\left|\frac{\frac{1}{c}(J_\theta B_\phi-J_\phi B_\theta)}{\partial_rp}\right|
\end{equation}

\begin{figure}
  \centering
  \includegraphics[width=1\linewidth]{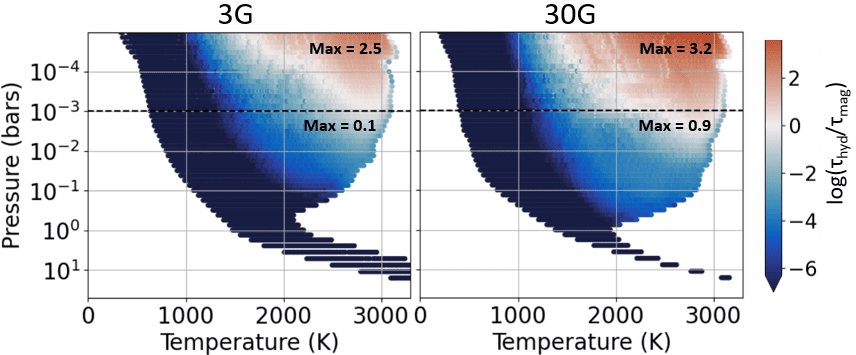}
\vspace{-0.3cm}
  \caption{\label{fig:hyd} Ratio of the hydrostatic timescale to the vertical magnetic drag timescale. {We also include the maximum (in log space) of this ratio above and below 1 millibar.} For most of the entire modelled atmosphere, the hydrostatic timescale is much shorter, and therefore a much more significant effect, than the vertical magnetic drag timescale. Thus, we anticipate that hydrostatic balance is a good approximation.}
\end{figure}

This timescale ratio is most strongly dependent on temperature and pressure, and is therefore maximised at low pressures and high temperatures. We plot ${\tau_\mathrm{pres}}/{\tau_{\mathrm{mag,}r}}$ in a similar fashion to $\mathrm{Re_m}$ in Figure \ref{fig:hyd}.

Hydrostatic equilibrium is a valid approximation for most of the atmosphere as seen by the blue regions of Fig. \ref{fig:hyd}, aside from the uppermost atmosphere (pressures \(\lesssim1\)mbar in the 30G case). In comparison to the $\mathrm{Re_m}$ plot, a greater source of error comes from neglecting the induced magnetic field due to the atmospheric currents than from neglecting the radial Lorentz force, especially in the middle atmosphere where $\mathrm{Re_m}$ has strong spatial variation. Radial Lorentz force could be implemented for greater completeness, but would likely have a minimal effect on the overall dynamics.

\subsubsection{Approximating the currents}
\label{sec:discussion_pert}

In equations (\ref{equ:J_phi_full}) and (\ref{equ:J_theta_full}), we neglected the integral terms under the thin atmosphere approximation. The \(u_r\) and \(J_r\) terms were neglected for this reason. We motivate these simplifications by deriving an approximate solution to the current more rigorously by finding the asymptotic expansion for the currents. This allows us, in postprocessing, to locate regions of divergence of the series and explain why the approximation breaks down. Here, we estimate the error resulting from approximating \(\nabla\times \mathbf{E}=0\) and \(\nabla\cdot \mathbf{J}=0\).

\paragraph*{Perturbative approximation of J:}

Here we perturbatively expand in terms of \(\Delta r/R<<1\), where $R$ is the total radius of the planet, and \(\Delta r\) is the thickness of the atmosphere, specifically from a given point to the top of the model atmosphere ($\Delta r = R-r$).\footnote{A similar expansion can alternatively be performed in terms of spherical harmonics, as was done in \cite{Vigan_2025_1d_deep_atmos_mhd}. This would be more useful for 1D models or GCMs that work in expansions of spherical harmonics.} Writing \(J_r\), the spherical polar radial current component, in terms of such an asymptotic expansion:

\begin{equation}
\label{equ:pertJ}
    J_r \sim j_r^{(0)} + \frac{\Delta r}{R}j_r^{(1)}+ \left(\frac{\Delta r}{R}\right)^2j_r^{(2)}+...\equiv J_r^{(0)} + J_r^{(1)}+ J_r^{(2)}+...
\end{equation}

Where \(J_r^{(i+1)}=o(J_r^{(i)})\) in the limit \(\Delta r/R\to 0\) is a necessary condition by definition of asymptotic series, otherwise the different orders start to mix. For convergence, we further require \(\left|J_r^{(i)}\right|>\left|J_r^{(i+1)}\right|\) to be true for finite $\Delta r/R$ that exist in the atmosphere. A similar series is constructed for \(J_\theta\) and \(J_\phi\). Assuming steady state, we have three equations we need to solve: two independent components of \(\nabla\times\mathbf{E}=0\) and \(\nabla\cdot\mathbf{J}=0\). Treating \(\Delta r/R\) as our perturbative parameter allows us to assume that \(r\sim R\) and for a general function \(f\) of position:




\begin{equation}
\int_r^Rf(r,\theta,\phi)\,dr=O(\Delta r\cdot f(R,\theta,\phi))
\end{equation}

And as before, we take \(u_r = O(u_\theta \,\Delta r/R)= O(u_\phi \,\Delta r/R)\). We assume simplified boundary conditions that horizontal currents vanish at the top of the model atmosphere \citep{LIU_2008,Perna2010a}, but generally consider that radial boundary currents are non-zero and \(O(J_r^{(0)})\). Substituting (\ref{equ:pertJ}) into \(\nabla\times\mathbf{E}=0\) and \(\nabla\cdot\mathbf{J}=0\) (equations (\ref{equ:MHD_ideal})) and matching like orders of \(\Delta r/R\), we get:

\begin{align}
\label{equ:J_asym_0order}
    J_r^{(0)}=\frac{J_r(R,\theta,\phi)}{r^2}\qquad
    J_\theta^{(0)}=\frac{cu_\phi B_r}{4\pi\eta}\qquad
    J_\phi^{(0)}=-\frac{cu_\theta B_r}{4\pi\eta}
\end{align}


\begin{subequations} 
\label{equ:J_asym_1order}
\begin{align}
    &J_r^{(1)}=\frac{1}{r^2\sin\theta}\int_r^Rr\left[\partial_\theta(J_\theta^{(0)}\sin\theta)+\partial_\phi J_\phi^{(0)}\right]dr\\
    &J_\theta^{(1)}=\frac{c}{4\pi\eta}\left[-u_rB_\phi+\frac{1}{r}\partial_\theta\int_r^R G\,dr\right]\\
    &J_\phi^{(1)}=\frac{c}{4\pi\eta}\left[u_rB_\theta+\frac{1}{r\sin\theta}\partial_\phi\int_r^R G\,dr\right]\\
    &\quad\mathrm{where:}\qquad G\equiv u_\theta B_\phi-u_\phi B_\theta - \frac{4\pi\eta}{c}J_r^{(0)}
\end{align}
\end{subequations}

In equation (\ref{equ:J_asym_0order}), the \(J_r^{(0)}\) term corresponds to the boundary current at the top of the atmosphere. In an ideal steady state, the integral of $J_r(R,\theta,\phi)$ over solid angle would be 0, corresponding to no net charge inflow. In our model, for simplicity, we impose the boundary condition that the radial currents vanish at the top model boundary, following previous work \citep{Perna2010a}. Under this assumption, the first non-zero radial current term in the asymptotic series would be \(J_r^{(1)}=O(J_\theta\,\cdot\Delta r/R)\). The remaining terms in the asymptotic expansion for n\(\geq 1\) are given by:
\begin{subequations}
\label{equ:J_asym_norder}
\begin{align}
    &J_r^{(n+1)}=\frac{1}{r^2\sin\theta}\int_r^Rr\left[\partial_\theta\left(J_\theta^{(n)}\sin\theta\right)+\partial_\phi J_\phi^{(n)}\right]dr\\
    &J_\theta^{(n+1)}=-\frac{1}{\eta\,  r}\partial_\theta\int_r^R\eta J_r^{(n)}dr\\
    &J_\phi^{(n+1)}=-\frac{1}{\eta\,  r\sin\theta}\partial_\phi\int_r^R\eta J_r^{(n)}dr
\end{align}
\end{subequations}

\begin{figure}
  \centering
  \includegraphics[width=1\linewidth]{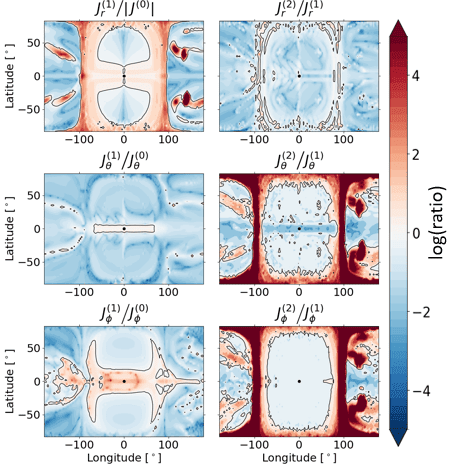}
\vspace{-0.3cm}
  \caption{\label{fig:J_asymapprox} Ratios of given terms in the current asymptotic expansion calculated for the 30G no tilt case at a pressure level of 10 mbars. The black contour represents when a given ratio is unity. The left side represents the ratio of first-order currents to zeroth-order currents and the right side is the ratio of second-order currents to first-order currents. The top row displays the radial currents, the second row the meridional, and the bottom row shows the zonal currents. At this pressure level, divergence (red colour) in the first-order terms occurs most clearly at the terminator and equator for the radial current.}
\end{figure}

Next, we will examine where approximating this expansion as the zeroth order term (Equation (\ref{equ:J_asym_0order})) is accurate, and where the expansion breaks down. Fig. \ref{fig:J_asymapprox} displays ratios of higher- to lower-order terms in the asymptotic expansion of the current (equations (\ref{equ:J_asym_0order}), (\ref{equ:J_asym_1order}), and (\ref{equ:J_asym_norder})) evaluated using atmospheric conditions from the 30G non-tilted model at the 10 mbar level. Ratios of less than unity (blue in the figures) indicate regions where the expansion is convergent and the approximation is valid, while ratios exceeding unity (red in the figures) indicate where the asymptotic expansion breaks down or becomes non-convergent. 

\paragraph*{Limitations of the thin atmosphere approximation:}
\label{sec:discussion_pert_breakdown}

There are three regions in which the thin-atmosphere approximation may not be appropriate, two of which are evident in Fig. \ref{fig:J_asymapprox}.

\begin{enumerate}[label=(\roman*), leftmargin=*, align=left, itemsep=1em]

\item At the magnetic equator, the radial magnetic field vanishes, causing all the zeroth-order currents to vanish as well (equations (\ref{equ:J_asym_0order})). As a result, the leading order term in the asymptotic expansion is first order, which we neglected in our approximation, as seen in the lower two subplots in the left column of Fig. \ref{fig:J_asymapprox}. 

\item Near the terminators, the ratio of first-order radial current to zeroth-order horizontal currents is much larger than unity. This behaviour propagates to the second-order horizontal currents, and subsequently appears as a correction every subsequent order, demonstrating a non-convergence of the series in these regions. Although asymptotic, the series diverges near the terminator because the perturbative parameter is not infinitesimally small, and the coefficients of the higher order terms are sufficiently large to violate the perturbative ordering. Physically, this reflects strong horizontal current divergence (\(\nabla_h\cdot \mathbf{J}\)) driven by sharp day-night contrasts in temperature, ionisation, and resistivity. Conservation of charge requires strong current upwelling or downwelling into adjacent pressure layers, and this interaction between pressure layers cannot be consistently captured within this thin-atmosphere expansion framework.

\end{enumerate}

\begin{figure}
  \centering
  \includegraphics[width=1\linewidth]{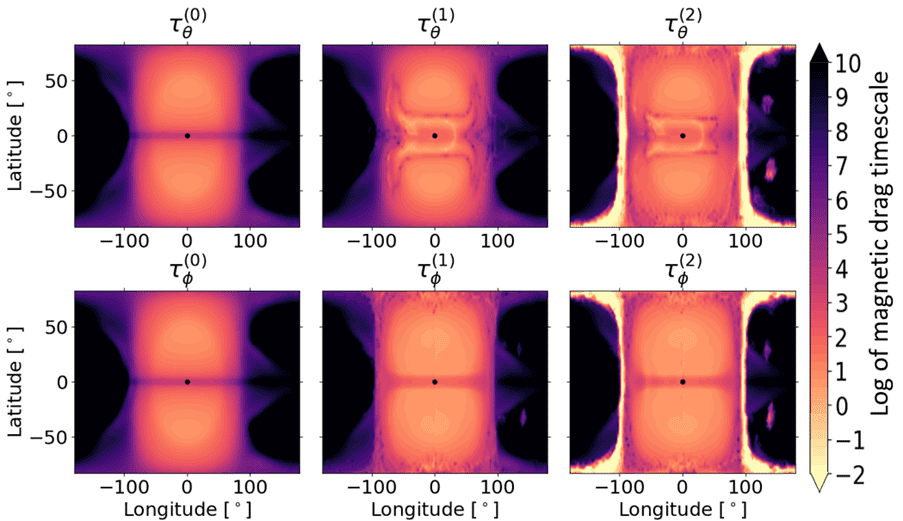}
\vspace{-0.3cm}
  \caption{\label{fig:tau_asymapprox} Magnetic drag timescale calculated at the 10 mbar level using the output of the 30G non-tilted run. The top row represents the timescale in the meridional direction \(\tau_\theta\), the bottom in the zonal direction \(\tau_\phi\), and the columns indicate that the drag timescale \(\tau^{(i)}\) was calculated using the first \(i\) terms in the current asymptotic expansion. The main differences from the zeroth order timescale occur at the equator, differing by about two orders of magnitude, and terminators, where \(\tau\) diverges completely.}
\end{figure}

The impact of higher-order currents on magnetic drag is illustrated in Fig. \ref{fig:tau_asymapprox}, which shows the horizontal magnetic drag timescales (\(\tau_\phi\) for zonal and \(\tau_\theta\) for meridional) computed using successive truncations of the current expansion (\(\tau^{(i)}\) indicating that \(\tau\) was calculated with the first \(i\) terms). Including first-order terms reduces \(\tau^{(1)}\) by a few orders of magnitude at the magnetic equator relative to \(\tau^{(0)}\) due to the effect (i) discussed above. Higher order contributions \(\tau^{(2)}\) introduce large changes in magnetic drag timescale near the terminators (ii) as a consequence of series non-convergence. Even the comparison between \(\tau^{(0)}\) and \(\tau^{(1)}\) near the terminators indicates that the first-order contributions to \(\tau^{(1)}\) are the same order, if not larger, than the zeroth-order terms. The second order terms then diverge significantly, demonstrating that there is no optimal truncation of the asymptotic series in these regions. This arises from sharp gradients at the terminators: velocity and resistivity increase, while temperature and ionisation decrease. Under the ideal frozen-in flux approximation, these effects cancel each other, so the Ohmic-to-advection ratio stays approximately constant \citep{ionisation_reconnection}. However, this approximation assumes homogeneity of the resistivity and therefore it cannot fully capture the effects due to the strong gradients at the terminators.  As a consequence, using a zeroth-order approximation is unlikely to be sufficient. Including these effects using more complete MHD models would cause the drag to be stronger at the equator and terminators, likely (i) slowing the equatorial jet further and weakening the hotspot splitting effect, and (ii) smoothing out the day-night contrast at the terminators.

\begin{figure}
  \centering
  \includegraphics[width=1\linewidth]{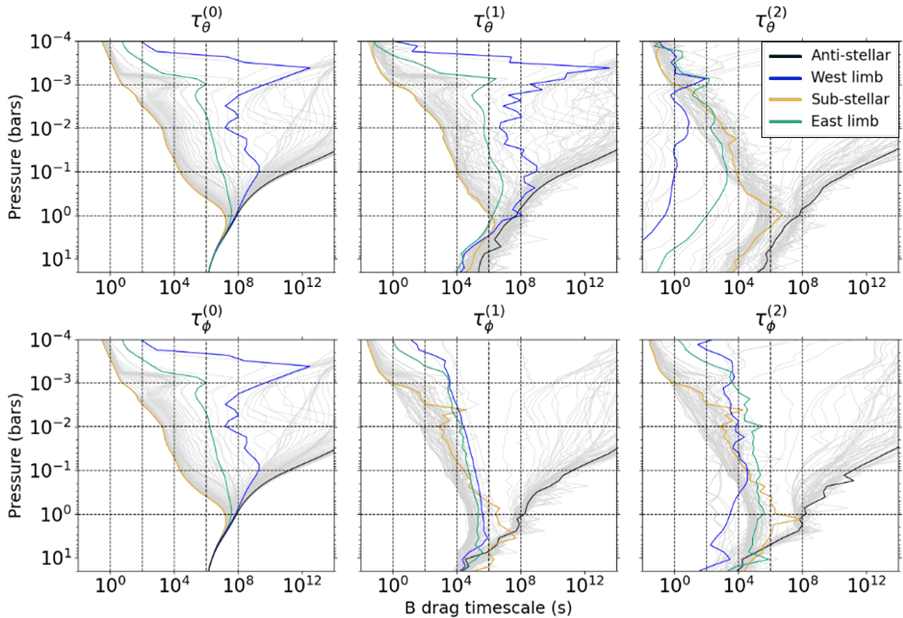}
\vspace{-0.3cm}
  \caption{\label{fig:tau_lonav_p} Latitudinally averaged magnetic drag timescale at different longitudes as a function of pressure, calculated using the output of the 30G non-tilted run. The top row represents the timescale in the meridional direction \(\tau_\theta\), the bottom in the zonal direction \(\tau_\phi\), and the columns indicate that the drag timescale \(\tau^{(i)}\) was calculated using the first \(i\) terms in the current asymptotic expansion.}
\end{figure}

Fig. \ref{fig:tau_lonav_p} presents latitudinally averaged drag timescales as a function of pressure. In addition to the effects of (ii) non-convergence of the series at the terminators, 

\begin{enumerate}[label=(\roman*), leftmargin=*, align=left, start=3]

\item the higher-order corrections become increasingly important at depth, where atmospheric thickness is no longer negligible relative to the planetary radius. This violates the assumptions necessary for using the asymptotic expansion: our perturbative parameter is no longer unity. Although drag in deeper layers remains weaker than in the upper atmosphere, the deeper, convective layers are where most heat is dissipated, therefore indicating that we are underestimating Ohmic heating due to the failure of the thin layer approximation to capture large scale effects of current closure.
\end{enumerate}

\subsection{Summary of Model Limitations and Future Work}
\label{sec:discussion_other}

Simplified magnetic drag models provide useful intuition for how magnetism may influence hot Jupiter atmospheres, but omit several important effects. Most importantly, such models neglect magnetic induction feedbacks and enforce current continuity only approximately (and/or locally), therefore underestimating propagation of Reynolds stresses into deeper layers. With these limitations in mind, we assess the extent to which our treatment is most reliable and place our results in the context of possible future work.

Our magnetic treatment is most reliable in the mid-to-upper atmosphere (0.5 bar~\(\gtrsim p\gtrsim\)~1 mbar), but fails at depth and at the top of the atmosphere. In deeper layers, the higher-order terms within the current expansion start to matter, indicating a breakdown of the asymptotic series discussed in (iii) in Section \ref{sec:discussion_pert_breakdown}. This indicates a fundamental limitation of models employing thin-atmosphere approximations such as our zeroth-order model or that of \cite{Beltz_2022} and \cite{christie2025geometricconsiderationshotjupiter}: they cannot self-consistently enforce 3D current continuity in deep layers. More complete semi-analytic treatments enforce current closure through expansions in different regimes \citep{batygin2010, Batygin_2011, Vigan_2025_1d_deep_atmos_mhd}, which simplify 3D dynamics to 1D to study the vertical structure of these atmospheres. \cite{Vigan_2025_1d_deep_atmos_mhd} n particular demonstrate that magnetic effects in deeper layers arise primarily from enforcing global current continuity. The Ohmic heating at deeper layers is more important in shaping the inflation of the planetary radii \citep{2018AJ....155..214T,HJ_radanom_mechanisms}, therefore highlighting the importance of capturing the effects of current closure at these deeper layers properly.

In the uppermost atmosphere (\(p \lesssim\) 1 mbar), our model results should also be treated with caution due to sensitivity to boundary conditions and additional physical processes not included in our model, such as ion-neutral drift and ambipolar diffusion (e.g., \citealp{Boening2025}), interactions with the thermosphere, as well as planet–stellar magnetospheric interaction. Close-in planets experience complex interactions between stellar and planetary magnetospheres \citep{bagheri2024freshlookinteractionexoplanets}, and induced atmospheric magnetism \citep{Boening2025} may further modify field geometries beyond a simple dipole, especially at high altitudes. Our imposed zero radial current and velocity boundary conditions at the top of the model atmosphere neglect the effects of atmospheric escape and current exchange with overlaying layers.  

Within the pressure range in which our magnetic model is more reliable (0.5 bar~\(\gtrsim p\gtrsim\)~1 mbar), the dominant source of error for ultra-hot Jupiters is due to neglecting induced magnetic effects, requiring small \(\mathrm{Re_m}\). In our simulation, Re\(_\mathrm{m}>O(10)\) for significant portions of the daysides at pressures p~\(\lesssim 100\) mbar, exceeding the analytically derived approximate threshold for significant induction effects \citep{rogers2017}, based on the critical Reynolds number criteria (e.g. \cite{Dynamo_review}). Our estimates of \(\mathrm{Re_m}\) are larger than those of \cite{Beltz_2022}, due to differences in our planetary parameter choices, model set-up, and assumptions, further reflecting the sensitivity to simulation properties. This suggests that neglecting induction likely constitutes the largest source of error in our model, outweighing secondary effects such as Hall and ambipolar diffusion, which are most important at low pressures {(e.g., \citealp{christie2025geometricconsiderationshotjupiter}, for ambipolar, and \citealp{sorianoguerrero2025nonidealmhdsimulationshot}, for Hall)}, or deviations from hydrostatic balance.

Other localized but significant errors are due to the limitations of the thin-atmosphere approximation we employed, which fails in three ways: (i) where leading-order terms vanish (e.g. near magnetic equator), (ii) where strong horizontal current divergence leads to strong current upwelling and downwelling (e.g. near the terminator), and (iii) at depth, where the perturbative parameter is no longer small. The error due to (i) indicates that our model, as well as other similar models (e.g. \citealp{Beltz_2022}) neglect the dominant currents in this region, the radial currents.  \cite{christie2025geometricconsiderationshotjupiter} noticed this limitation and addressed this issue using a formulation adapted from ionospheric physics parametrizations for currents (e.g., \citealp{zhu2004}). As a result, \cite{christie2025geometricconsiderationshotjupiter} find a stronger drag at the magnetic equator compared to our work, but still weaker than at higher latitudes contributing to hotspot splitting. Our model underestimates this drag near the equator and exaggerates the hotspot splitting and wind deflections. (ii) and (iii) are limitations not properly accounted for by simple models including \cite{christie2025geometricconsiderationshotjupiter}. For (ii), if this upwelling/downwelling is strong enough that it acts on a scale of the order of an atmospheric scale height, it will not simply deform pressure layers but impact the dynamics at the terminators significantly, and must be accounted for properly on a global scale. This effect may be missed if a model does not enforce 3D large scale current closure, whether failing to account for global closure poses an issue. The error due to (iii) can be mediated by working in a different magnetic regime, such as \cite{Vigan_2025_1d_deep_atmos_mhd}, and applying it to 3D modelling.


This indicates that our findings are likely exaggerated but illustrative of what may be expected due to a tilted dipole. Secondly, we caution against overinterpretation of the diagnostic ratios we considered to assess our assumptions. Large ratios of neglected-to-retained terms indicate the breakdown of assumptions, but do not quantify the true magnitude of the neglected terms. Furthermore, a steady state that is reached by neglecting some of these effects with favourable diagnostic ratios may only be artificially stable: for example, a non-induction model may find a steady state configuration with small effective Re\(\mathrm{_m}\), but such a state may only exist because induction effects were artificially suppressed during initialisation of the model.

It is also important to consider how robustly the impacts of the magnetic field geometry could be observed and whether they could be distinguished from different geometries or other atmospheric processes. The simulated phase curves presented in Section \ref{sec:results_phase_curves} showed that magnetic field strength does impact the phase curve amplitude and offset, consistent with earlier work (e.g. \citealp{Beltz_2022}). However, varying dipole tilt has little impact on the shape of these simulated phase curves. This is because tilt does not significantly impact the total heat content in the atmosphere, but primarily introduces North–South asymmetries in the temperature and velocity fields which phase curves are unable to pick up on. Instead, observational techniques that can reveal these North–South asymmetries, such as eclipse mapping, may be more promising in revealing differences between different tilt cases, as these have been shown to detect North–South asymmetries \citep{challener2024latitudinalasymmetrydaysideatmosphere, hammond2024twodimensionaleclipsemappinghot}. {Future work should additionally explore spectroscopic phase curves from tilted dipole models, which may allow for better constraints than white-light models. }

In order to fully implement magnetic effects accurately into ultra-hot Jupiter atmospheres, a semiconsistent coupled MHD and GCM framework is necessary. This model would assume a fixed background magnetic field while evolving the induced field equation (\ref{equ:induction_expanded}) while enforcing \(\mathbf{\nabla\cdot\delta B}=0\), alongside the fluid dynamical equations, as explored in \cite{rogers2017}. Such a model would have the benefit of evolving towards a steady state that is enforced to satisfy current continuity, rather than having to invert the complex steady-state differential equation system and solve for the current this way. However, such semicoupled MHD models would neglect the feedback from the atmospheric dynamo to the internal dynamo, which could further alter field geometry.

\section{Conclusions}
\label{sec:conclusion}

In this work, we extend previous models of hot Jupiter atmospheric magnetism (e.g., \citealp{Perna2010a, Rauscher_Menou_2013, Beltz_2022}) to examine the impact of an obliquely oriented dipolar magnetic field on the resulting atmospheric structure. Working within a simplified MHD framework, we further developed a more rigorous mathematical `check' for the validity of the steady state, induction-free, MHD approximation. Using this formulation, we conducted an array of simulations for a range of magnetic field strengths and dipole orientations, using WASP-121b as a case study for our planetary parameter inputs.

\smallskip

Our simulations yield several key results:
\begin{enumerate}
\item Increasing the magnetic field strength increases the day-night temperature contrast and reduces the east-west asymmetry, consistent with previous work \citep{Rauscher_Menou_2013,Beltz_2022,christie2025geometricconsiderationshotjupiter}. For fixed field strength, increasing the dipole obliquity, especially in the substellar direction, introduces north–south asymmetries into the temperature profile, resulting in a meridional hotspot offset.

\item Stronger magnetic fields substantially weaken the superrotating equatorial jet in the mid-atmosphere, in agreement with \cite{Beltz_2022} and \cite{ christie2025geometricconsiderationshotjupiter}. Increasing the dipole tilt perturbs this jet and further weakens its zonal-mean speed, supporting the analytic prediction of \cite{Batygin_2014}. This effect is more significant for stronger magnetic field strengths.

\item The phase curve amplitude increases and the hotspot offset decreases as the magnetic field strength is increased. This agrees with previous work (e.g., \citealp{Rauscher_Menou_2013, Beltz_2022}). The differences in spectroscopic white-light phase curves between dipole tilt scenarios are minimal, with the main effects being to the nightside emission. We expect that constraining varying magnetic dipole tilt scenarios with white light only phase curves would therefore be infeasible.

\end{enumerate}

Our model relies on several classic assumptions used in previous work \citep{Rauscher_Menou_2013,Beltz_2022,christie2025geometricconsiderationshotjupiter}, such as the thin atmosphere approximation and a small magnetic Reynolds number, the most critical of which we re-examined in this work. We find that neglecting induced magnetic fields and fully-3D MHD feedbacks constitutes the most significant limitation of our model, especially when working in the regime of UHJs. Secondary sources of error include omitting the Hall and ambipolar effects, and neglecting radial magnetic forces, which may become non-negligible in the uppermost atmosphere (pressures \(\lesssim\)1 mbar).

We reason that such simplified models can therefore provide a qualitative picture of how magnetism impacts the atmospheric dynamics of these planets, especially in the regions where the errors discussed in Section \ref{sec:discussion} are smaller. However, we caution against drawing quantitative conclusions regarding magnetic field strengths or detailed flow structures from current models. A fully coupled MHD treatment could reveal features in the atmosphere due to the non-linear, non-analytic feedbacks that are challenging to otherwise anticipate. While such fully coupled GCM and MHD models may be computationally demanding, their development may be a necessary step to improving the reliability of quantitative predictions of these magnetic effects (e.g. for varying magnetic field strengths and geometries). 

\section*{Acknowledgements}

We thank the anonymous reviewer for a careful and constructive review which improved this manuscript. This work was performed using the Cambridge Service for Data Driven Discovery (CSD3), part of which is operated by the University of Cambridge Research Computing on behalf of the STFC DiRAC HPC Facility (\url{www.dirac.ac.uk}). Resources supporting this work were further provided by the NASA High-End Computing (HEC) Program through the NASA Advanced Supercomputing (NAS) Division at Ames Research Center. We thank Tom Evans-Soma for sharing \textit{JWST}/NIRSpec reductions of WASP-121b. TDK thanks Duncan Christie for helpful discussions, {and JSF acknowledges Sarah Rugheimer for her support.}

\section*{Data Availability}

All simulation outputs from this work are available on Zenodo at \url{https://doi.org/10.5281/zenodo.21511412}.



\bibliographystyle{mnras}
\bibliography{example} 

@ARTICLE{2021PNAS..11822705H,
       author = {{Hammond}, Mark and {Lewis}, Neil T.},
        title = "{The rotational and divergent components of atmospheric circulation on tidally locked planets}",
      journal = {Proc. Natl Acad. Sci.},
         year = 2021,
        month = mar,
       volume = {118},
       number = {13},
          eid = {e2022705118},
        pages = {e2022705118},
          doi = {10.1073/pnas.2022705118},
archivePrefix = {arXiv},
       eprint = {2102.11760},
 primaryClass = {astro-ph.EP},
       adsurl = {https://ui.adsabs.harvard.edu/abs/2021PNAS..11822705H}
}

@ARTICLE{2018AJ....155..214T,
       author = {{Thorngren}, Daniel P. and {Fortney}, Jonathan J.},
        title = "{Bayesian Analysis of Hot-Jupiter Radius Anomalies: Evidence for Ohmic Dissipation?}",
      journal = {\aj},
         year = 2018,
        month = may,
       volume = {155},
       number = {5},
          eid = {214},
        pages = {214},
          doi = {10.3847/1538-3881/aaba13},
archivePrefix = {arXiv},
       eprint = {1709.04539},
 primaryClass = {astro-ph.EP},
       adsurl = {https://ui.adsabs.harvard.edu/abs/2018AJ....155..214T}
}

@article{Arcangeli:2019aa,
	adsurl = {https://ui.adsabs.harvard.edu/abs/2019A&A...625A.136A},
	archiveprefix = {arXiv},
	author = {{Arcangeli}, Jacob and {D{\'e}sert}, Jean-Michel and {Parmentier}, Vivien and {Stevenson}, Kevin B. and {Bean}, Jacob L. and {Line}, Michael R. and {Kreidberg}, Laura and {Fortney}, Jonathan J. and {Showman}, Adam P.},
	doi = {10.1051/0004-6361/201834891},
	eid = {A136},
	eprint = {1904.02069},
	journal = {\aap},
	month = may,
	pages = {A136},
	primaryclass = {astro-ph.EP},
	title = {{Climate of an ultra hot Jupiter. Spectroscopic phase curve of WASP-18b with HST/WFC3}},
	volume = {625},
	year = 2019}

@article{Perna2010a,
   title={MAGNETIC DRAG ON HOT JUPITER ATMOSPHERIC WINDS},
   volume={719},
   ISSN={1538-4357},
   url={http://dx.doi.org/10.1088/0004-637X/719/2/1421},
   DOI={10.1088/0004-637x/719/2/1421},
   number={2},
   journal={ApJ},
   publisher={American Astronomical Society},
   author={Perna, Rosalba and Menou, Kristen and Rauscher, Emily},
   year={2010},
   month=jul, pages={1421–1426} }

@article{Perna2010b,
   title={OHMIC DISSIPATION IN THE ATMOSPHERES OF HOT JUPITERS},
   volume={724},
   ISSN={1538-4357},
   url={http://dx.doi.org/10.1088/0004-637X/724/1/313},
   DOI={10.1088/0004-637x/724/1/313},
   number={1},
   journal={ApJ},
   publisher={American Astronomical Society},
   author={Perna, Rosalba and Menou, Kristen and Rauscher, Emily},
   year={2010},
   month=nov, pages={313–317} }

@article{HJ_review_origins_properties_Tad,
   title={Hot Jupiters: Origins, Structure, Atmospheres},
   volume={126},
   ISSN={2169-9100},
   url={http://dx.doi.org/10.1029/2020JE006629},
   DOI={10.1029/2020je006629},
   number={3},
   journal={J. Geophys. Res. Planets},
   publisher={American Geophysical Union (AGU)},
   author={Fortney, Jonathan J. and Dawson, Rebekah I. and Komacek, Thaddeus D.},
   year={2021},
   pages = {e06629},
   month=mar }

@article{w121_PhaseCurve,
   title={A JWST NIRSpec Phase Curve for WASP-121b: Dayside Emission Strongest Eastward of the Substellar Point and Nightside Conditions Conducive to Cloud Formation},
   year = {2023},
   volume={943},
   ISSN={2041-8213},
   url={http://dx.doi.org/10.3847/2041-8213/acb049},
   DOI={10.3847/2041-8213/acb049},
   number={2},
   journal={ApJ},
   publisher={American Astronomical Society},
   author={Mikal-Evans, Thomas and Sing, David K. and Dong, Jiayin and Foreman-Mackey, Daniel and Kataria, Tiffany and Barstow, Joanna K. and Goyal, Jayesh M. and Lewis, Nikole K. and Lothringer, Joshua D. and Mayne, Nathan J. and Wakeford, Hannah R. and Christie, Duncan A. and Rustamkulov, Zafar},
   year={2023},
   month=jan, pages={L17} }

@article{Rauscher_Menou_2013,
   title={THREE-DIMENSIONAL ATMOSPHERIC CIRCULATION MODELS OF HD 189733b AND HD 209458b WITH CONSISTENT MAGNETIC DRAG AND OHMIC DISSIPATION},
   volume={764},
   ISSN={1538-4357},
   url={http://dx.doi.org/10.1088/0004-637X/764/1/103},
   DOI={10.1088/0004-637x/764/1/103},
   number={1},
   journal={ApJ},
   publisher={American Astronomical Society},
   author={Rauscher, Emily and Menou, Kristen},
   year={2013},
   month=jan, pages={103} }

@ARTICLE{first_hotJupiter,
       author = {{Mayor}, Michel and {Queloz}, Didier},
        title = "{A Jupiter-mass companion to a solar-type star}",
      journal = {\nat},
         year = 1995,
        month = nov,
       volume = {378},
       number = {6555},
        pages = {355-359},
          doi = {10.1038/378355a0},
       adsurl = {https://ui.adsabs.harvard.edu/abs/1995Natur.378..355M}
}

@article{Beltz_2022,
   title={Exploring the Effects of Active Magnetic Drag in a General Circulation Model of the Ultrahot Jupiter WASP-76b},
   volume={163},
   ISSN={1538-3881},
   url={http://dx.doi.org/10.3847/1538-3881/ac3746},
   DOI={10.3847/1538-3881/ac3746},
   number={1},
   journal={AJ},
   publisher={American Astronomical Society},
   author={Beltz, Hayley and Rauscher, Emily and Roman, Michael T. and Guilliat, Abigail},
   year={2021},
   month=dec, pages={35} }

@article{Komacek_2022_Nightside_Clouds,
   title={Patchy Nightside Clouds on Ultra-hot Jupiters: General Circulation Model Simulations with Radiatively Active Cloud Tracers},
   volume={934},
   ISSN={1538-4357},
   url={http://dx.doi.org/10.3847/1538-4357/ac7723},
   DOI={10.3847/1538-4357/ac7723},
   number={1},
   journal={ApJ},
   publisher={American Astronomical Society},
   author={Komacek, Thaddeus D. and Tan, Xianyu and Gao, Peter and Lee, Elspeth K. H.},
   year={2022},
   month=jul, pages={79} }

@article{Showman_2011_EastWest_Equatorial_Jet,
   title={EQUATORIAL SUPERROTATION ON TIDALLY LOCKED EXOPLANETS},
   volume={738},
   ISSN={1538-4357},
   url={http://dx.doi.org/10.1088/0004-637X/738/1/71},
   DOI={10.1088/0004-637x/738/1/71},
   number={1},
   journal={ApJ},
   publisher={American Astronomical Society},
   author={Showman, Adam P. and Polvani, Lorenzo M.},
   year={2011},
   month=aug, pages={71} }

@article{Showman_2020_Review,
   title={Atmospheric Dynamics of Hot Giant Planets and Brown Dwarfs},
   volume={216},
   ISSN={1572-9672},
   url={http://dx.doi.org/10.1007/s11214-020-00758-8},
   DOI={10.1007/s11214-020-00758-8},
   number={8},
   journal={Space Sci. Rev.},
   publisher={Springer Science and Business Media LLC},
   author={Showman, Adam P. and Tan, Xianyu and Parmentier, Vivien},
   year={2020},
   pages = {139},
   month=dec }

@article{Tan_Komacek_2019,
   title={The Atmospheric Circulation of Ultra-hot Jupiters},
   volume={886},
   ISSN={1538-4357},
   url={http://dx.doi.org/10.3847/1538-4357/ab4a76},
   DOI={10.3847/1538-4357/ab4a76},
   number={1},
   journal={ApJ},
   publisher={American Astronomical Society},
   author={Tan, Xianyu and Komacek, Thaddeus D.},
   year={2019},
   month=nov, pages={26} }

@inbook{Parmentier_2018_Phase_Curves,
   title={Exoplanet Phase Curves: Observations and Theory},
   ISBN={9783319553337},
   DOI={10.1007/978-3-319-55333-7_116},
   booktitle={Handbook of Exoplanets},
   publisher={Springer International Publishing},
   author={Parmentier, Vivien and Crossfield, Ian J. M.},
   year={2018},
   pages={1419–1440},
   }

@article{Parmentier_2018,
   title={From thermal dissociation to condensation in the atmospheres of ultra hot Jupiters: WASP-121b in context},
   volume={617},
   ISSN={1432-0746},
   url={http://dx.doi.org/10.1051/0004-6361/201833059},
   DOI={10.1051/0004-6361/201833059},
   journal={A&A},
   publisher={EDP Sciences},
   author={Parmentier, Vivien and Line, Mike R. and Bean, Jacob L. and Mansfield, Megan and Kreidberg, Laura and Lupu, Roxana and Visscher, Channon and Désert, Jean-Michel and Fortney, Jonathan J. and Deleuil, Magalie and Arcangeli, Jacob and Showman, Adam P. and Marley, Mark S.},
   year={2018},
   month=sep, pages={A110} }

@article{Batygin_2014,
   title={NON-AXISYMMETRIC FLOWS ON HOT JUPITERS WITH OBLIQUE MAGNETIC FIELDS},
   volume={794},
   ISSN={1538-4357},
   url={http://dx.doi.org/10.1088/0004-637X/794/1/10},
   DOI={10.1088/0004-637x/794/1/10},
   number={1},
   journal={ApJ},
   publisher={American Astronomical Society},
   author={Batygin, Konstantin and Stanley, Sabine},
   year={2014},
   month=sep, pages={10} }

@article{LIU_2008,
   title={Constraints on deep-seated zonal winds inside Jupiter and Saturn},
   volume={196},
   ISSN={0019-1035},
   url={http://dx.doi.org/10.1016/j.icarus.2007.11.036},
   DOI={10.1016/j.icarus.2007.11.036},
   number={2},
   journal={Icarus},
   publisher={Elsevier BV},
   author={Liu, J and Goldreich, P and Stevenson, D},
   year={2008},
   month=aug, pages={653–664} }

@article{Yadav_2017,
   title={Estimating the Magnetic Field Strength in Hot Jupiters},
   volume={849},
   ISSN={2041-8213},
   url={http://dx.doi.org/10.3847/2041-8213/aa93fd},
   DOI={10.3847/2041-8213/aa93fd},
   number={1},
   journal={ApJ},
   publisher={American Astronomical Society},
   author={Yadav, Rakesh K. and Thorngren, Daniel P.},
   year={2017},
   month=oct, pages={L12} }

@article{Vigan_2025_1d_deep_atmos_mhd,
   title={Inflated hot Jupiters: Inferring average atmospheric velocity via Ohmic models coupled with internal dynamo evolution},
   volume={701},
   ISSN={1432-0746},
   url={http://dx.doi.org/10.1051/0004-6361/202555219},
   DOI={10.1051/0004-6361/202555219},
   journal={A&A},
   publisher={EDP Sciences},
   author={Viganò, Daniele and Sengupta, Soumya and Soriano-Guerrero, Clàudia and Perna, Rosalba and Elias-López, Albert and Kumar, Sandeep and Akgün, Taner},
   year={2025},
   month=aug, pages={A8} }

@article{sorianoguerrero2025nonidealmhdsimulationshot,
    author = {Soriano-Guerrero, Clàudia and Viganò, Daniele and Perna, Rosalba and Elias-López, Albert and Beltz, Hayley},
    title = {Non-ideal MHD simulations of hot Jupiter atmospheres},
    journal = {MNRAS},
    volume = {540},
    number = {2},
    pages = {1827-1843},
    year = {2025},
    month = {06},
    issn = {0035-8711},
    doi = {10.1093/mnras/staf842},
    url = {https://doi.org/10.1093/mnras/staf842},
}

@ARTICLE{Draine_resistivity,
       author = {{Draine}, B.~T. and {Roberge}, W.~G. and {Dalgarno}, A.},
        title = "{Magnetohydrodynamic shock waves in molecular clouds.}",
      journal = {\apj},
         year = 1983,
        month = jan,
       volume = {264},
        pages = {485-507},
          doi = {10.1086/160617},
       adsurl = {https://ui.adsabs.harvard.edu/abs/1983ApJ...264..485D}
}

@ARTICLE{KnaepenReynoldsnumber,
       author = {{Knaepen}, Bernard and {Moreau}, Ren{\'e}},
        title = "{Magnetohydrodynamic Turbulence at Low Magnetic Reynolds Number}",
      journal = {Annu. Rev. Fluid Mech.},
         year = 2008,
        month = jan,
       volume = {40},
       number = {1},
        pages = {25-45},
          doi = {10.1146/annurev.fluid.39.050905.110231},
       adsurl = {https://ui.adsabs.harvard.edu/abs/2008AnRFM..40...25K}
}

@article{Komacek_2016_constant_drag,
   title={ATMOSPHERIC CIRCULATION OF HOT JUPITERS: DAYSIDE–NIGHTSIDE TEMPERATURE DIFFERENCES},
   volume={821},
   ISSN={1538-4357},
   url={http://dx.doi.org/10.3847/0004-637X/821/1/16},
   DOI={10.3847/0004-637x/821/1/16},
   number={1},
   journal={ApJ},
   publisher={American Astronomical Society},
   author={Komacek, Thaddeus D. and Showman, Adam P.},
   year={2016},
   month=apr, pages={16}}

@article{ionisation,
	author = {{Helling, Ch.} and {Worters, M.} and {Samra, D.} and {Molaverdikhani, K.} and {Iro, N.}},
	title = {Understanding the atmospheric properties and chemical composition of the ultra-hot Jupiter HAT-P-7b - III. Changing ionisation and the emergence of an ionosphere},
	DOI= "10.1051/0004-6361/202039699",
	url= "https://doi.org/10.1051/0004-6361/202039699",
	journal = {A\&A},
	year = 2021,
	volume = 648,
	pages = "A80",
}

@article{Roth_2021,
   title={Pseudo-2D modelling of heat redistribution through H2 thermal dissociation/recombination: consequences for ultra-hot Jupiters},
   volume={505},
   ISSN={1365-2966},
   url={http://dx.doi.org/10.1093/mnras/stab1256},
   DOI={10.1093/mnras/stab1256},
   number={3},
   journal={MNRAS},
   publisher={Oxford University Press (OUP)},
   author={Roth, A and Drummond, B and Hébrard, E and Tremblin, P and Goyal, J and Mayne, N},
   year={2021},
   month=may, pages={4515–4530} }

@article{guillot1996,
doi = {10.1086/309935},
url = {https://doi.org/10.1086/309935},
year = {1996},
month = {mar},
publisher = {},
volume = {459},
number = {1},
pages = {L35},
author = {Guillot, T. and Burrows, A. and Hubbard, W. B. and Lunine, J. I. and Saumon, D.},
title = {Giant Planets at Small Orbital Distances},
journal = {ApJ},
}

@article{Sing_2024,
   title={An Absolute Mass, Precise Age, and Hints of Planetary Winds for WASP-121A and b from a JWST NIRSpec Phase Curve},
   volume={168},
   ISSN={1538-3881},
   url={http://dx.doi.org/10.3847/1538-3881/ad7fe7},
   DOI={10.3847/1538-3881/ad7fe7},
   number={6},
   journal={AJ},
   publisher={American Astronomical Society},
   author={Sing, David K. and Evans-Soma, Thomas M. and Rustamkulov, Zafar and Lothringer, Joshua D. and Mayne, Nathan J. and Schlaufman, Kevin C.},
   year={2024},
   month=nov, pages={231} }

@article{Liu_Showman_2013,
   title={ATMOSPHERIC CIRCULATION OF HOT JUPITERS: INSENSITIVITY TO INITIAL CONDITIONS},
   volume={770},
   ISSN={1538-4357},
   url={http://dx.doi.org/10.1088/0004-637X/770/1/42},
   DOI={10.1088/0004-637x/770/1/42},
   number={1},
   journal={ApJ},
   publisher={American Astronomical Society},
   author={Liu, Beibei and Showman, Adam P.},
   year={2013},
   month=may, pages={42} }

@article{Kataria_2016,
doi = {10.3847/0004-637X/821/1/9},
url = {https://doi.org/10.3847/0004-637X/821/1/9},
year = {2016},
month = {apr},
publisher = {The American Astronomical Society},
volume = {821},
number = {1},
pages = {9},
author = {Kataria, Tiffany and Sing, David K. and Lewis, Nikole K. and Visscher, Channon and Showman, Adam P. and Fortney, Jonathan J. and Marley, Mark S.},
title = {THE ATMOSPHERIC CIRCULATION OF A NINE-HOT-JUPITER SAMPLE: PROBING CIRCULATION AND CHEMISTRY OVER A WIDE PHASE SPACE},
journal = {ApJ}
}

@article{Steinrueck2021,
    author = {Steinrueck, Maria E and Showman, Adam P and Lavvas, Panayotis and Koskinen, Tommi and Tan, Xianyu and Zhang, Xi},
    title = {3D simulations of photochemical hazes in the atmosphere of hot Jupiter HD 189733b},
    journal = {MNRAS},
    volume = {504},
    number = {2},
    pages = {2783-2799},
    year = {2021},
    month = {04},
    issn = {0035-8711},
    doi = {10.1093/mnras/stab1053},
    url = {https://doi.org/10.1093/mnras/stab1053},
    eprint = {https://academic.oup.com/mnras/article-pdf/504/2/2783/37789274/stab1053.pdf},
}

@article{Murphy_2025,
   title={A Panchromatic Characterization of the Evening and Morning Atmosphere of WASP-107 b: Composition and Cloud Variations, and Insight into the Effect of Stellar Contamination},
   volume={170},
   ISSN={1538-3881},
   url={http://dx.doi.org/10.3847/1538-3881/addf38},
   DOI={10.3847/1538-3881/addf38},
   number={1},
   journal={AJ},
   publisher={American Astronomical Society},
   author={Murphy, Matthew M. and Beatty, Thomas G. and Schlawin, Everett and Bell, Taylor J. and Radica, Michael and Kennedy, Thomas D. and Mehta, Nishil and Welbanks, Luis and Line, Michael R. and Parmentier, Vivien and Greene, Thomas P. and Mukherjee, Sagnick and Fortney, Jonathan J. and Ohno, Kazumasa and Wiser, Lindsey and Triantafillides, Anastasia and Rauscher, Emily and Edelman, Isaac R. and Rieke, Marcia J.},
   year={2025},
   month=jun, pages={61} }

@article{Christensen2010,
  author    = {Christensen, U. R.},
  title     = {Dynamo Scaling Laws and Applications to the Planets},
  journal   = {Space Sci. Rev.},
  year      = {2010},
  volume    = {152},
  number    = {1},
  pages     = {565--590},
  doi       = {10.1007/s11214-009-9553-2},
  url       = {https://doi.org/10.1007/s11214-009-9553-2},
  issn      = {1572-9672},
  month     = may
}

@article{ionisation_reconnection,
  author = {Liu, Yi-Hsin and Hesse, Michael and Genestreti, Kevin and Nakamura, Rumi and Burch, James L. and Cassak, Paul A. and Bessho, Naoki and Eastwood, Jonathan P. and Phan, Tai and Swisdak, Marc and Toledo-Redondo, Sergio and Hoshino, Masahiro and Norgren, Cecilia and Ji, Hantao and Nakamura, Takuma K. M.},
  title = {Ohm’s Law, the Reconnection Rate, and Energy Conversion in Collisionless Magnetic Reconnection},
  journal = {Space Sci. Rev.},
  year = {2025},
  volume = {221},
  number = {1},
  pages = {16},
  doi = {10.1007/s11214-025-01142-0},
  url = {https://doi.org/10.1007/s11214-025-01142-0}
}

@article{wardenier2024phaseresolvingabsorptionsignatureswater,
doi = {10.1088/1538-3873/ad5c9f},
url = {https://doi.org/10.1088/1538-3873/ad5c9f},
year = {2024},
month = {aug},
publisher = {The Astronomical Society of the Pacific},
volume = {136},
number = {8},
pages = {084403},
author = {Wardenier, Joost P. and Parmentier, Vivien and Line, Michael R. and Weiner Mansfield, Megan and Tan, Xianyu and Tsai, Shang-Min and Bean, Jacob L. and Birkby, Jayne L. and Brogi, Matteo and Désert, Jean-Michel and Gandhi, Siddharth and Lee, Elspeth K. H. and Levens, Colette I. and Pino, Lorenzo and Smith, Peter C. B.},
title = {Phase-resolving the Absorption Signatures of Water and Carbon Monoxide in the Atmosphere of the Ultra-hot Jupiter WASP-121b with GEMINI-S/IGRINS},
journal = {PASP},
}

@ARTICLE{Lee2022,
       author = {{Lee}, Elspeth K.~H. and {Prinoth}, Bibiana and {Kitzmann}, Daniel and {Tsai}, Shang-Min and {Hoeijmakers}, Jens and {Borsato}, Nicholas W. and {Heng}, Kevin},
        title = "{The Mantis Network II: examining the 3D high-resolution observable properties of the UHJs WASP-121b and WASP-189b through GCM modelling}",
      journal = {\mnras},
         year = 2022,
        month = nov,
       volume = {517},
       number = {1},
        pages = {240-256},
          doi = {10.1093/mnras/stac2246},
archivePrefix = {arXiv},
       eprint = {2210.11986},
 primaryClass = {astro-ph.EP},
       adsurl = {https://ui.adsabs.harvard.edu/abs/2022MNRAS.517..240L}
}

@article{Seidel2025,
  author = {Seidel, Julia V. and Prinoth, Bibiana and Pino, Lorenzo and dos Santos, Leonardo A. and Chakraborty, Hritam and Parmentier, Vivien and Sedaghati, Elyar and Wardenier, Joost P. and Farret Jentink, Casper and Zapatero Osorio, Maria Rosa and Allart, Romain and Ehrenreich, David and Lendl, Monika and Roccetti, Giulia and Damasceno, Yuri and Bourrier, Vincent and Lillo-Box, Jorge and Hoeijmakers, H. Jens and Pallé, Enric and Santos, Nuno and Suárez Mascareño, Alejandro and Sousa, Sergio G. and Tabernero, Hugo M. and Pepe, Francesco A.},
  title   = {Vertical structure of an exoplanet’s atmospheric jet stream},
  journal = {Nature},
  year    = {2025},
  volume  = {639},
  number  = {8056},
  pages   = {902--908},
  doi     = {10.1038/s41586-025-08664-1},
  url     = {https://doi.org/10.1038/s41586-025-08664-1},
  issn    = {1476-4687}
}

@article{magneticsim,
   title={Magnetic field evolution of hot exoplanets},
   volume={535},
   ISSN={1365-2966},
   url={http://dx.doi.org/10.1093/mnras/stae2505},
   DOI={10.1093/mnras/stae2505},
   number={4},
   journal={MNRAS},
   publisher={Oxford University Press (OUP)},
   author={Kilmetis, K and Vidotto, A A and Allan, A and Kubyshkina, D},
   year={2024},
   month=nov, pages={3646–3655} }

@article{Cauley_2019,
   title={Magnetic field strengths of hot Jupiters from signals of star–planet interactions},
   volume={3},
   ISSN={2397-3366},
   url={http://dx.doi.org/10.1038/s41550-019-0840-x},
   DOI={10.1038/s41550-019-0840-x},
   number={12},
   journal={Nat. Astron.},
   publisher={Springer Science and Business Media LLC},
   author={Cauley, P. Wilson and Shkolnik, Evgenya L. and Llama, Joe and Lanza, Antonino F.},
   year={2019},
   month=jul, pages={1128–1134} }

@article{blocker2026inhomogeneousmagneticcouplingexoplanets,
      title={Inhomogeneous magnetic coupling in exoplanets: the stop and go of WASP-18 b's atmospheric flows}, 
      author={Aljona Blöcker and Ludmila Carone and Christiane Helling},
      year={2026},
      eprint={2602.18101},
      archivePrefix={arXiv},
      primaryClass={astro-ph.EP},
      url={https://arxiv.org/abs/2602.18101}, 
      doi = {10.1051/0004-6361/202558210},
      journal = {A&A},
      volume = {708},
      pages = {A84}
}

@ARTICLE{Fromang2016,
       author = {{Fromang}, S{\'e}bastien and {Leconte}, Jeremy and {Heng}, Kevin},
        title = "{Shear-driven instabilities and shocks in the atmospheres of hot Jupiters}",
      journal = {\aap},
         year = 2016,
        month = jul,
       volume = {591},
          eid = {A144},
        pages = {A144},
          doi = {10.1051/0004-6361/201527600},
archivePrefix = {arXiv},
       eprint = {1603.02794},
 primaryClass = {astro-ph.EP},
       adsurl = {https://ui.adsabs.harvard.edu/abs/2016A&A...591A.144F}
}

@ARTICLE{SkinnerCho2022,
       author = {{Skinner}, J.~W. and {Cho}, J.~Y.-K.},
        title = "{Modons on tidally synchronized extrasolar planets}",
      journal = {\mnras},
         year = 2022,
        month = apr,
       volume = {511},
       number = {3},
        pages = {3584-3601},
          doi = {10.1093/mnras/stab2809},
archivePrefix = {arXiv},
       eprint = {2109.06568},
 primaryClass = {astro-ph.EP},
       adsurl = {https://ui.adsabs.harvard.edu/abs/2022MNRAS.511.3584S}
}

@article{Ehrenreich_2020,
   title={Nightside condensation of iron in an ultrahot giant exoplanet},
   volume={580},
   ISSN={1476-4687},
   url={http://dx.doi.org/10.1038/s41586-020-2107-1},
   DOI={10.1038/s41586-020-2107-1},
   number={7805},
   journal={Nature},
   publisher={Springer Science and Business Media LLC},
   author={Ehrenreich, David and Lovis, Christophe and Allart, Romain and Zapatero Osorio, María Rosa and Pepe, Francesco and Cristiani, Stefano and Rebolo, Rafael and Santos, Nuno C. and Borsa, Francesco and Demangeon, Olivier and Dumusque, Xavier and González Hernández, Jonay I. and Casasayas-Barris, Núria and Ségransan, Damien and Sousa, Sérgio and Abreu, Manuel and Adibekyan, Vardan and Affolter, Michael and Allende Prieto, Carlos and Alibert, Yann and Aliverti, Matteo and Alves, David and Amate, Manuel and Avila, Gerardo and Baldini, Veronica and Bandy, Timothy and Benz, Willy and Bianco, Andrea and Bolmont, Émeline and Bouchy, François and Bourrier, Vincent and Broeg, Christopher and Cabral, Alexandre and Calderone, Giorgio and Pallé, Enric and Cegla, H. M. and Cirami, Roberto and Coelho, João M. P. and Conconi, Paolo and Coretti, Igor and Cumani, Claudio and Cupani, Guido and Dekker, Hans and Delabre, Bernard and Deiries, Sebastian and D’Odorico, Valentina and Di Marcantonio, Paolo and Figueira, Pedro and Fragoso, Ana and Genolet, Ludovic and Genoni, Matteo and Génova Santos, Ricardo and Hara, Nathan and Hughes, Ian and Iwert, Olaf and Kerber, Florian and Knudstrup, Jens and Landoni, Marco and Lavie, Baptiste and Lizon, Jean-Louis and Lendl, Monika and Lo Curto, Gaspare and Maire, Charles and Manescau, Antonio and Martins, C. J. A. P. and Mégevand, Denis and Mehner, Andrea and Micela, Giusi and Modigliani, Andrea and Molaro, Paolo and Monteiro, Manuel and Monteiro, Mario and Moschetti, Manuele and Müller, Eric and Nunes, Nelson and Oggioni, Luca and Oliveira, António and Pariani, Giorgio and Pasquini, Luca and Poretti, Ennio and Rasilla, José Luis and Redaelli, Edoardo and Riva, Marco and Santana Tschudi, Samuel and Santin, Paolo and Santos, Pedro and Segovia Milla, Alex and Seidel, Julia V. and Sosnowska, Danuta and Sozzetti, Alessandro and Spanò, Paolo and Suárez Mascareño, Alejandro and Tabernero, Hugo and Tenegi, Fabio and Udry, Stéphane and Zanutta, Alessio and Zerbi, Filippo},
   year={2020},
   month=mar, pages={597–601} }

@article{Batygin_2011,
doi = {10.1088/0004-637X/738/1/1},
url = {https://doi.org/10.1088/0004-637X/738/1/1},
year = {2011},
month = {aug},
publisher = {The American Astronomical Society},
volume = {738},
number = {1},
pages = {1},
author = {Batygin, Konstantin and Stevenson, David J. and Bodenheimer, Peter H.},
title = {EVOLUTION OF OHMICALLY HEATED HOT JUPITERS},
journal = {ApJ}
}

@article{batygin2010,
  author    = {Konstantin Batygin and David J. Stevenson},
  title     = {Inflating Hot Jupiters with Ohmic Dissipation},
  journal   = {ApJ},
  volume    = {714},
  number    = {2},
  pages     = {L238--L243},
  year      = {2010},
  doi       = {10.1088/2041-8205/714/2/L238},
  url       = {https://doi.org/10.1088/2041-8205/714/2/L238}
}

@article{rogers2017,
  author    = {T. M. Rogers and J. N. McElwaine},
  title     = {The Hottest Hot Jupiters May Host Atmospheric Dynamos},
  journal   = {ApJ},
  volume    = {841},
  number    = {2},
  pages     = {L26},
  year      = {2017},
  doi       = {10.3847/2041-8213/aa72da},
  url       = {https://doi.org/10.3847/2041-8213/aa72da}
}

@article{lee2021,
  author    = {Elspeth K. H. Lee and Vincent Parmentier and Matthew Hammond and Sebastian L. Grimm and Daniel Kitzmann and Xianyu Tan and Shang-Min Tsai and Raymond T. Pierrehumbert},
  title     = {Simulating gas giant exoplanet atmospheres with Exo-FMS: comparing semigrey, picket fence, and correlated-k radiative-transfer schemes},
  journal   = {MNRAS},
  volume    = {506},
  number    = {2},
  pages     = {2695--2711},
  year      = {2021},
  doi       = {10.1093/mnras/stab1851},
  url       = {https://doi.org/10.1093/mnras/stab1851}
}

@article{zhu2004,
AUTHOR = {Zhu, X. and Talaat, E. R. and Baker, J. B. H. and Yee, J.-H.},
TITLE = {A self-consistent derivation of ion drag and Joule heating for atmospheric dynamics in the thermosphere},
JOURNAL = {Annales Geophysicae},
VOLUME = {23},
YEAR = {2005},
NUMBER = {10},
PAGES = {3313--3322},
URL = {https://angeo.copernicus.org/articles/23/3313/2005/},
DOI = {10.5194/angeo-23-3313-2005}
}

@article{Elias_L_pez_2025,
   title={Planetary dynamos in evolving cold gas giants},
   volume={696},
   ISSN={1432-0746},
   url={http://dx.doi.org/10.1051/0004-6361/202453372},
   DOI={10.1051/0004-6361/202453372},
   journal={A&A},
   publisher={EDP Sciences},
   author={Elias-López, Albert and Del Sordo, Fabio and Viganò, Daniele and Soriano-Guerrero, Clàudia and Akgün, Taner and Reboul-Salze, Alexis and Cantiello, Matteo},
   year={2025},
   month=apr, pages={A161} }

@article{connerney2018,
author = {Connerney, J. E. P. and Kotsiaros, S. and Oliversen, R. J. and Espley, J. R. and Joergensen, J. L. and Joergensen, P. S. and Merayo, J. M. G. and Herceg, M. and Bloxham, J. and Moore, K. M. and Bolton, S. J. and Levin, S. M.},
title = {A New Model of Jupiter's Magnetic Field From Juno's First Nine Orbits},
journal = {Geophys. Res. Lett.},
volume = {45},
number = {6},
pages = {2590-2596},
doi = {https://doi.org/10.1002/2018GL077312},
url = {https://agupubs.onlinelibrary.wiley.com/doi/abs/10.1002/2018GL077312},
eprint = {https://agupubs.onlinelibrary.wiley.com/doi/pdf/10.1002/2018GL077312},
year = {2018}
}

@article{saturn,
author = {Michele K. Dougherty  and Hao Cao  and Krishan K. Khurana  and Gregory J. Hunt  and Gabrielle Provan  and Stephen Kellock  and Marcia E. Burton  and Thomas A. Burk  and Emma J. Bunce  and Stanley W. H. Cowley  and Margaret G. Kivelson  and Christopher T. Russell  and David J. Southwood },
title = {Saturn’s magnetic field revealed by the Cassini Grand Finale},
journal = {Science},
volume = {362},
number = {6410},
pages = {eaat5434},
year = {2018},
doi = {10.1126/science.aat5434},
URL = {https://www.science.org/doi/abs/10.1126/science.aat5434},
eprint = {https://www.science.org/doi/pdf/10.1126/science.aat5434}
}

@article{Connerney2017,
  author    = {Connerney, J. E. P. and Benn, M. and Bjarno, J. B. and Denver, T. and Espley, J. and Jorgensen, J. L. and Jorgensen, P. S. and Lawton, P. and Malinnikova, A. and Merayo, J. M. and Murphy, S. and Odom, J. and Oliversen, R. and Schnurr, R. and Sheppard, D. and Smith, E. J.},
  title     = {The Juno Magnetic Field Investigation},
  journal   = {Space Sci. Rev.},
  year      = {2017},
  volume    = {213},
  number    = {1},
  pages     = {39--138},
  doi       = {10.1007/s11214-017-0334-z},
  url       = {https://doi.org/10.1007/s11214-017-0334-z},
  issn      = {1572-9672}
}

@article{Menou2012,
  author  = {Menou, Kristen},
  title   = {Magnetic Scaling Laws for the Atmospheres of Hot Giant Exoplanets},
  journal = {ApJ},
  year    = {2012},
  volume  = {745},
  number  = {2},
  pages   = {138},
  doi     = {10.1088/0004-637X/745/2/138}
}

@article{showman2009,
  author = {Showman, Adam P. and Fortney, Jonathan J. and Lian, Yuan and Marley, Mark S. and Freedman, Richard S. and Knutson, Heather A. and Charbonneau, David},
  title = {Atmospheric circulation of hot Jupiters: Coupled radiative-dynamical general circulation model simulations of HD 189733b and HD 209458b},
  journal = {ApJ},
  volume = {699},
  number = {1},
  pages = {564--584},
  year = {2009},
  doi = {10.1088/0004-637X/699/1/564}
}

@article{adcroft2004,
author = {Adcroft, Alistair and Hill, Chris and Campin, Jean-Michel and Marshall, John and Heimbach, Patrick},
year = {2004},
month = {01},
pages = {2882},
volume = {132},
doi = {10.1175/MWR2835.1},
title = {Overview of the formulation and numerics of the MIT GCM},
journal = {Monthly Weather Review},
url = {http://mitgcm.org/pdfs/ECMWF2004-Adcroft.pdf}
}

@article{challener2024latitudinalasymmetrydaysideatmosphere,
doi = {10.3847/2041-8213/ad5958},
url = {https://doi.org/10.3847/2041-8213/ad5958},
year = {2024},
month = {jul},
publisher = {The American Astronomical Society},
volume = {969},
number = {2},
pages = {L32},
author = {Challener, Ryan C. and Rustamkulov, Zafar and Lee, Elspeth K. H. and Lewis, Nikole and Sing, David K. and Birkmann, Stephan M. and Crouzet, Nicolas and Espinoza, Néstor and Manjavacas, Elena and Oliveros-Gomez, Natalia and Valenti, Jeff A. and Yang, Jingxuan},
title = {Latitudinal Asymmetry in the Dayside Atmosphere of WASP-43b},
journal = {ApJ},
}

@article{Tsai_2014,
   title={THREE-DIMENSIONAL STRUCTURES OF EQUATORIAL WAVES AND THE RESULTING SUPER-ROTATION IN THE ATMOSPHERE OF A TIDALLY LOCKED HOT JUPITER},
   volume={793},
   ISSN={1538-4357},
   url={http://dx.doi.org/10.1088/0004-637X/793/2/141},
   DOI={10.1088/0004-637x/793/2/141},
   number={2},
   journal={ApJ},
   publisher={American Astronomical Society},
   author={Tsai, Shang-Min and Dobbs-Dixon, Ian and Gu, Pin-Gao},
   year={2014},
   month=sep, pages={141} }

@article{Hammond_2018,
   title={Wave-mean Flow Interactions in the Atmospheric Circulation of Tidally Locked Planets},
   volume={869},
   ISSN={1538-4357},
   url={http://dx.doi.org/10.3847/1538-4357/aaec03},
   DOI={10.3847/1538-4357/aaec03},
   number={1},
   journal={ApJ},
   publisher={American Astronomical Society},
   author={Hammond, Mark and Pierrehumbert, Raymond T.},
   year={2018},
   month=dec, pages={65} }

@ARTICLE{W121b_params_2016,
       author = {{Delrez}, L. and {Santerne}, A. and {Almenara}, J.-M. and {Anderson}, D.~R. and {Collier-Cameron}, A. and {D{\'\i}az}, R.~F. and {Gillon}, M. and {Hellier}, C. and {Jehin}, E. and {Lendl}, M. and {Maxted}, P.~F.~L. and {Neveu-VanMalle}, M. and {Pepe}, F. and {Pollacco}, D. and {Queloz}, D. and {S{\'e}gransan}, D. and {Smalley}, B. and {Smith}, A.~M.~S. and {Triaud}, A.~H.~M.~J. and {Udry}, S. and {Van Grootel}, V. and {West}, R.~G.},
        title = "{WASP-121 b: a hot Jupiter close to tidal disruption transiting an active F star}",
      journal = {\mnras},
         year = 2016,
        month = jun,
       volume = {458},
       number = {4},
        pages = {4025-4043},
          doi = {10.1093/mnras/stw522},
archivePrefix = {arXiv},
       eprint = {1506.02471},
 primaryClass = {astro-ph.EP},
       adsurl = {https://ui.adsabs.harvard.edu/abs/2016MNRAS.458.4025D}
}

@ARTICLE{W121b_params_2020,
       author = {{Bourrier}, V. and {Ehrenreich}, D. and {Lendl}, M. and {Cretignier}, M. and {Allart}, R. and {Dumusque}, X. and {Cegla}, H.~M. and {Su{\'a}rez-Mascare{\~n}o}, A. and {Wyttenbach}, A. and {Hoeijmakers}, H.~J. and {Melo}, C. and {Kuntzer}, T. and {Astudillo-Defru}, N. and {Giles}, H. and {Heng}, K. and {Kitzmann}, D. and {Lavie}, B. and {Lovis}, C. and {Murgas}, F. and {Nascimbeni}, V. and {Pepe}, F. and {Pino}, L. and {Segransan}, D. and {Udry}, S.},
        title = "{Hot Exoplanet Atmospheres Resolved with Transit Spectroscopy (HEARTS). III. Atmospheric structure of the misaligned ultra-hot Jupiter WASP-121b}",
      journal = {\aap},
         year = 2020,
        month = mar,
       volume = {635},
          eid = {A205},
        pages = {A205},
          doi = {10.1051/0004-6361/201936640},
archivePrefix = {arXiv},
       eprint = {2001.06836},
 primaryClass = {astro-ph.EP},
       adsurl = {https://ui.adsabs.harvard.edu/abs/2020A&A...635A.205B}
}

@article{May_2022,
   title={A New Analysis of Eight Spitzer Phase Curves and Hot Jupiter Population Trends: Qatar-1b, Qatar-2b, WASP-52b, WASP-34b, and WASP-140b},
   volume={163},
   ISSN={1538-3881},
   url={http://dx.doi.org/10.3847/1538-3881/ac6261},
   DOI={10.3847/1538-3881/ac6261},
   number={6},
   journal={AJ},
   publisher={American Astronomical Society},
   author={May, E. M. and Stevenson, K. B. and Bean, Jacob L. and Bell, Taylor J. and Cowan, Nicolas B. and Dang, Lisa and Desert, Jean-Michel and Fortney, Jonathan J. and Keating, Dylan and Kempton, Eliza M.-R. and Komacek, Thaddeus D. and Lewis, Nikole K. and Mansfield, Megan and Morley, Caroline and Parmentier, Vivien and Rauscher, Emily and Swain, Mark R. and Zellem, Robert T. and Showman, Adam},
   year={2022},
   month=may, pages={256} }

@article{Bell_2021,
   title={A comprehensive reanalysis of Spitzer’s 4.5 μm phase curves, and the phase variations of the ultra-hot Jupiters MASCARA-1b and KELT-16b},
   volume={504},
   ISSN={1365-2966},
   url={http://dx.doi.org/10.1093/mnras/stab1027},
   DOI={10.1093/mnras/stab1027},
   number={3},
   journal={MNRAS},
   publisher={Oxford University Press (OUP)},
   author={Bell, Taylor J and Dang, Lisa and Cowan, Nicolas B and Bean, Jacob and Désert, Jean-Michel and Fortney, Jonathan J and Keating, Dylan and Kempton, Eliza and Kreidberg, Laura and Line, Michael R and Mansfield, Megan and Parmentier, Vivien and Stevenson, Kevin B and Swain, Mark and Zellem, Robert T},
   year={2021},
   month=apr, pages={3316–3337} }

@article{hammond2024twodimensionaleclipsemappinghot,
doi = {10.3847/1538-3881/ad434d},
url = {https://doi.org/10.3847/1538-3881/ad434d},
year = {2024},
month = {jun},
publisher = {The American Astronomical Society},
volume = {168},
number = {1},
pages = {4},
author = {Hammond, Mark and Bell, Taylor J. and Challener, Ryan C. and Lewis, Neil T. and Weiner Mansfield, Megan and Malsky, Isaac and Rauscher, Emily and Bean, Jacob L. and Carone, Ludmila and Mendonça, João M. and Teinturier, Lucas and Tan, Xianyu and Crouzet, Nicolas and Kreidberg, Laura and Morello, Giuseppe and Parmentier, Vivien and Blecic, Jasmina and Désert, Jean-Michel and Helling, Christiane and Lagage, Pierre-Olivier and Molaverdikhani, Karan and Nixon, Matthew C. and Rackham, Benjamin V. and Yang, Jingxuan},
title = {Two-dimensional Eclipse Mapping of the Hot-Jupiter WASP-43b with JWST MIRI/LRS},
journal = {AJ},
}

@article{Rogers_2014,
   title={MAGNETIC EFFECTS IN HOT JUPITER ATMOSPHERES},
   volume={794},
   ISSN={1538-4357},
   url={http://dx.doi.org/10.1088/0004-637X/794/2/132},
   DOI={10.1088/0004-637x/794/2/132},
   number={2},
   journal={ApJ},
   publisher={American Astronomical Society},
   author={Rogers, T. M. and Komacek, T. D.},
   year={2014},
   month=oct, pages={132} }

@article{Rogers:2014aa,
	author = {T.M. Rogers and A.P. Showman},
	doi = {10.1088/2041-8205/782/1/L4},
	journal = {ApJ},
	pages = {L4},
	title = {Magnetohydrodynamic simulations of the atmosphere of {HD}209458b},
	volume = {782},
	year = {2014}}

@article{christie2025geometricconsiderationshotjupiter,
    author = {Christie, D A and Evans-Soma, T M and Mayne, N J and Kohary, K},
    title = {Geometric considerations in hot Jupiter magnetic drag models},
    journal = {MNRAS},
    volume = {541},
    number = {3},
    pages = {2773-2789},
    year = {2025},
    month = {08},
    issn = {0035-8711},
    doi = {10.1093/mnras/staf1146},
    url = {https://doi.org/10.1093/mnras/staf1146},
    eprint = {https://academic.oup.com/mnras/article-pdf/541/3/2773/63734443/staf1146.pdf},
}

@article{WASP121b_Splinter_2025,
   title={Precise Constraints on the Energy Budget of WASP-121 b from Its JWST NIRISS/SOSS Phase Curve},
   volume={170},
   ISSN={1538-3881},
   url={http://dx.doi.org/10.3847/1538-3881/ae0e52},
   DOI={10.3847/1538-3881/ae0e52},
   number={6},
   journal={AJ},
   publisher={American Astronomical Society},
   author={Splinter, Jared and Coulombe, Louis-Philippe and Frazier, Robert C. and Cowan, Nicolas B. and Rauscher, Emily and Dang, Lisa and Radica, Michael and Collins, Sean and Pelletier, Stefan and Allart, Romain and MacDonald, Ryan J. and Lafrenière, David and Loïc Albert and Benneke, Björn and Doyon, René and Jayawardhana, Ray and Johnstone, Doug and Krishnamurthy, Vigneshwaran and Piaulet-Ghorayeb, Caroline and Kaltenegger, Lisa and Meyer, Michael R. and Taylor, Jake and Turner, Jake D.},
   year={2025},
   month=nov, pages={323} }

@article{Heng_2012,
   title={THE INFLUENCE OF ATMOSPHERIC SCATTERING AND ABSORPTION ON OHMIC DISSIPATION IN HOT JUPITERS},
   volume={748},
   ISSN={2041-8213},
   url={http://dx.doi.org/10.1088/2041-8205/748/1/L17},
   DOI={10.1088/2041-8205/748/1/l17},
   number={1},
   journal={ApJ},
   publisher={American Astronomical Society},
   author={Heng, Kevin},
   year={2012},
   month=mar, pages={L17} }

@article{Keating_2020,
doi = {10.3847/1538-3881/ab83f4},
url = {https://doi.org/10.3847/1538-3881/ab83f4},
year = {2020},
month = {apr},
publisher = {The American Astronomical Society},
volume = {159},
number = {5},
pages = {225},
author = {Keating, Dylan and Stevenson, Kevin B. and Cowan, Nicolas B. and Rauscher, Emily and Bean, Jacob L. and Bell, Taylor and Dang, Lisa and Deming, Drake and Désert, Jean-Michel and Feng, Y. Katherina and Fortney, Jonathan J. and Kataria, Tiffany and Kempton, Eliza M.-R. and Lewis, Nikole and Line, Michael R. and Mansfield, Megan and May, Erin and Morley, Caroline and Showman, Adam P.},
title = {Smaller than Expected Bright-spot Offsets in Spitzer Phase Curves of the Hot Jupiter Qatar-1b},
journal = {AJ},
}

@article{Hindle_2021,
   title={The Magnetic Mechanism for Hotspot Reversals in Hot Jupiter Atmospheres},
   volume={922},
   ISSN={1538-4357},
   url={http://dx.doi.org/10.3847/1538-4357/ac0e2e},
   DOI={10.3847/1538-4357/ac0e2e},
   number={2},
   journal={ApJ},
   publisher={American Astronomical Society},
   author={Hindle, A. W. and Bushby, P. J. and Rogers, T. M.},
   year={2021},
   month=nov, pages={176} }

@article{Boening2025,
  author       = {Böning, Vincent G. A. and Dietrich, Wieland and Wicht, Johannes},
  title        = {Reduced or westward hotspot offset explained by dynamo action in atmospheres of ultrahot Jupiters},
  journal      = {A&A},
  volume       = {699},
  pages        = {A339},
  year         = {2025},
  doi          = {10.1051/0004-6361/202553695},
  url          = {https://doi.org/10.1051/0004-6361/202553695}
}

@article{Dietrich2022,
    author = {Dietrich, Wieland and Kumar, Sandeep and Poser, Anna Julia and French, Martin and Nettelmann, Nadine and Redmer, Ronald and Wicht, Johannes},
    title = {Magnetic induction processes in hot Jupiters, application to KELT-9b},
    journal = {MNRAS},
    volume = {517},
    number = {3},
    pages = {3113-3125},
    year = {2022},
    month = {10},
    issn = {0035-8711},
    doi = {10.1093/mnras/stac2849},
    url = {https://doi.org/10.1093/mnras/stac2849},
    eprint = {https://academic.oup.com/mnras/article-pdf/517/3/3113/46596021/stac2849.pdf}
}

@article{HJ_radanom_mechanisms,
	author = {{Sarkis, P.} and {Mordasini, C.} and {Henning, Th.} and {Marleau, G. D.} and {Mollière, P.}},
	title = {Evidence of three mechanisms explaining the radius anomaly of hot Jupiters},
	DOI= "10.1051/0004-6361/202038361",
	url= "https://doi.org/10.1051/0004-6361/202038361",
	journal = {A&A},
	year = 2021,
	volume = 645,
	pages = "A79",
}

@article{bagheri2024freshlookinteractionexoplanets,
      title={A Fresh Look into the Interaction of Exoplanets Magnetosphere with Stellar Winds using MHD Simulations}, 
      author={Fatemeh Bagheri and Ramon E. Lopez and Kevin Pham},
      year={2024},
      eprint={2404.14377},
      journal = {Front. Astron. Space Sci.},
      volume = {11},
      pages = {1398379},
      archivePrefix={arXiv},
      primaryClass={astro-ph.EP},
      url={https://arxiv.org/abs/2404.14377}, 
      doi = {10.3389/fspas.2024.1398379}
}

@article{Bell_2018,
   title={Increased Heat Transport in Ultra-hot Jupiter Atmospheres through H2 Dissociation and Recombination},
   volume={857},
   ISSN={2041-8213},
   url={http://dx.doi.org/10.3847/2041-8213/aabcc8},
   DOI={10.3847/2041-8213/aabcc8},
   number={2},
   journal={ApJ},
   publisher={American Astronomical Society},
   author={Bell, Taylor J. and Cowan, Nicolas B.},
   year={2018},
   month=apr, pages={L20} }

@book{plasma_book,
    author = "J. A. Bittencourt",
    title = "Fundamentals of Plasma Physics",
    publisher = "Springer New York, NY",
    year = "2004",
    edition = "3",
    doi = {10.1007/978-1-4757-4030-1}
}

@article{Ballester2018Plasmas,
author = {Ballester, Jose and Alexeev, Igor and Collados, Manuel and Downes, Turlough and Pfaff, Robert and Gilbert, Holly and Khodachenko, Maxim and Khomenko, Elena and Shaikhislamov, Ildar and Soler, Roberto and Vazquez-Semadeni, Enrique and Zaqarashvili, T.},
year = {2018},
month = {03},
pages = {58},
title = {Partially Ionized Plasmas in Astrophysics},
volume = {214},
journal = {Space Sci. Rev.},
doi = {10.1007/s11214-018-0485-6}
}

@book{rudiger2006magnetic,
  title={The Magnetic Universe: Geophysical and Astrophysical Dynamo Theory},
  author={R{\"u}diger, G. and Hollerbach, R.},
  isbn={9783527605002},
  url={https://books.google.co.uk/books?id=akxiW6ELAngC},
  year={2006},
  publisher={Wiley}
}

@book{Davidson_2016, place={Cambridge}, edition={2}, series={Cambridge Texts in Applied Mathematics}, title={Introduction to Magnetohydrodynamics}, publisher={Cambridge University Press}, author={Davidson, P. A.}, year={2016}, collection={Cambridge Texts in Applied Mathematics}, doi = {10.1119/1.1482065}}

@article{wicht2010,
  author    = {Johannes Wicht and Andreas Tilgner},
  title     = {Theory and Modeling of Planetary Dynamos},
  journal   = {Space Sci. Rev.},
  volume    = {152},
  number    = {1-4},
  pages     = {501--542},
  year      = {2010},
  doi       = {10.1007/s11214-010-9638-y},
  url       = {https://doi.org/10.1007/s11214-010-9638-y}
}

@article{Dynamo_review,
   title={Dynamo theories},
   volume={85},
   ISSN={1469-7807},
   url={http://dx.doi.org/10.1017/S0022377819000539},
   DOI={10.1017/s0022377819000539},
   number={4},
   journal={Journal of Plasma Physics},
   publisher={Cambridge University Press (CUP)},
   author={Rincon, François},
   year={2019},
   pages = {205850401},
   month=aug }

@inbook{energy_density_txtbk, place={Cambridge}, title={Principles of Magnetohydrodynamics: With Applications to Laboratory and Astrophysical Plasmas}, publisher={Cambridge University Press}, author={Goedbloed, J. P. Hans and Poedts, Stefaan}, year={2004}, pages={131–185}, doi = {10.1017/S0022112004001466}}

@ARTICLE{2022ELeegcmcrt,
doi = {10.3847/1538-4357/ac61d6},
url = {https://doi.org/10.3847/1538-4357/ac61d6},
year = {2022},
month = {apr},
publisher = {The American Astronomical Society},
volume = {929},
number = {2},
pages = {180},
author = {Lee, Elspeth K. H. and Wardenier, Joost P. and Prinoth, Bibiana and Parmentier, Vivien and Grimm, Simon L. and Baeyens, Robin and Carone, Ludmila and Christie, Duncan and Deitrick, Russell and Kitzmann, Daniel and Mayne, Nathan and Roman, Michael and Thorsbro, Brian},
title = {3D Radiative Transfer for Exoplanet Atmospheres. gCMCRT: A GPU-accelerated MCRT Code},
journal = {ApJ},
}

@ARTICLE{2024rodrigofilterthroughput,
       author = {{Rodrigo}, Carlos and {Cruz}, Patricia and {Aguilar}, John F. and {Aller}, Alba and {Solano}, Enrique and {G{\'a}lvez-Ortiz}, Maria Cruz and {Jim{\'e}nez-Esteban}, Francisco and {Mas-Buitrago}, Pedro and {Bayo}, Amelia and {Cort{\'e}s-Contreras}, Miriam and {Murillo-Ojeda}, Raquel and {Bonoli}, Silvia and {Cenarro}, Javier and {Dupke}, Renato and {L{\'o}pez-Sanjuan}, Carlos and {Mar{\'\i}n-Franch}, Antonio and {de Oliveira}, Claudia Mendes and {Moles}, Mariano and {Taylor}, Keith and {Varela}, Jes{\'u}s and {Rami{\'o}}, H{\'e}ctor V{\'a}zquez},
        title = "{Photometric segregation of dwarf and giant FGK stars using the SVO Filter Profile Service and photometric tools}",
      journal = {\aap},
         year = 2024,
        month = sep,
       volume = {689},
          eid = {A93},
        pages = {A93},
          doi = {10.1051/0004-6361/202449998},
archivePrefix = {arXiv},
       eprint = {2406.03310},
 primaryClass = {astro-ph.SR},
       adsurl = {https://ui.adsabs.harvard.edu/abs/2024A&A...689A..93R}
}




\appendix

\section{Supplemental numerics}

\subsection{Topology of the \texttt{MITgcm}}
\label{app:topology}

The \texttt{MITgcm} uses a cube-sphere topology within each pressure level, and the pressure grid consists of these concentric cube-sphere meshes stacked on top of each other. A cube-sphere is a spherical mesh that is sub-divided by straight lines into six faces, and then each face is sub-divided further into similar-sized similar-shaped square-like tiles. The motivation in using this topology rather than, say, a normal spherical polar grid system is that each tile remains roughly the same size and shape, as opposed to spherical polar tiles which would consist of larger square parcels near the equator, and small triangular slices near the poles. A cube-sphere grid also avoids the singularities at each pole of a spherical-polar grid, allowing for better modelling of polar regions.

Within the \texttt{MITgcm}, the velocities are defined at the interface of two tiles, whereas the temperatures, densities, diffusivity, the specific gas constants etc. are calculated at the centre of the tile. As a result, when calculating \(\tau_\mathrm{mag}\) in the momentum equation, it is necessary to take the average calculated at the centre of the tile to the left and right of the boundary in question before applying it as a drag term to the velocity at the boundary.

The local $x$ and $y$ velocities are an artefact of the cube-sphere grid and are not the actual meridional and zonal coordinates. For our zeroth order simplified magnetic forcing model, the force is parallel to the velocity (equation (\ref{equ:GCM_force})), but this is not generally true. Were we to adopt a different approximation, the velocities and force would need to be rotated into the cube-sphere coordinate system to properly calculate the force in the $x$ and $y$ directions.

The longitudinal and latitudinal change between the centres of the tiles is stored in the matrix variables \(\Delta x_c,\, \Delta y_c\) (in radians). The longitude and latitude locations are given \((x_c,\,y_c)\).  Therefore the proper coordinate transform to the \(\hat{i}\) direction pointing from one tile to an adjacent one is:

\begin{equation}
\label{equ:varichange}
    \hat{i}=\frac{\Delta x_c\cos(y_c)\cdot\hat{x}+\Delta y_c\cdot \hat{y}}{((\Delta x_c\cos(y_c))^2+(\Delta y_c)^2)^{1/2}}
\end{equation}

\(\hat{x}\) is the unit vector in the zonal direction, and \(\hat{y}\) is the unit vector in the meridional direction. \(\hat{j}\) could be similarly found by considering the tiles in the other direction.


\subsection{Varying the minimum magnetic drag timescale}

\label{app:min_timescale}
\begin{table}
\centering
\begin{tabular}{|c|cc|}
\hline
 Name & Magnetic field strength & Minimum drag timescale\\
 \hline
 Baseline & 0G & ---\\
 Stronger field & 3G & 1000s\\
 Small min timescale & 3G & 100s\\
 Large min timescale & 3G & $10^5$ s\\
 Constant drag & --- & $10^4$ s\\
 \hline
\end{tabular}
\caption{
\label{tab:B-field_params_all}Non-tilted magnetic field run parameters. The constant drag case implements a constant drag timescale of $10^4$ s everywhere.}
\end{table}

\begin{figure*}
\vspace{0cm}
  \centering
  \makebox[\textwidth][c]{%
    \includegraphics[
        width=\textwidth,
        trim=5.1cm 12cm 5.4cm 22.5cm,   
        clip
    ]{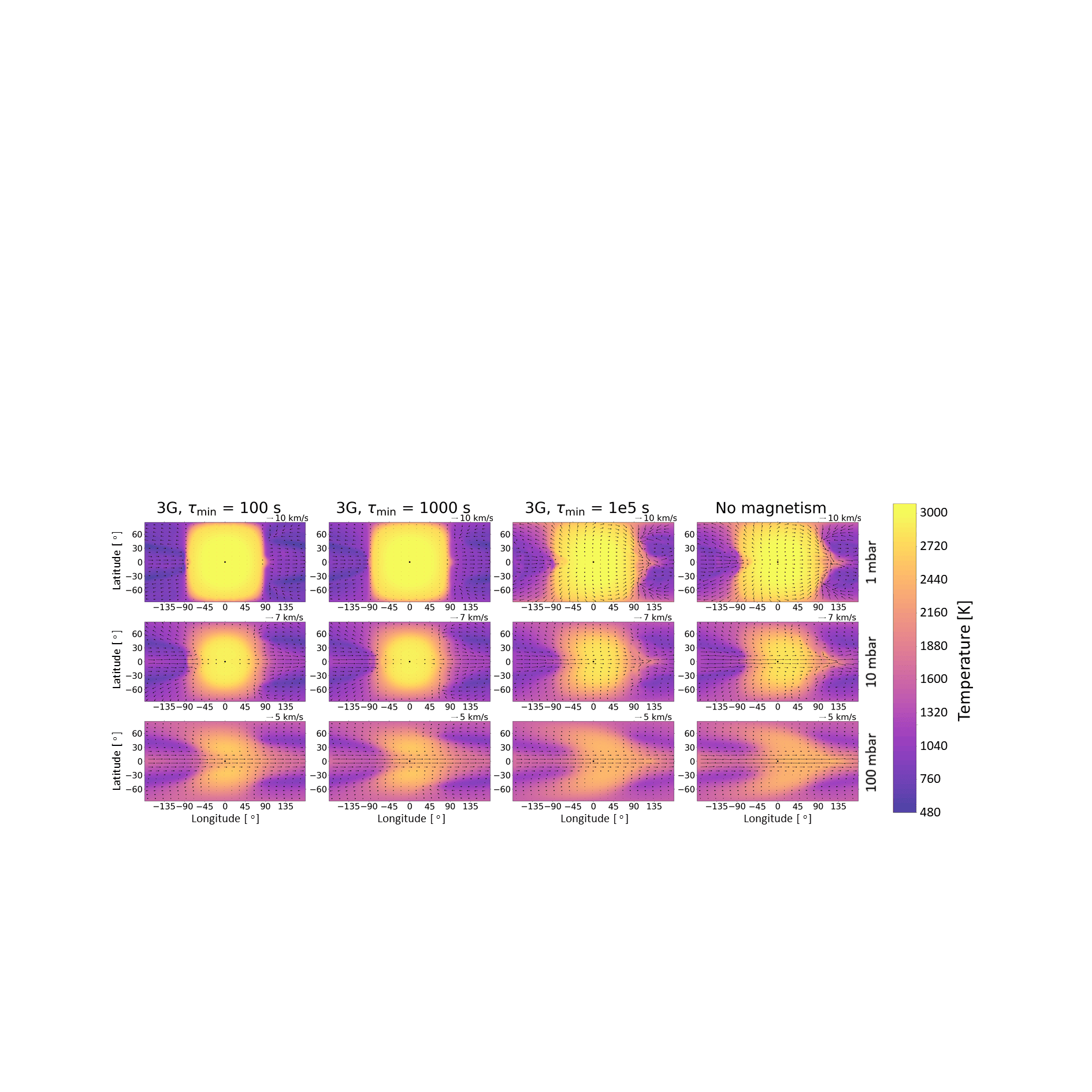}}
  \caption{\label{fig:min_drag_constants} Temperature and velocity field maps for non-tilted 3G magnetic field runs (left 3 plots) and non-magnetic field run (right). The 3G runs each have a different minimum drag timescale ($\tau_{\mathrm{min}}$ = 100s, 1000s, 1e5s). As $\tau_{\mathrm{min}}$ is increased, the temperature and velocity profiles of the modelled atmosphere become more similar to the non-magnetism run.}
\end{figure*}

In Section \ref{sec:methods_algorithm}, we briefly discussed the numerical reason to set a minimum drag timescale in order to avoid numerical instabilities that crop up due to discrete timestepping. A minimum drag timescale should be chosen such that it is large enough to avoid instability, but small enough to accurately portray the effects of the magnetic drag. In Fig. \ref{fig:min_drag_constants}, we plot the temperature and velocity maps for 3G non-tilted cases with different minimum drag timescales.

The difference between the 100 s and the 1000 s minimum drag timescale runs is quite minimal near the photosphere (100 mbars), and most significant in the upper atmosphere. The winds are more strongly damped on the daysides of the upper atmosphere of the 100 s case, whereas the { $10^5$} s case looks quite similar to the no-magnetism baseline runs. This is because at these longer timescales, the minimum magnetic drag timescale is approximately the order of magnitude of the advective timescale. There will always be errors added due to the minimum drag timescale, and they should be kept in mind when investigating the qualitative behaviour.

\section{Extra figures}

\begin{figure}
  \centering
  \includegraphics[width=0.6\linewidth]{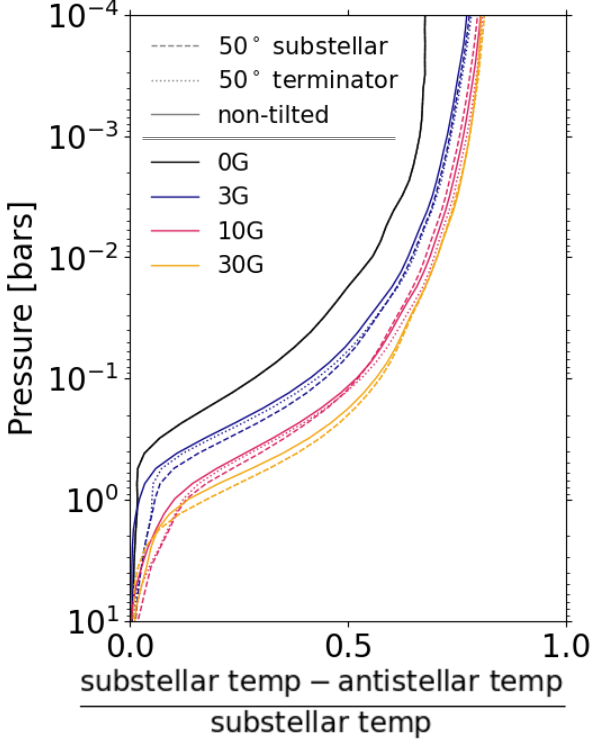}
  \caption{\label{fig:TP_diff_daynight_tilted} Substellar-antistellar temperature contrast normalized by the substellar temperature as a function of pressure. Magnetic field strength is distinguished by line colour, with the different lines of the same colour representing different tilts. This profile is most strongly dependent on magnetic field strength, and not significantly impacted by the orientation. }
\end{figure}

\begin{figure}[H]
  \centering
  \includegraphics[width=0.8\linewidth]{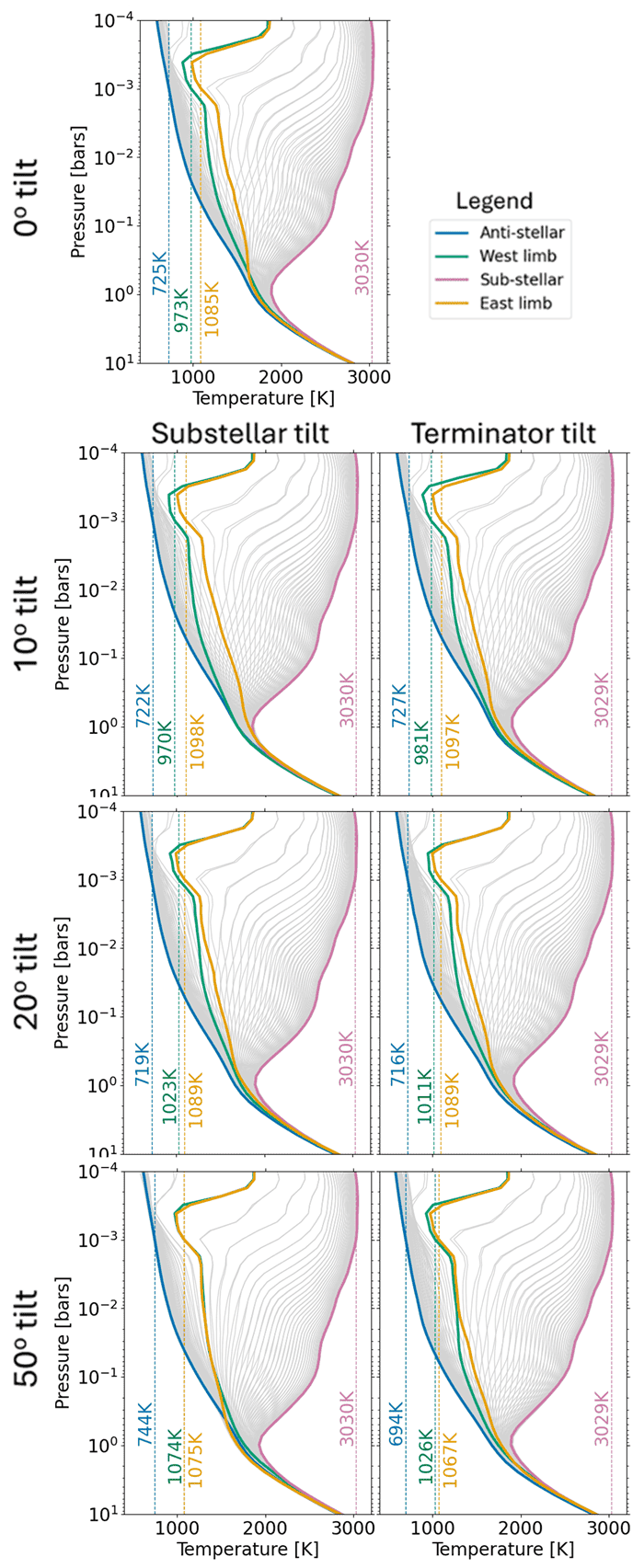}
  \caption{\label{fig:TP_10G_extra} Temperature vs. pressure profile for runs with magnetic field strengths of {10}G of a variety of orientations.}
  \vspace{20pt}
\end{figure}

\begin{figure*}
\vspace*{0cm}
  \centering
  \makebox[\textwidth][c]{%
    \includegraphics[width=1.1\textwidth]{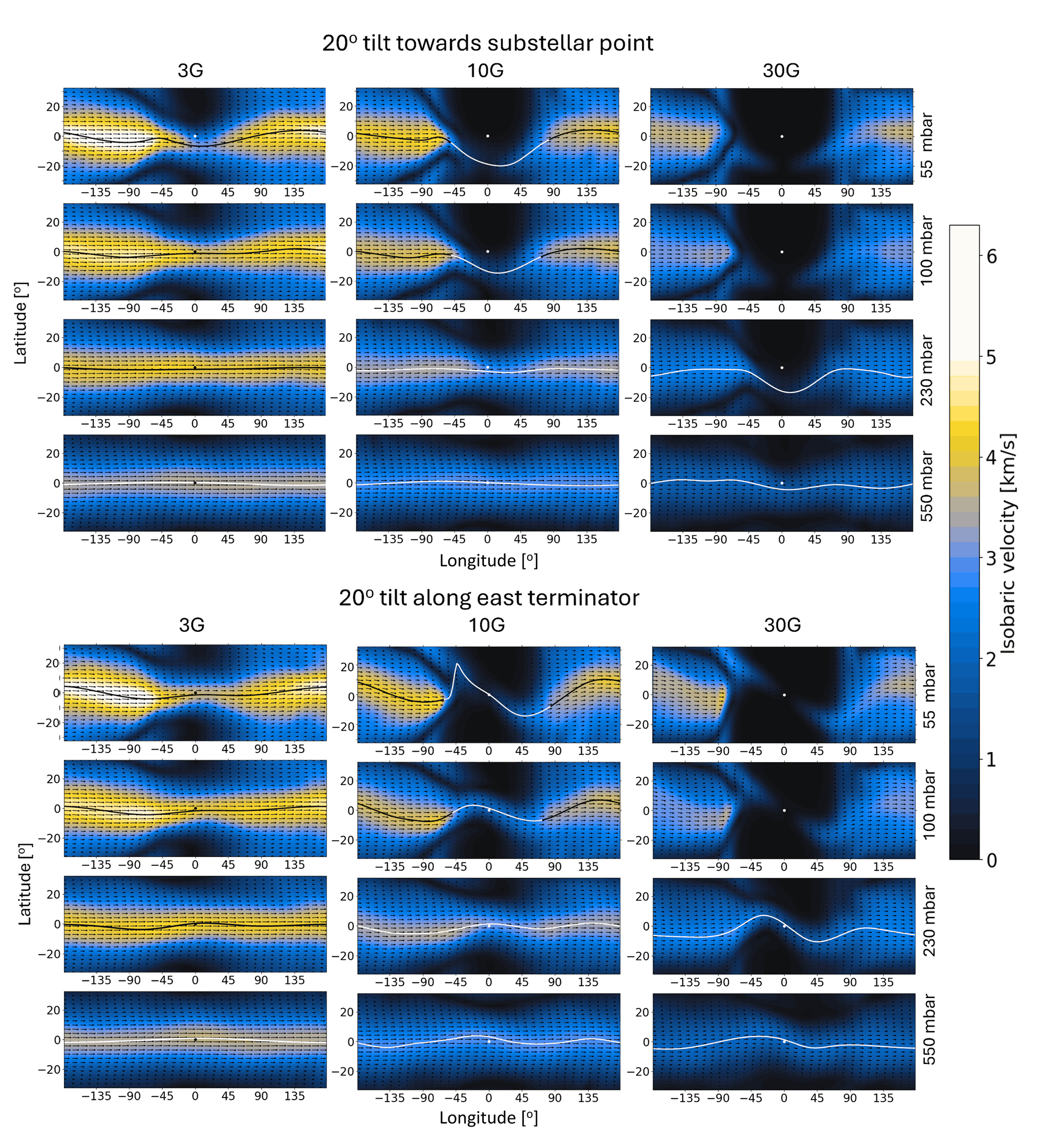}}
  \caption{\label{fig:vel_defl_20deg} Atmospheric velocity on isobaric surfaces for different runs with magnetic dipoles oriented at 20\(^\circ\) tilts. The velocity quivers (overlaid in black) show the direction of the winds, and the plotted line is the location of the latitude where this velocity is greatest at a given pressure level and longitude.}
\end{figure*}

\begin{figure*}
\vspace*{-1cm}
\makebox[\textwidth][c]{%
    \includegraphics[
        width=.95\textwidth,
        trim=0cm 0cm 0cm 0cm,   
        clip
    ]{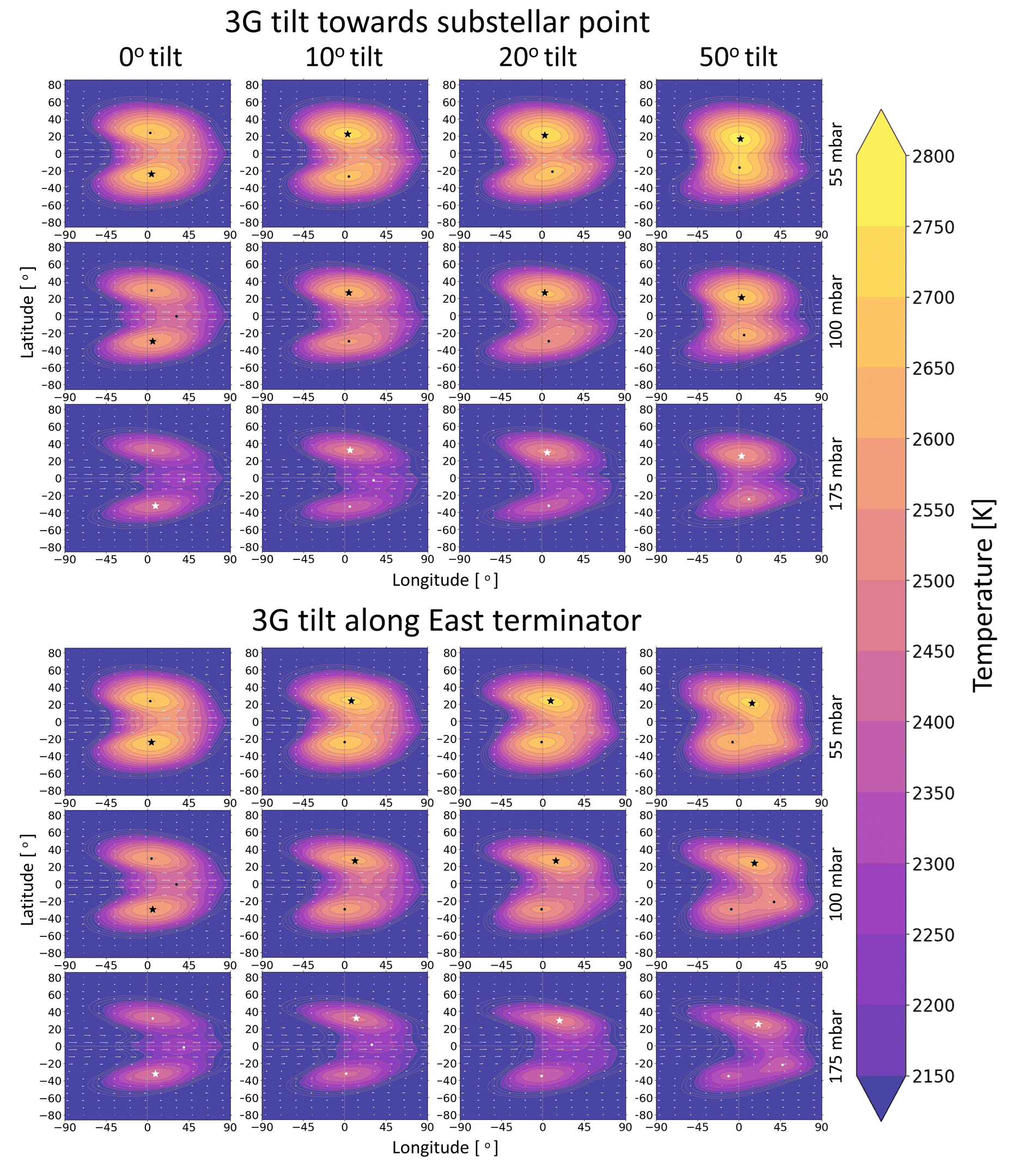}}
\vspace*{-0.5cm}
  \caption{\label{fig:hot_3G_big} High-contrast temperature contour maps for simulations with 3G magnetic field strength at various dipole orientations. Multiple pressure levels are shown between 50-300mbar, with the isobaric global maxima indicated by the star, and local maxima by smaller dots. The top half displays runs with dipole tilts directed towards the substellar point, and the bottom half shows runs with dipole tilts along the east terminator. With the inclusion of a dipole tilt, the hotspot location can move latitudinally for some combinations of field strength and tilts.}
\end{figure*}

\begin{figure*}
\vspace*{-1cm}
\makebox[\textwidth][c]{%
    \includegraphics[
        width=.95\textwidth,
        trim=4cm 10.5cm 19cm 3.5cm,   
        clip
    ]{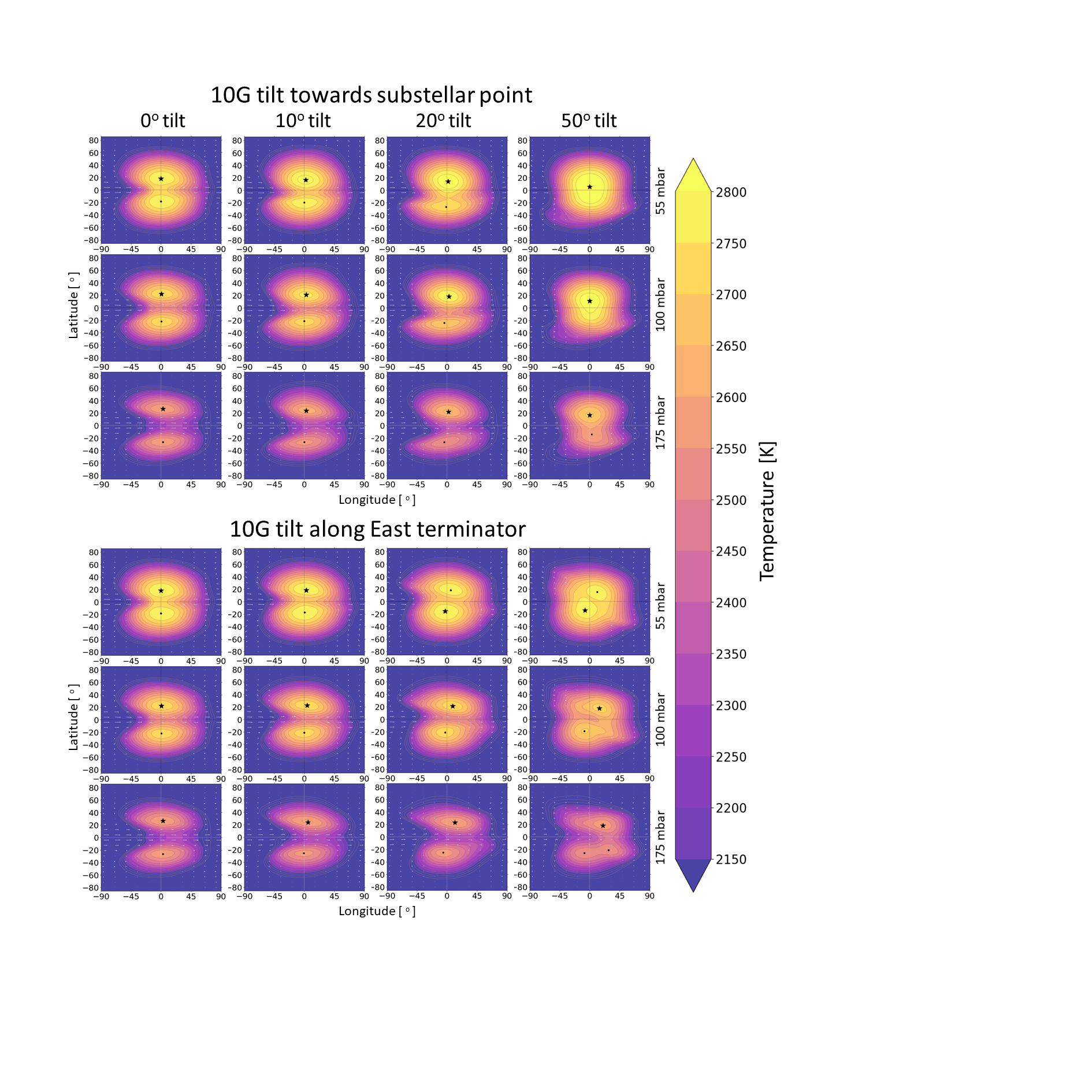}}
\vspace*{-0.5cm}
\caption{\label{fig:hot_10G_big} High-contrast temperature contour maps for simulations with 10G magnetic field strength at various dipole orientations. Multiple pressure levels are shown between 50-300mbar, with the isobaric global maxima indicated by the star, and local maxima by smaller dots. The top half displays runs with dipole tilts directed towards the substellar point, and the bottom half shows runs with dipole tilts along the east terminator. With the inclusion of a dipole tilt, the hotspot location can move latitudinally for some combinations of field strength and tilts.}
\end{figure*}

Here we include several additional results from our simulations for completeness.  Fig. \ref{fig:TP_diff_daynight_tilted} shows the day-night temperature contrast for our tilted dipole models. {Figure \ref{fig:TP_10G_extra} shows the temperature-pressure profiles from our 10G models}. Fig. \ref{fig:vel_defl_20deg} is similar to Fig. \ref{fig:vel_defl_10G}, but for a tilt strength of 20$^{\circ}$. Figs \ref{fig:hot_3G_big} and \ref{fig:hot_10G_big} show high contrast temperature {contour} plots for our 3G and 10G models respectively. 
\label{app:figs}

\bsp	
\label{lastpage}
\end{document}